\documentclass[draft,final,oldfontcommands]{vutinfth} 

\usepackage{lmodern}        
\usepackage[T1]{fontenc}    
\usepackage[utf8]{inputenc} 
\usepackage[square,comma,numbers,sort&compress]{natbib}
\usepackage{paralist}

\usepackage{floatrow}
\DeclareOldFontCommand{\bf}{\normalfont\bfseries}{\mathbf}
\usepackage{adjustbox}
\usepackage{color,soul}
\usepackage{bibentry}
\nobibliography*
\usepackage{afterpage}

\usepackage[T1]{fontenc}
\usepackage[utf8]{inputenc}
\usepackage{microtype}
\usepackage{libertine}
\usepackage{epigraph}
\renewcommand{\added}[1]{\textcolor{black}{#1}}

\definecolor{lightblue}{rgb}{.80,.95,1}
\sethlcolor{lightblue}

\usepackage[usenames,dvipsnames,table]{xcolor} 

\usepackage{amsmath}    
\usepackage{algorithmic}
\usepackage{verbatim}
\usepackage{amsfonts}
\usepackage{amssymb}
\usepackage{array,multirow}
\usepackage{booktabs}
\usepackage{graphicx}
\usepackage{todonotes}
\presetkeys{todonotes}{inline}{}
\usepackage{listings}
\usepackage{amsthm}
\usepackage{amssymb}    
\usepackage{mathtools}  
\usepackage[inline]{enumitem} 
\usepackage{multirow}   
\usepackage{booktabs}   
\usepackage{subcaption} 
\usepackage[ruled,linesnumbered,algochapter]{algorithm2e} 
\usepackage{incgraph,tikz}

\usepackage{nag}       
\usepackage{todonotes} 
\usepackage{hyperref}  
\usepackage{mdframed}
\usepackage{lscape}

\usepackage{url}
\usepackage[skip=0pt]{caption}
\usepackage{verbatim}
\usepackage{amsfonts}
\usepackage{array}
\usepackage{graphicx}
\usepackage{listings}
\usepackage{textcomp}

\newmdtheoremenv{definition}{Definition}[chapter]

\newcommand{\commentA}[2][.5\linewidth]{%
  \leavevmode\hfill\makebox[#1][l]{$\triangleright$~#2}}

\usepackage{amsfonts}
\usepackage{amssymb}

\newcommand{\authorname}{Svitlana Vakulenko} 
\newcommand{\thesistitle}{Knowledge-based\break Conversational Search}
\newcommand{\smalltt}[1]{\texttt{\small #1}}

\makeatletter
\def\blfootnote{\xdef\@thefnmark{}\@footnotetext}
\makeatother

\hypersetup{
    pdfpagelayout   = TwoPageRight,           
    linkbordercolor = {Melon},                
    pdfauthor       = {\authorname},          
    pdftitle        = {\thesistitle},         
    pdfsubject      = {Subject},              
    pdfkeywords     = {a, list, of, keywords} 
}

\setpnumwidth{2.5em}        
\setsecnumdepth{subsection} 

\nonzeroparskip             
\setauthor{}{\authorname}{MSc}{female}
\setadvisor{Prof.\ dr.}{Axel Polleres}{}{male}
\setsecondadvisor{Prof.\ dr.}{Maarten de Rijke}{}{male}

\setfirstreviewer{Prof.\ dr.}{Mark Sanderson}{}{male}
\setsecondreviewer{Prof.\ dr.}{Jens Lehmann}{}{male}

\setaddress{Address}
\setregnumber{1525172}
\setdate{18}{11}{2019} 
\settitle{\thesistitle}{Wissensbasierte Konversationssuche}

\setthesis{doctor}
\setdoctordegree{techn.}
\setfirstreviewerdata{Affiliation, Country}
\setsecondreviewerdata{Affiliation, Country}

\newcommand{\OurApproach}{QAmp}

\begin{document}

\incgraph[documentpaper]
  [width=\paperwidth,height=\paperheight]{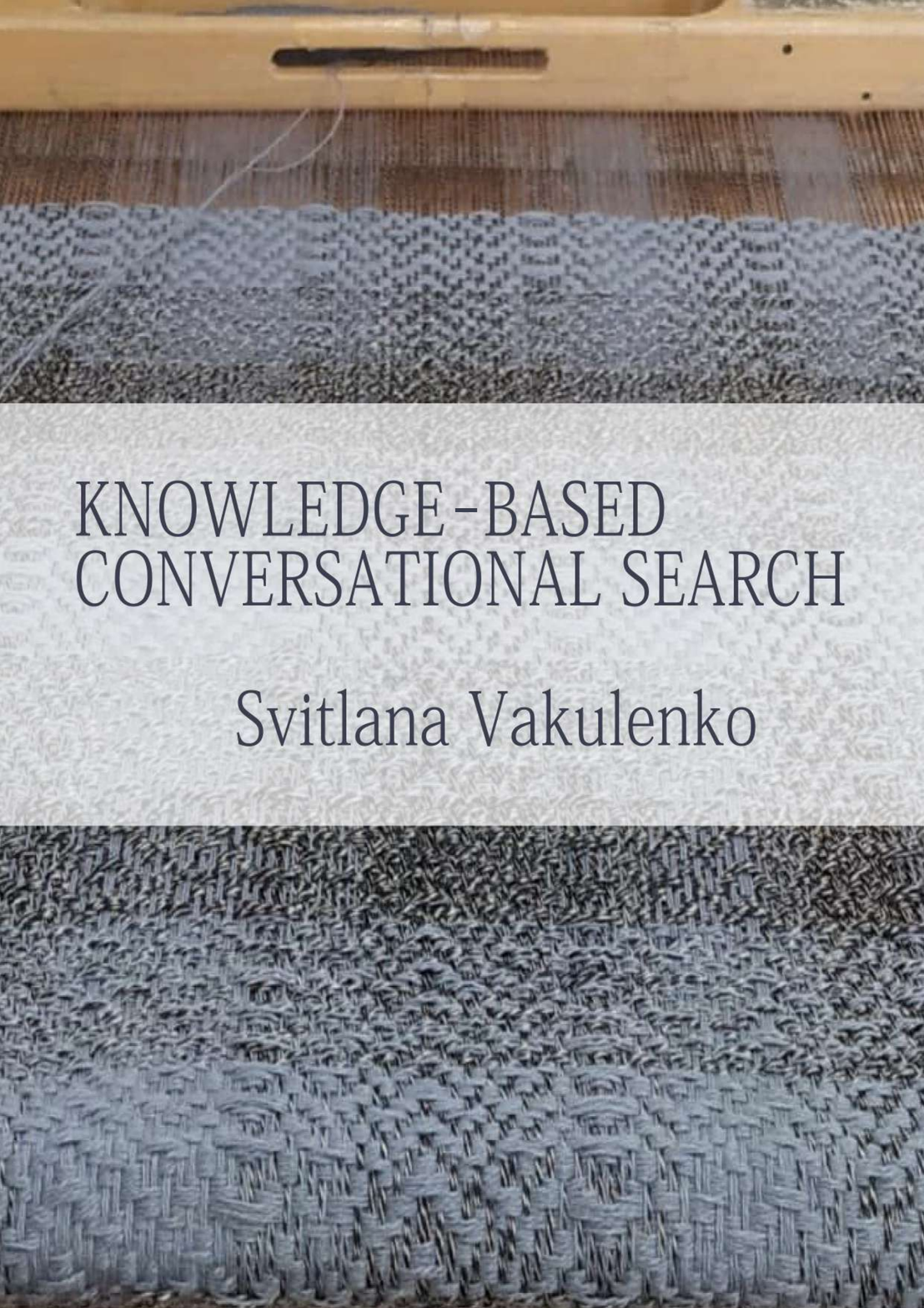}

\frontmatter 
\addtitlepage{english} 

\noindent%
\textbf{Evaluation committee} \\\\
\begin{tabular}{@{}l l l}
Advisors: \\
& Prof.\ dr.\ A.\ Polleres & TU Wien \\  
& Prof.\ dr.\ M.\ de Rijke & University of Amsterdam \\  
Other members: \\
& Prof.\ dr.\ H.\ Tompits & TU Wien \\ 
& Prof.\ dr.\ M. Schedl\  & Johannes Kepler University Linz \\  
& Prof.\ dr.\ M.\ Sanderson & RMIT University \\  
& Prof.\ dr.\ J.\ Lehmann & University of Bonn \\  
\end{tabular}

\bigskip\noindent%
Faculty of Informatics\\

\vfill

\noindent
The research was supported by the Austrian Research Promotion Agency (FFG) under project number 855407 and the EU H2020 programme under the MSCA-RISE agreement 645751 ({RISE\_BPM}).  \\

\bigskip

\noindent
Copyright \copyright~2019 Svitlana Vakulenko, Vienna, Austria\\
Cover by Kami Brandst\"atter




\clearpage

\addstatementpage

\begin{acknowledgements*}
Doing a PhD is like going up a glacier: attempting to make big steps while exploring an unfamiliar landscape in an unpredictable environment, trying to catch up in rhythm with the community pacing on the horizon. You need to choose your way wisely. Walking on ice requires learning a technique but most important are the people who are next to you when you are out there, those people you can trust, who stand still and make sure to catch you every time you fall. They are also the ones who give you the energy and motivation to keep going forward.

There was a person who gave me a chance to learn and the first ticket to academia.
Marlon was the reason I considered starting a PhD in the fist place because he said that I can. It was not obvious to me that I could even apply.
He also showed me how a stimulating working environment can be like, where everyone’s contribution is welcomed and appreciated, even of the most junior team members.
His group did set a very high standard for me and it was very hard to find a space like it afterwards.

Axel was there for me every time I was loosing ground. He managed to pull me back up very gracefully and with a smile.
Maarten gave me his time, patience and attention, an opportunity to develop and contemplate a variety of compelling ideas, while stirring me back once in a while when I wondered too far off the track.
I could not have done this work without these strong people being on my side.
A simple advice seems trivial but it works like a warm blanket gently wrapped around your shoulders at times when you feel tired and exhausted. It helps to believe that you can eventually get somewhere if you just keep on going.
I wish I could also be a ``lion with wings''~\cite{van2017remedies} for someone one day.

I would like to acknowledge my colleagues, who at different times and at different universities made my working days such a delight with an opportunity to catch up and have a good laugh together: Adrian, Alessio, Amr, Ana, Anton, Anya, Artem, Basil, Basti, Bob, Christophe, Claudio, Cristina, Daniel, Denny, Djordje, Erwin, Fabi, Florian, Giray, Hans, Harrie, Hosein, Ilya, Jan, Javi, Johannes, Jose, J{\"u}rgen, Kate, Kathrin, Martin, Maribel, Maximilian, Max, Monika, Philipp, Rebecca, Rod, Roman, Sabrina, Saimir, Sebastian, Shu, Steven, Thomas, Tobi, Tom, Vadim, Xinyi.
It was never possible to anticipate the next topic that could turn up during the next lunch or coffee break and where our discussion could ultimately lead us. Thank you for all those wonderful days we had together!
I learned a lot from you.

A special thanks goes to Vadim for supporting me along the way and letting me explore alternative directions.

I am very grateful to my family and friends, who were taking care of me saving my body and soul so many times during these years: Adina, Aldo, Alex, Artem, Barbara, Christoph, Daniel, David, Florian, Iya, Jakob, Jakub, Jelke, Kami, Markus, Matthias, Maryna, Michal, Olya, Olena, Oleksandra, Patrik, Peter, Roger, Sasha, Stefan, Sue, Yuliya.
You are all very dear to me and I hope you know that well.

It is due to my parents that I could take the time and the risk to continue learning and make mistakes.
It takes me often some time to go in a wrong direction at first to realise where I actually want to be.
My parents never gave me a feeling that I was not able or meant to do something, which is a very precious gift to have.

There were so many people in my life who made an impact on the way I am that it would not be possible to list all of them here. I hope you can recognise yourself when reading these lines. Even so far away you are so close to me.
\end{acknowledgements*}


\begin{abstract}
Conversational interfaces that allow for intuitive and comprehensive access to digitally stored information remain an ambitious goal.
In this thesis, we lay foundations for designing conversational search systems by analyzing the requirements and proposing concrete solutions for automating some of the basic components and tasks that such systems should support.
We describe several interdependent studies that were conducted to analyse the design requirements for more advanced conversational search systems able to support complex human-like dialogue interactions and provide access to vast knowledge repositories.
In the first two research chapters, we focus on analyzing the structures common to information-seeking dialogues by capturing recurrent patterns in terms of both domain-independent functional relations between utterances as well as domain-specific implicit semantic relations from shared background knowledge.

Our results show that question answering is one of the key components required for efficient information access but it is not the only type of dialogue interactions that a conversational search system should support.
In the third research chapter, we propose a novel approach for complex question answering from a knowledge graph that surpasses the current state-of-the-art results in terms of both efficacy and efficiency.
In the last research chapter, we turn our attention towards an alternative interaction mode, which we termed conversational browsing, in which, unlike question answering, the conversational system plays a more pro-active role in the course of a dialogue interaction.
We show that this approach helps users to discover relevant items that are difficult to retrieve using only question answering due to the vocabulary mismatch problem.

\end{abstract}

\selectlanguage{english}

\setcounter{tocdepth}{3}
\tableofcontents 

\mainmatter

\chapter{Introduction}
\label{chap:intro}

\epigraph{Progress is not made by finding the ``right answers'', but by asking meaningful questions -- ones that evoke an openness to new ways of being. \\ --- Terry Winograd and Fernando Flores, Understanding computers and cognition: A new foundation for design, 1986}

People are overwhelmed with screens and the amount of visual information they project~\cite{FVIN}.
Both desktop and mobile devices engage their users predominantly through visual and tactile interactions~\cite{DBLP:journals/ahci/PunchoojitH17,DBLP:conf/chi/JangKTIF16}.
Human-computer interfaces have a major impact on many aspects of personal and professional life including consequences for individual physical health due to long-hours sitting in front of a desktop screen and the dynamics of social interactions, such as attending to mobile phone screens in transport, meetings and in the street~\cite{DBLP:books/daglib/0029131}.
The ``virtual life'' within the information space has become richer and more dynamic but at the same time more demanding by drawing more attention and energy, which are also required for balanced functioning within the physical space~\cite{haeuslschmid2017presenting,DBLP:conf/hci/KaratasYSO15}.

Conversational interfaces aim at addressing the issues of usability and balancing information load more evenly across different senses by engaging the human auditory perception system more actively by means of speech-based and multimodal interactions~\cite{melichar2008design}.
Dialogue interactions, both voice-based and text-chat, are minimalistic and concise by design.
Enabling interactive retrieval, contextual semantics and robust summarisation as the key technologies can provide flexibility and efficiency, which are characteristic of a human dialogue~\cite{black1991pragmatics}.

Another major benefit that the idea of a conversational interface based on a natural language provides, is a universal multipurpose communication protocol, which should be intuitively accessible for every language speaker without any additional training required.
In practice, however, all modern conversational systems are able to interpret only a bounded set of commands, which effectively makes them reminiscent of the traditional command-line interfaces enhanced with some paraphrasing and speech-recognition features.
Users need to know the commands that the system supports to be able to use it efficiently.
This type of interaction resembles a dialogue, but it is very far from a real conversation, which humans are capable of.
More complex conversations tend to be multi-turn, coherent, contextualized and grounded in a shared background knowledge, such as common sense and situational awareness~\cite{horton2017theories,DBLP:conf/emnlp/TandonDGYBC18}.

One of the major applications for dialogue-based interfaces, which we consider in this work, is the role of a knowledge communication medium that is designed to facilitate information exchange between people~\cite{DBLP:books/daglib/0029131}.
The traditional broadcasting model of main-stream mass media, in which informational content is produced exclusively by a few central nodes and transmitted to the rest of the network, was recently replaced by a more decentralized communication model enabled by social media platforms, in which every consumer-node can play the role of an information producer as well~\cite{pascu2007potential}.
This paradigm shift contributed to the massive surge of the information stream, which is being constantly shared on-line.
Whilst the current level of technological developments allows us to scale information publishing and accumulation, it does not yet provide any reliable means to gain a comprehensive overview and discover relevant content~\cite{DBLP:journals/ijinfoman/GandomiH15}.
Human cognition effectively serves as a bottleneck in the information consumption and knowledge sharing process~\cite{DBLP:conf/chi/MunteanuP17}.
Thus, our goal will be to develop intelligent systems with human-computer interfaces designed to account for the cognitive limitations of their users and able to assist them, when navigating and consuming shared knowledge sources.
Such intelligent assistants, however, do not constitute autonomous agents, but are mere mediators that facilitate communication between the autonomous members of the community by providing a high-speed bus for sharing knowledge.
Efficient communication is the key allowing people to act as community members, align their knowledge about the shared environment and collaborate in common creative projects.

\begin{figure}[!t]
\centering
\includegraphics[width=0.7\textwidth]{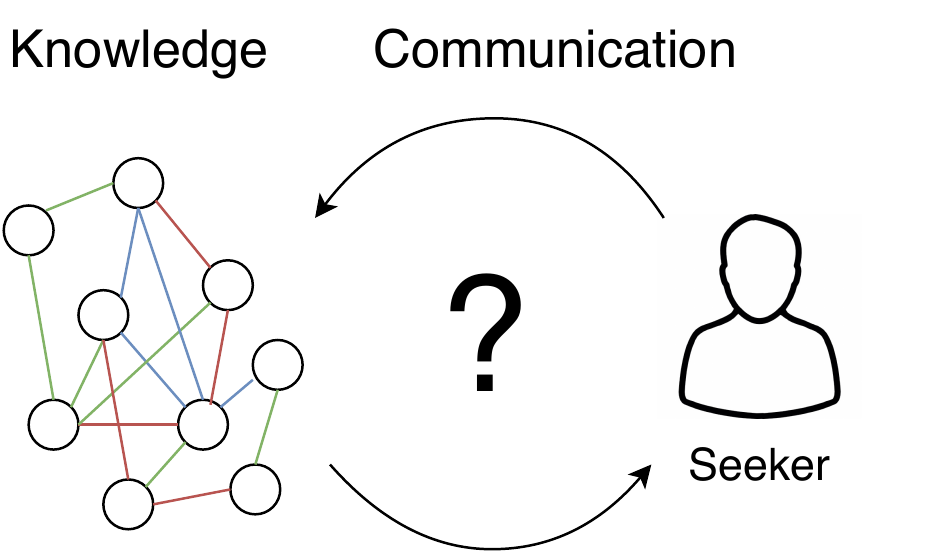}
\caption{How to design an efficient communication interface to a knowledge repository?}
\label{fig:research_question}
\end{figure}

Thus, we see a clear need for a more efficient access mechanism to shared digital information spaces, such as the Web.
This motivation leads us to the main question ``How to design an efficient communication interface to a vast knowledge repository?'' (Figure~\ref{fig:research_question}).
Thereby, we are confronting two complex issues at the same time: (1)~knowledge representation and (2)~communication interfaces.
Our motivation for such a systematic take on this problem as a whole is that we consider knowledge representation approaches as a means to a goal, which for us is enabling an efficient communication medium through interactive natural-language interfaces inspired by human dialogue.
Our main question is further decomposed into more concrete research questions that are outlined in the next section.

\section{Research Outline and Questions}

We tackle the overall composite problem along four directions formulated as research questions, which we aim to answer in this thesis.
We begin with an analysis of information-seeking dialogue transcripts to extract basic structural patterns that hold true across specific domains and datasets.

\textbf{RQ 1} What is the general structure of an information-seeking dialogue?

Furthermore, we hypothesize that besides having certain structural properties, which are common across different information-seeking conversations, each conversation also has an innate semantic structure, meaning that the set of semantic concepts and the order in which they are used in the conversation are not random but determined by the structure of the underlying semantic space, i.e., the background knowledge of the conversation participants.

\textbf{RQ 2} What are the relations between concepts mentioned in the course of a conversation and how can we detect them?

Question answering is one of the main components of a conversational search system.
Knowledge graphs provide a convenient data structure that allow for modeling, sharing and accessing explicit knowledge representations.

\textbf{RQ 3} How to design a system able to answer complex questions using information stored in a knowledge graph?

Finally, we turn our attention towards the other functionality expected from a conversational search system beyond the question answering task.
Query formulation has been shown to constitute a bottleneck hindering effectiveness of information retrieval systems~\cite{kelly2009comparison}.
A similar problem has been reported for conversational search as well~\cite{DBLP:journals/corr/abs-1709-05298}.

\textbf{RQ 4} How to design a conversational system able to support information retrieval without the need to explicitly formulate a search query?

\section{Main Contributions}
In this section, we summarize the main theoretical, algorithmic and empirical contributions of this thesis.

\subsection{Theoretical contributions}
\begin{enumerate}
    \item We propose a novel data-driven model of information-seeking dialogues addressing \textbf{RQ 1}. Cf.\ Chapter~\ref{chap:structure}.
    \item We design the task of measuring semantic coherence of a conversation to evaluate the ability of a model to distinguish between original human conversations and adversarial examples generated by perturbing the original conversations addressing \textbf{RQ 2}. Cf.\ Chapter~\ref{chap:coherence}.
    \item We introduce and formally define the task of conversational browsing addressing \textbf{RQ 4}. Cf.\ Chapter~\ref{chap:browsing}.
\end{enumerate}

\subsection{Algorithmic contributions}
\begin{enumerate}[resume]

    \item We propose a message-passing algorithm for confidence score propagation and aggregation for answering complex questions from knowledge graphs addressing \textbf{RQ 3}. Cf.\ Chapter~\ref{chap:qa}.
    \item We propose the first approach to the conversational browsing task based on information theoretic criteria of optimality addressing \textbf{RQ 4}. Cf.\ Chapter~\ref{chap:browsing}.
\end{enumerate}

\subsection{Empirical contributions}
\begin{enumerate}[resume]

    \item We show empirically that our model of information-seeking dialogues generalizes to unseen dialogues and is able to detect dialogue break-downs addressing \textbf{RQ 1}. Cf.\ Chapter~\ref{chap:structure}.
    \item We test the effectiveness of different knowledge representation models: word embeddings, knowledge graphs and knowledge graph embeddings, for the task of measuring semantic coherence of a conversation addressing \textbf{RQ 2}. Cf.\ Chapter~\ref{chap:coherence}.
    \item We evaluate the proposed message-passing algorithm on a large-scale benchmark for complex question answering from knowledge graphs and show a significant performance improvement over the previously proposed approach, in terms of both effectiveness and efficiency, addressing \textbf{RQ 3}. Cf.\ Chapter~\ref{chap:qa}.
    \item We conduct a fist user study to collect human conversations as examples for the conversational browsing task addressing \textbf{RQ 4}. Cf.\ Chapter~\ref{chap:browsing}.
    \item We conduct a second user study to evaluate our approach to conversational browsing addressing \textbf{RQ 4}. Cf.\ Chapter~\ref{chap:browsing}.
\end{enumerate}

\section{Thesis Overview}

We outline the research described in this thesis in Chapter~\ref{chap:intro}.
Next, Chapter~\ref{chap:background} lays out the background with the main concepts addressed in this thesis, namely analysis of the communication process that supports the design of natural-language dialogue interfaces, as well as a brief description of the main characteristics and the structure of the prevalent knowledge models that were used in this thesis for conversation analysis and knowledge retrieval.

\begin{figure}[!ht]
\centering
\includegraphics[width=0.5\textwidth]{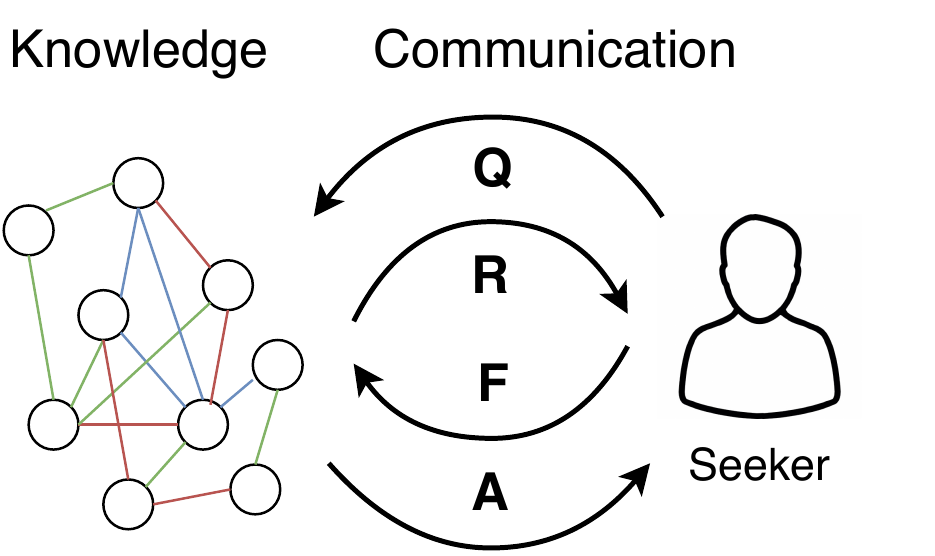}
\caption{Chapter~\ref{chap:structure} analyzing the structure of the communication process.}
\label{fig:chap1}
\end{figure}

In Chapter~\ref{chap:structure} we focus on analyzing the structure of communication processes in terms of the interaction types that are characteristic for conversations in which one of the participants is seeking access to a part of the knowledge that may be available to the other conversation participant (Figure~\ref{fig:chap1}).
We present the resulting QRFA model that was empirically derived from a systematic semi-automated study conducted over four independent datasets and describes the general structure of an information-seeking dialogue.
It is the first step that helps us to argue for the functions that a conversational search system should support.

\begin{figure}[!ht]
\centering
\includegraphics[width=0.5\textwidth]{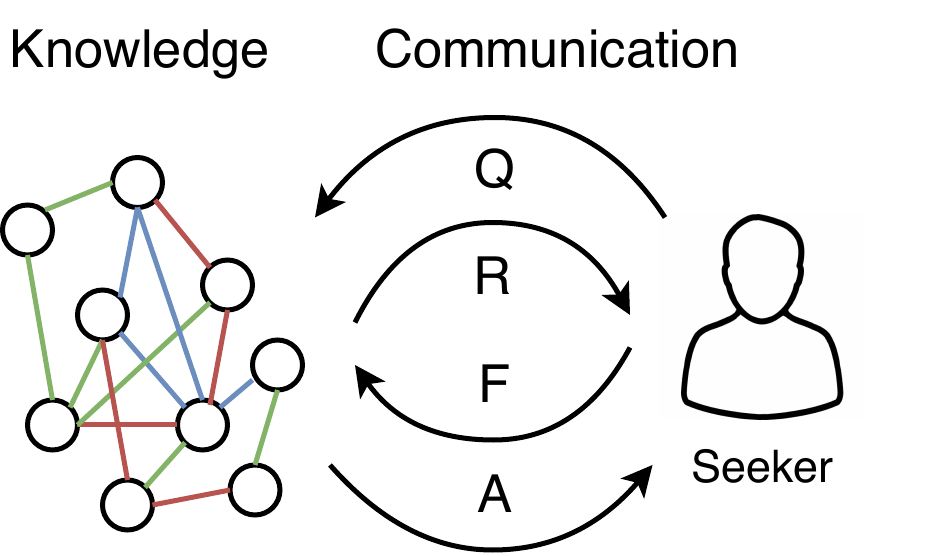}
\caption{Chapter~\ref{chap:coherence} showcases that semantic relations from the knowledge model are reflected in the conversation structure.}
\label{fig:chap2}
\end{figure}

In Chapter~\ref{chap:coherence} we confirm our hypothesis that relations between concepts in the knowledge models are also reflected in the structure of the conversation (Figure~\ref{fig:chap2}).
The obtained results are empirical in nature, showing that it is possible to predict the dialogue coherence based on relations stored in the knowledge model.
But the major implications are of a more theoretical nature:
(1) there is an immediate connection between the dialogue structure and the structure of the background knowledge even if the dialogues were not explicitly designed to communicate parts of this background knowledge;
(2) the existing knowledge models are incomplete; and
(3) one of the major challenges is aligning natural language to semantic concepts, i.e., linking and disambiguation tasks.

\begin{figure}[!ht]
\centering
\includegraphics[width=0.5\textwidth]{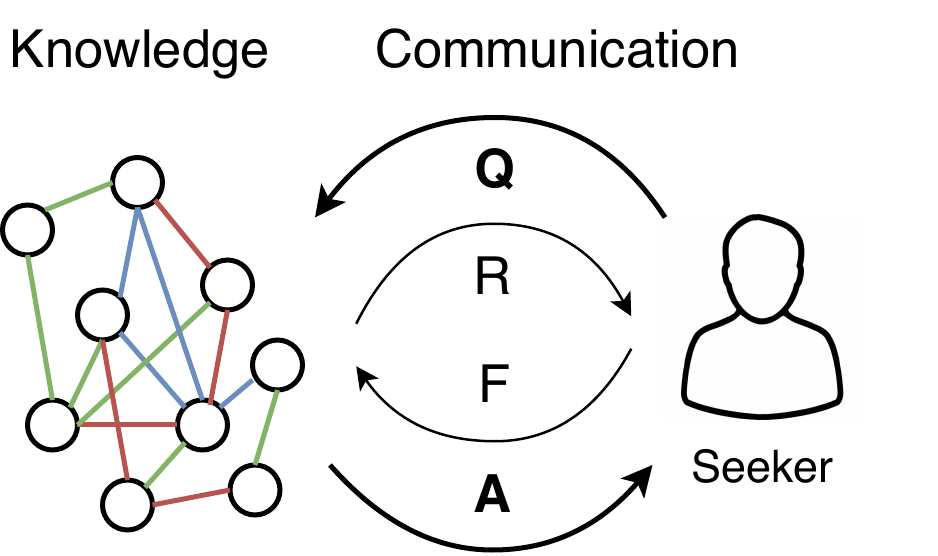}
\caption{Chapter~\ref{chap:qa} describes our novel approach to complex question answering over knowledge graphs.}
\label{fig:chap3}
\end{figure}

In Chapter~\ref{chap:qa} we follow up and integrate the previous two lines of research that were started in parallel by focusing on the key interaction identified in Chapter~\ref{chap:structure}, namely question answering (QA), and applying it to the knowledge graph as one of the main knowledge models motivated in Chapter~\ref{chap:coherence} (Figure~\ref{fig:chap3}).
Our approach outperforms the baseline, achieving state-of-the-art results on a complex QA benchmark, and demonstrates the limitations of the ground-truth sampling method, while improving recall even over the original answer sets provided by human annotators.

\begin{figure}[!ht]
\centering
\includegraphics[width=0.5\textwidth]{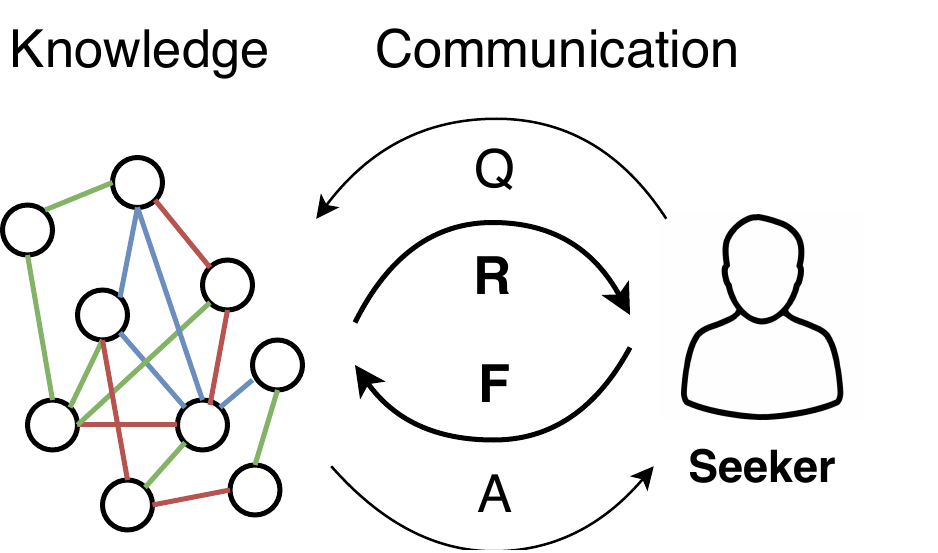}
\caption{Chapter~\ref{chap:browsing} introduces the novel task of conversational browsing and the first approach that allows one to iteratively search and discover the content of a knowledge source.}
\label{fig:chap4}
\end{figure}

There are natural limits to how complex a single question can get.
A conversational search system should provide a good balance between complexity and simplicity by supporting interactive knowledge retrieval.
In Chapter~\ref{chap:browsing} we describe conversational browsing as an alternative interaction mode, which allows one to access a knowledge source iteratively without the need to formulate a single complete question-query at once (Figure~\ref{fig:chap4}).
Conversational browsing was designed to complement the question answering mode described in the previous chapter by completing the requirements prescribed by the QRFA model.
Moreover, we further extend the interaction model by considering not only the model of the knowledge source but also modeling the information seeker (user) to automate some parts of the system evaluation and algorithm tuning via a user simulation.

Finally, in Chapter~\ref{chap:conclusion}, we conclude the thesis and discuss the limitations and promising directions for future work.

\section{Origins}


The chapters of the thesis are based on the following publications, all of which were first-authored by Vakulenko as the main contributor.

\textbf{Chapter~\ref{chap:structure}} is based on Svitlana Vakulenko, Kate Revoredo, Claudio Di Ciccio and Maarten de Rijke. QRFA: A Data-Driven Model of Information Seeking Dialogues. In ECIR. 2019. \textit{Best User Paper Award}. Vakulenko designed the main algorithm. Vakulenko and Di Ciccio conducted the initial experiments. Vakulenko and Revoredo annotated the datasets and performed the evaluation. All authors contributed to the text, Vakulenko did most of the writing.

\textbf{Chapter~\ref{chap:coherence}} is based on Svitlana Vakulenko, Maarten de Rijke, Michael Cochez, Vadim Savenkov and Axel Polleres. Measuring Semantic Coherence of a Conversation. In ISWC. 2018. \textit{Spotlight Paper}. Vakulenko designed the main algorithm and ran most of the experiments. Savenkov contributed to the implementation. Cochez ran additional experiments. All other authors contributed to writing the paper.

\textbf{Chapter~\ref{chap:qa}} is based on Svitlana Vakulenko, Javier Fernandez, Axel Polleres, Maarten de Rijke and Michael Cochez. Message Passing for Complex Question Answering over Knowledge Graphs. In CIKM. 2019. Vakulenko designed the main algorithm and conducted the main experiments. Fernandez contributed to the implementation and evaluation design. All other authors contributed to the writing.

\textbf{Chapter~\ref{chap:browsing}} is based on Svitlana Vakulenko, Vadim Savenkov and Maarten de Rijke. Conversational browsing: Dialog-based Access to Structured Information Sources. \textit{Under review}. Vakulenko designed the main algorithm, wrote code and ran the experiments. All authors contributed to the development of the theoretical model and writing the paper.

The thesis also indirectly builds on the following publications that helped to form a general understanding of the problem space and shaped the ideas presented in this thesis:

\begin{itemize}
    \item \bibentry{KeynerSV19}
    \item \bibentry{DBLP:journals/corr/abs-1709-05298} 
    \item \bibentry{DBLP:conf/i-semantics/VakulenkoS17} 
    \item \bibentry{DBLP:conf/esws/NeumaierSV17} 
\end{itemize}

\chapter{Background}
\label{chap:background}

\epigraph{The world's in trouble. There's no communication. \\ --- Joan Jett, Bad Reputation, 1981 \\ \\A human being, after all, is only what it is defined to be. \\ --- Isaac Asimov, Robots and Empire, 1985}

This chapter introduces the main concepts concerning knowledge and communication that form the basis for understanding previous work and highlight the gaps that the research contributions we describe in the next chapters fill.
We provide a foundation covering existing viewpoints and current research directions.
Additionally, we will discuss specific related work within the following chapters directly when we explain our approaches to tackle the specific sub-problems addressed in this thesis.

Communication is an important social phenomenon necessary for establishing relationship, trust and successful collaboration\footnote{\url{https://ncase.me/trust/}}~\cite{evolution}.
The study of communication is one of the cornerstones that falls within the scope of several scientific disciplines: psychology, sociology and linguistics.
Understanding the structure, dynamics and roles that communication plays in private and social lives, are among their main research goals.
We are interested in communication as a process that enables knowledge exchange.

\section{Communication Process}
\label{sec:communication}

There are different types of communication, e.g., synchronous/asynchronous, verbal/non-verbal~\cite{littlejohn2009encyclopedia}.
In this thesis we focus specifically on verbal communication, i.e., the one conducted by means of a natural language, as a system of symbols with assigned semantics.
Human language is an effective communication tool, flexible and able to evolve so as to adopt to new domains and purposes~\cite{fitch2010social,jablonka2012co}.

\begin{definition}
A \textbf{conversation} is a sequence of natural language expressions (utterances) made by several conversation participants in turns.
\end{definition}


We consider sample conversations as individual instances of a general communication process.
Our goal is to design a human-computer interface that resembles a human dialogue.
We believe the first legit step in this direction is understanding the properties and functions of human conversations, which we are trying to reproduce, as the patterns of language usage that support human communication.

\subsection{Understanding conversations}

We briefly review the main theories of linguistic communication and link them to the studies of information-seeking behavior.
Further on, we describe how Chapters~\ref{chap:structure} and \ref{chap:coherence} of this thesis extend and empirically evaluate these theories using large datasets of conversational transcripts.
We use these insights to propose concrete solutions to question answering and conversational browsing in Chapters~\ref{chap:qa} and \ref{chap:browsing}, respectively.





The main bottleneck in traditional approaches to discourse analysis, especially ones grounded in social and psychological theories \cite[e.g.,][]{goffman1988erving}, is context-dependence of the conversational semantics. 
Many studies rely on a handful of sampled or even artificially constructed conversations to illustrate and advocate their discourse theories, which limits their potential for generalization.

\paragraph{Discourse analysis}
Interviews and observation of naturally occurring social interactions that involve recording of conversation transcripts are the traditional research methods of anthropologists and other social scientists used to collect data for their analysis~\cite{mack2005qualitative,mohajan2018qualitative}.
This research motivation also stimulated early attempts at developing methodologies for systematic analysis of conversational data, such as standardized categorization schemes used as tools for manual annotations, which help not only to summarize and generalize findings across different studies but also record an extra dimension of the expression, such as an intent or an attitude of the speaker~\cite{steinzor1949development}.

A plethora of approaches to analyse language use have been proposed over decades, focusing on different aspects of the communication process~\cite{schiffrin1994approaches}.
Some of them are grounded in social and psychological theories \cite[e.g.,][]{goffman1988erving}, while relying only on a handful of illustrative conversations or artificially constructed examples.
Discourse analysis further evolved as a study of regularities in language composition, such as frequent structures and relations between individual sentences.
For example, Conversational Analysis (CA)~\cite{schegloff1968sequencing} proposed to analyze regularities such as adjacency pairs and turn-taking in conversational structures, and Speech Act Theory~\cite{searle1969speech,austin1976things} to identify utterances with functions enabled through language (speech acts).
Among the more recent developments are attempts at standardizing annotation schemes for conversations (dialogue acts) and statistical models for automatic annotations~\cite{stolcke2000dialogue} aimed at dramatically scaling the analysis of conversation transcripts.

Discourse analysis also leads to formulating theories and models of discourse structure, which attempt to explain how a sequence of sentences forms a coherent discourse~\cite{grosz1986attention,mann1988rhetorical}.
Discourse structure theories were successfully applied to characterize a dialogue phenomena and inform the design of dialogue systems~\cite{rotaru2009applications}.

\paragraph{Information-seeking dialogues}
Another large research area devoted to the study of information-seeking behavior has unfolded in parallel to discourse analysis.
Information-seeking behavior is a complex process that has been extensively studied in the literature and several models have been proposed as an attempt to describe its structure and characteristics~\citep{bates1979information,ellis1989behavioural,kuhlthau1991inside}.
A search session is an instance of this process, which may include several interaction turns (information exchanges) between the user and a search engine as an information source~\citep{hagen2011query}.

``Naturally occurring'' conversations have often served as evidence and a source of inspiration for developing theoretical models and prototypical implementations in the area of information retrieval (IR)~\citep[see, e.g.,][]{trippas2016how,DBLP:journals/corr/abs-1812-10720,DBLP:conf/interspeech/SegundoMCGRP01}.
For example, observations of information-seeking dialogues with a reference librarian, which is a classic example of a help-desk information service, suggest that a process of negotiation is an important component that helps to better understand and adjust the information need when discussing it with an expert~\citep{taylor1968}.
The observations derived from these studies were extensively discussed in the research literature.
``There has been a great deal written about information seeking behavior, but there is little consensus as to a specific conceptual model. This presents difficulties for a deeper understanding of the process''~\cite{krikelas1983information}.

The key properties of an information-seeking dialogue are asymmetry of roles and cooperation --- the information seeker and provider (intermediary) cooperate to better understand and satisfy the information need of the seeker.
An information provider has access to the information source and plays the role of an intermediary for the information seeker who has an information need.
Let us have a closer look at a sample information-seeking dialogue.
It is an excerpt from a real chat-based conversation that occurred between a pair of students during a lab study that we conducted, which is described in more detail in Chapter~\ref{chap:browsing}.
One of the students seeks information (S -- Seeker) and the other one is trying to help using the Austrian Open Data portal (I -- Intermediary):

\textbf{(I)} - \url{opendataportal.at} contains more than 417 data sets.\\
\textbf{(S)} - Is there data about universities?\\
\textbf{(I)} - Yes, especially about WU.\\
\textbf{(S)} - How many courses are there at WU?\\
\textbf{(I)} - For which semester?\\
\textbf{(S)} - I want to have the latest data possible.\\
\textbf{(I)} - There were 2,576 courses offered at WU in the SS17~\footnote{SS17 refers to Summer Semester 2017. The background knowledge necessary to correctly interpret natural language is discussed in Section~\ref{sec:knowledge}.}.\\
\textbf{(S)} - Thank you.

This dialogue exemplifies the question negotiation process.
The students exchange knowledge about the content of the data sets for the details of the information need.
This language exchange is governed by certain conversational rules, which is evident from the relations between the utterances in conversation, e.g., questions are either followed by answers or clarifying questions in-between~\cite{schegloff1968sequencing}.

\subsection{Theoretical models of information-seeking dialogues}
\label{sec:models}

Several models have been proposed as an attempt to describe the structure and characteristics of the information seeking process as a whole~\cite{bates1979information,ellis1989behavioural,kuhlthau1991inside}, such as the berrypicking model~\cite{bates1989design}.
The berrypicking model of search behavior suggests that the understanding of an information need evolves during the search process as more information becomes available to the seeker and browsing interfaces are the key enablers for supporting this kind of a dynamic search process~\citep{bates1989design}.
These models, however, are too high-level and cannot be applied to describe the pattern of interactions within a single information-seeking dialogue~\cite{DBLP:journals/jd/Savolainen18,trippas2018informing}.

\begin{figure}[!t]
\centering
\includegraphics[width=0.65\columnwidth]{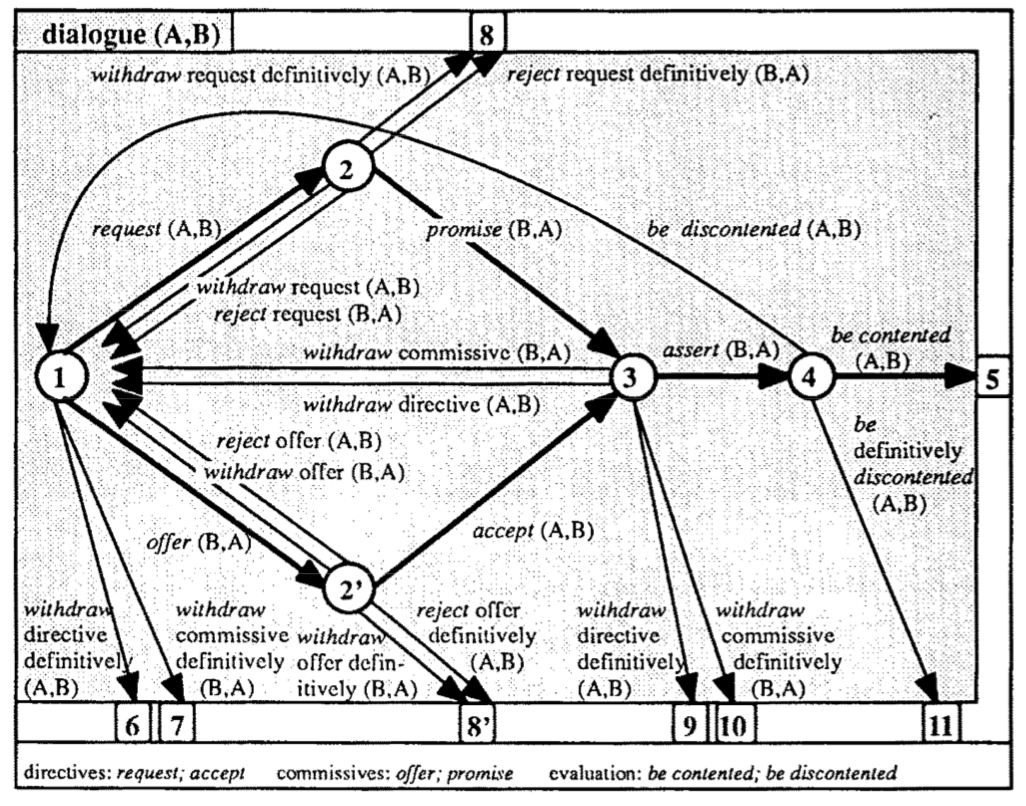}
\caption{COnversational Roles (COR) model of information-seeking dialogues proposed by \citet{sitter1992modeling}.}
\label{fig:model_sitter}
\end{figure}

The first theoretical model of an information-seeking dialogue has been proposed by \citet{winograd1986understanding} and further extended by \citet{sitter1992modeling} to form the COnversational Roles (COR) model, which has remained de facto the only established model of an information-seeking conversation until today.
The authors envision an implementation of a human-computer dialogue system that could support necessary functionality to provide efficient information access and illustrate it as a transition network over a set of speech acts (Figure~\ref{fig:model_sitter}).
This model describes a use case of a ``conversation for action'' and is mainly focused on tracking commitments rather than analyzing language variations.
The COR model describes a conversation in terms of commitments and operates the corresponding set of utterance labels: request, promise, offer, accept, be contended etc. (see Figure~\ref{fig:model_sitter}).
In Chapter~\ref{chap:structure} we show in an empirical evaluation on four publicly available datasets that this model does not adequately reflect the structure of an information-seeking dialogue, such as the example provided above, and propose a new QRFA model as an alternative.

Belkin et al.~\cite{belkin1995cases} argue for a modular structure of an interactive information retrieval (IR) system that would be able to support various dialogue interactions. 
The system should be able to compose interactions using a set of scripts, which provide for various information-seeking strategies (ISSs) that can be described using the COR model. 
The authors introduce four dimensions to describe different ISSs and propose to collect cases for each of the ISSs to design the scripts. 

\begin{figure}[!t]
\centering
\includegraphics[width=0.9\textwidth]{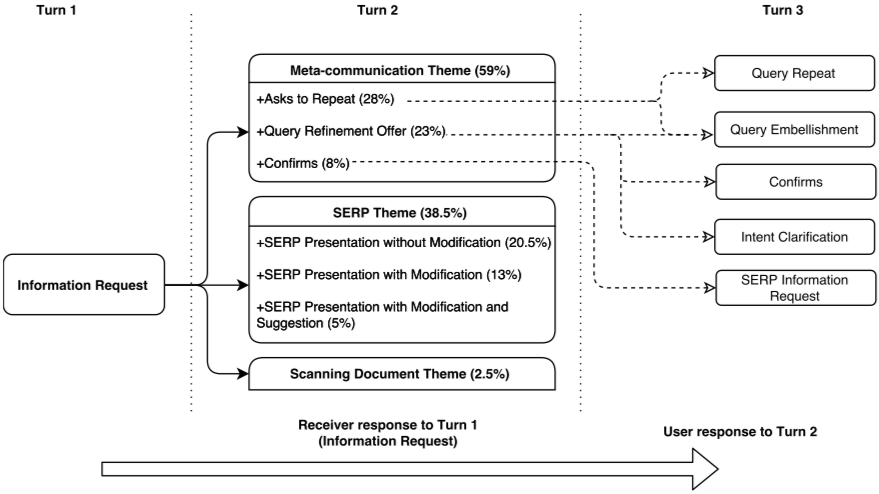}
\caption{Spoken Conversational Search (SCS) model by \citet{trippas2016how} (SERP is short for Search Engine Results Page.)}
\label{fig:model_trippas}
\end{figure}

More recently, \citet{DBLP:conf/chiir/RadlinskiC17} have also proposed a theoretical model for a conversational search system. 
They propose a set of five actions available to the agent and a set of five possible user responses to describe the user-system interactions. 
However, they do not describe in detail the conversation flow between these actions. 
In contrast, \citet{trippas2016how} empirically derived a model of Spoken Conversational Search (SCS) from conversational transcripts collected in a Wizard of Oz study for informing the design of speech-only interfaces to a web search engine (Figure~\ref{fig:model_trippas}).
The SCS model captures patterns of interactions that generalize across the individual dialogues: information requests are followed by result presentation or clarifying questions.
However, the SCS model is limited to sequences of three conversational turns.
Inspired by the data-driven approach of \citet{trippas2016how} we continue this line of work by proposing a methodology that allows us to scale analysis of conversational transcripts in Chapter~\ref{chap:structure}.
We also extend the analysis to other publicly available conversational datasets and define a new label set integrating different annotation schemes.
In Chapter~\ref{chap:structure} we show how such an empirical approach to analyzing and structuring conversation transcripts into conversation models can be performed at scale on multiple conversational datasets (including SCS) using process mining techniques.
The resulting QRFA model is a grounded theory of information-seeking dialogues that motivates and lays the foundation for the follow-up studies described in the next chapters.
This model should not be treated as a final result but rather as an initial hypothesis and the basis for the development of a conversational knowledge model, which can be shared and discussed within the community, refuted or further extended by providing additional sources of evidence, i.e., more conversational datasets.
Thereby we also closely follow the line of work proposed by Belkin et al.~\cite{belkin1995cases} by accumulating empirical evidence from publicly available conversational datasets to validate both the COR model and the ISS dimensions proposed by Sitter and Stein~\cite{sitter1992modeling} that form the basis for accumulating the body of sample scripts describing various ISSs.
Chapter~\ref{chap:coherence} describes the study that analyses dialogues by annotating them with entities from an external knowledge source and explains the phenomenon of dialogue coherence by the regularities in the structure of semantic relations manifested in the background knowledge.

The search engines that are being used today do not reflect the complexity and challenges inherent in human information seeking behavior, and the research community continues to work on aligning the priorities and developing viable alternatives~\cite{DBLP:journals/sigir/Culpepper0S18}.
In the next section we briefly review the main approaches proposed to design conversational interfaces and dialogue systems, as an alternative tool for enabling more advanced human-machine interactions, and motivate their adoption for automating information-seeking dialogues as well, in the context of conversational search.

\subsection{Conversational systems}

Due to recent advances in speech recognition and natural language generation technologies conversational interfaces experience a surge of interest from both industry and academia~\cite[see, e.g.,][]{scai,DBLP:conf/acl/DhingraLLGCAD17,DBLP:journals/dad/WilliamsRH16a}.
There are two major types of dialogue system application considered now: chit-chat dialogues and task-oriented dialogues~\citep{DBLP:journals/ftir/GaoGL19}.
The idea of an automated system being able to communicate by means of a natural language goes back to the beginning of computing.
The seminal paper by Alan Turing~\cite{alan1950turing}, which introduced a human-machine dialogue as a test for artificial intelligence, is still regarded as an ambitious goal and drives the development of systems capable to hold an open-domain (chit chat) conversation.
The chatbot history began with ELIZA~\cite{DBLP:journals/cacm/Weizenbaum66}, a rule-based system that followed a predefined script of a psychotherapy interview.
The next famous chatbot simulated a schizophrenic patient\footnote{\url{https://web.stanford.edu/group/SHR/4-2/text/dialogues.html}}~\cite{dewdney1985artificial}.
In practice, this kind of conversational systems are being used predominantly for entertainment~\cite{DBLP:journals/corr/abs-1801-03604}, in more radical cases even aiming to replace a human companion\footnote{\url{https://replika.ai}}~\cite{DBLP:journals/corr/abs-1712-05626}.
However, none of the systems proposed to date has so far proven to be capable of ``maintaining a coherent and engaging conversation for 20 minutes''~\cite{hakkani_tur_2018}.

Another direction for conversational systems are task-oriented dialogues, in which one of the main measures of conversation success is the task completion ratio that makes evaluation more straight-forward.
Classic examples of task-oriented dialogue systems are in the restaurant reservation and trip planning domains~\cite{DBLP:journals/dad/WilliamsRH16a}.
In this case the dialogues that the system is able to support are more specialized and domain-specific, in comparison with their chitchatting counterparts.
Task-oriented dialogues rely on a domain specification, which can be defined in terms of an ontology, a table or a set of annotated dialogue samples that provide a frame for linking slots and intents to possible replies.
This ontology enumerates all concepts and attributes (slots) that a user can specify or request information for~\citep{DBLP:conf/acl/MrksicSWTY17}.
The dialogue models are then designed to jointly perform the tasks of parsing the input utterances, slot matching/filling and belief state tracking~\cite{williams2013dialog}.
Today, most of the small-scale commercial chatbots are task-oriented and hard-coded to support a very limited functionality fitted to a specific use case, such as make a purchase or check the weather forecast~\cite{DBLP:conf/ACMdis/JainKKP18,DBLP:conf/chi/JacquesFGGLMW19}.
Digital assistants embedded in smartphones and smart home devices, such as Siri, Alexa and Cortana~\cite{DBLP:journals/nle/Dale16b}, play the role of a hub that can serve different chatbots (skills) from the same interface.

Every conversational system both for open-domain and task-oriented dialogues relies on knowledge, defined in terms of the information stored to assist in understanding the user input and formulating a response.
This information can be dynamically updated and should be efficiently accessible from the dialogue model.

\begin{definition}
The task of retrieving relevant information using a conversational interface is termed \textbf{conversational search}.
\end{definition}

The main goal in conversational search, which is the same as for the discipline of information retrieval in general, is to satisfy an information need, which makes it distinct from both open-domain and task-oriented dialogues.
Conversational search systems have a diverse set of applications, including such challenging scenarios as recommendation and education~\cite{DBLP:conf/chiir/RadlinskiC17,dalton2018vote}.

Lately, conversational search systems are becoming increasingly popular~\citep{DBLP:conf/chi/VtyurinaSAC17}.
So far, however, such systems mainly focus on question answering and simple search tasks, those that are to a large extent solved by web search engines.
Question answering (QA) is one of the central functionalities required for conversational search (interactive question answering)~\cite{konstantinova2013interactive}.
Effectively it is mirroring the already existing querying mechanism of the traditional search engines into a new conversational interface coupled with more advanced speech-recognition and natural language processing technologies, such as machine reading comprehension and summarisation.
QA systems have been evolving since the early 1960s with the efforts in the database community to support natural language queries by translating them into structured queries~\citep[see, e.g.,][]{green-automatic-1963,woods-lunar-1977,bronnenberg-question-1980}.
In Chapter~\ref{chap:qa} we describe our contribution as a novel approach to complex QA that pushes the state-of-the-art in this direction on a recent benchmark dataset.
Instead of attempting to translate questions into queries, our approach applies probabilistic reasoning directly on the graph structure of the background knowledge.

As already discussed in the previous section, QA is not the only type of interaction that a conversational search system has to provide in order to support comprehensive information-seeking dialogues~\citep{DBLP:journals/corr/abs-1812-10720}.
Vocabulary mismatch (aka vocabulary or semantic gap) is a common phenomenon that occurs when two parties, e.g. user in queries and system in collection documents, use different words to describe the same concepts~\cite{van2017remedies}.
Morover, due to the anomalous state of knowledge (ASK)~\cite{belkin1982ask} an information seeker is unable to precisely and unambiguously formulate a question when searching for something previously unknown.
We argue that conversational agents and search systems should also provide support for exploratory search and provide a detailed description of this relatively under-explored direction in Chapter~\ref{chap:browsing} reporting the results of our first attempt when applying it to design an alternative conversational interface to a database.
Our empirical study of the information-seeking dialogues and strategies that humans employ to communicate content of an information source informs the design of a \emph{conversational browsing model}.
Our experiments show that a dialogue system design based on this model is effective in providing basic exploratory search (browsing) functionality.

In summary, we consider efficient information access, communication and knowledge transfer to be the key functions of a conversational search system.
Knowledge, which is required to interpret natural language, understanding the context and formulating an appropriate answer to support a human conversation --- all these have to be at the center of such systems.

\section{Knowledge Models}
\label{sec:knowledge}

There are several alternative representation methods to encode, store and accumulate information over time.
The purpose of information gathering is to be able to use and re-use it at a later stage, and also share it with others as a knowledge repository.
We consider two of the most popular types of knowledge models that differ in terms of their sources and structure: (1) \emph{language models} that learn co-occurrence patterns from unstructured text corpora, such as word embeddings; and (2) \emph{knowledge bases} as curated structures that explicitly define relations between concepts, such as tables, databases, ontologies and knowledge graphs.

\subsection{Language models}
Language is a major tool for communication and knowledge transfer.
Massive amounts of information expressed in natural language are being recorded on a daily basis as text and speech.
This information is split between separate individual narratives of different kinds, such as stories (monologues) and discussions (conversations)~\cite{barthes1975introduction}.

The usage patterns of language recorded in texts reflect the structure of the background knowledge used to generate them.
Statistical models are able to capture and extract such patterns from a text corpus~\cite{kulkarni2015statistically}.
Textual documents contain sequences of symbols.
Therefore, the basic language patterns that can be extracted from text are frequent co-occurrences of groups of symbols (n-grams).
Knowledge of such regular patterns can be accumulated, compressed and stored as a language model, which encodes conditional probabilities of the next symbol given the preceding ones~\cite{DBLP:books/lib/JurafskyM09}.

\begin{definition}
A \textbf{language model} is a probability distribution over word sequences.
\end{definition}

Current state-of-the-art approaches to language modeling rely on neural network embeddings~\cite{mikolov}, such as word2vec~\cite{DBLP:conf/nips/MikolovSCCD13}, which are trained to predict each word given its neighbouring words as context.
Co-occurrence relations are encoded as word vectors that also reflect similarities between words.
More recently, transformer-based architectures, which encode patterns observed in texts into the weights of multiple hidden layers of a neural network, were shown to boost performance on a variety of natural language processing tasks, as an alternative to the word embeddings input layer~\cite{devlin2018bert}.

Though natural language is an often-used medium to express human knowledge that can be condensed using state-of-the-art pattern recognition techniques, it has certain limitations in its practical applications.
Firstly, texts are not self-explanatory and usually require some prerequisites for adequate interpretation, such as commonsense knowledge~\cite{DBLP:books/daglib/0066824}.
Secondly, it is hard to control which patterns are being learned by language models~\cite{DBLP:conf/nips/BolukbasiCZSK16,DBLP:conf/ranlp/SalimbajevsS15}.
An effect prominent in dialogue systems is the difficulty to maintain the speaker consistency (persona) in neural models trained on diverse conversation samples~\cite{DBLP:conf/acl/LiGBSGD16}.
Unawareness of the model structure may also lead to undesirable biases being propagated from the input distribution and reflected in the model predictions, such as gender-based discrimination~\cite{DBLP:conf/nips/BolukbasiCZSK16}.
These challenges in training neural language models motivate the need for careful sampling and data curation.
The need to perform data engineering leads us directly to the motivation for constructing and maintaining knowledge bases that help to accommodate and edit explicit knowledge models.

\subsection{Knowledge bases}

Beyond learning patterns from naturally occurring communication traces it is also possible to explicitly specify models that can encode distinct concepts from a vocabulary and semantic relations between them.
This knowledge can be stored in manually or semi-automatically constructed data structures, such as tables or databases that can connect several related tables~\cite{jansen2018worldtree}.

\begin{definition}
A \textbf{knowledge base} is a structured information repository used for knowledge sharing and management.
\end{definition}

The practice of organizing knowledge into meaningful structures goes back to Ancient Greece.
Ontology is a field of philosophy concerned with defining ``a particular system of categories accounting for a certain vision of the world''~\cite{guarino1998formal}.
In computer science, ontologies refer to models that enumerate and describe entities in terms of their properties, classes and relations~\cite{suchanek2009automated}.
Tree-like structures, called taxonomies, list entities grouped due to similar properties into a single hierarchy and are commonly used for building browsing interfaces, e.g. product catalogs in e-commerce applications.

DBpedia~\cite{DBLP:journals/semweb/LehmannIJJKMHMK15} and Wikidata~\cite{DBLP:journals/cacm/VrandecicK14} are famous examples of large-scale graph-structured knowledge bases, also called Knowledge Graphs (KGs).
DBpedia describes entities that correspond to the Wikipedia pages.
It is automatically populated from Wikipedia, while Wikidata imitates the editing mechanism of Wikipedia but for structured data.
All entities are assigned Unique Resource Identifiers (URIs) that allow to unambiguously reference them and link across different datasets thereby bridging the semantic (vocabulary) gap.
The entities of DBpedia and Wikidata are interlinked.
Both graphs form an important part of the Linked Open Data (LOD) cloud\footnote{\url{https://lod-cloud.net}} that groups all the datasets published in the RDF format~\cite{hayes2014rdf}.

KGs were successfully applied for disambiguating natural language text in a variety of tasks, such as information retrieval~\cite{citeulike:14281342,DBLP:conf/sigir/HasibiBGZ17} and textual entailment~\cite{silva2018recognizing}. 
They serve an important role by providing additional relations that help to bridge the lexical gap and gain a more complete understanding of the context in comparison with shallow approaches based on lexical features alone.
There was also a recent surge in development of question answering interfaces to KGs~\cite{DBLP:conf/esws/UsbeckNHKRN17,DBLP:conf/www/LukovnikovFLA17,ramngongausbeck2018}.

The term ``Knowledge Graph'' was proposed as a medium to accumulate and integrate knowledge from heterogeneous data sources~\cite{DBLP:journals/dagstuhl-reports/BonattiDPP18}.
In contrast with LOD, KG is an abstraction, which does not prescribe any particular format or standard on handling data.
Thereby, tables, taxonomies and texts can be seen as a special kind of relational knowledge and transformed into graph-like structures~\cite{wilcke2017knowledge}.
Then, graph algorithms can be applied on this structure directly or the nodes can be embedded into a vector space for neural network approaches~\cite{10.1007/978-3-319-46523-4_30}.

\begin{figure}[!t]
\centering
\includegraphics[width=\textwidth]{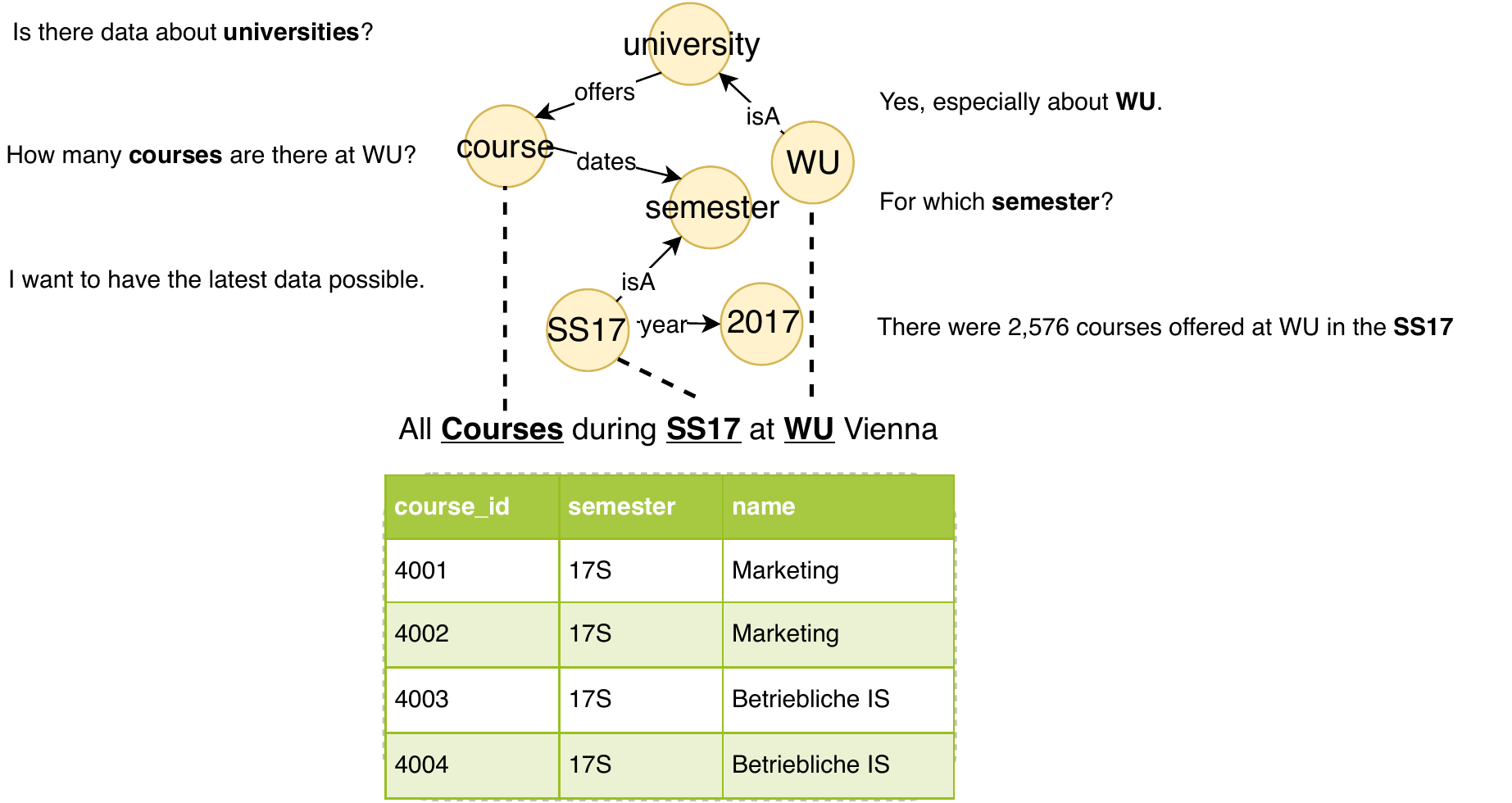}
\caption{Aligning knowledge and conversation.}
\label{fig:wuODgraph}
\end{figure}

Let us go back to the information-seeking dialogue example that we introduced before.
Both conversation participants share common knowledge about the conversation domain (Figure~\ref{fig:wuODgraph}).
This shared knowledge allows them to recognize the semantic relations between the concepts mentioned in the conversation, such as universities, courses and semesters and produce appropriate responses.
For someone who does not possess the prerequisite knowledge necessary for a correct interpretation this conversation will appear nonmeaningful or even misleading.
Figure~\ref{fig:wuODgraph} includes a snippet of the real Open Data table released by the Economic University of Vienna (WU).\footnote{\url{https://data.wu.ac.at/portal/dataset/all_course_events_2017s}}

Clear advantages of using knowledge models with explicit structure and relation types lie mostly in interpretability and transparency.
KGs give control over the representation and reasoning process that can be traced directly to the visited nodes in case of graph traversal algorithms, which is important for explainability~\cite{DBLP:journals/corr/abs-1811-04540,DBLP:journals/corr/abs-1903-03714}.
The downside is the difficulty to scale curated knowledge resources in comparison with neural language models that can be trained on existing text corpora~\cite{DBLP:conf/acl/LiTTG16,lee2018scaling}.
Semi-automated knowledge base construction is an on-going research direction useful for integration of heterogeneous information sources~\cite{DBLP:conf/wsdm/RenKWS18}.

In Chapter~\ref{chap:coherence} we evaluate the potential of different knowledge models (word embeddings, KGs and KG embeddings) to predict coherent dialogues.
This analysis shows the limitations of current knowledge representation approaches and identifies implications for conversational system design.
There is a clear need for hybrid approaches that could leverage knowledge encoded in both symbolic and neural representations.
In Chapter~\ref{chap:qa} we present a novel approach that leverages a language model to answer complex questions using a knowledge graph.
The language model helps to disambiguate natural language questions and retrieve relevant relations that are stored in a knowledge graph.
In Chapter~\ref{chap:browsing} we show how basic information-theoretic measures can be applied to the graph structure of a background knowledge to create a novel conversational browsing interface for exploratory search.
This idea transcends the boundaries of the standard query-response paradigm and is based on the analysis of real information-seeking dialogues reported in Chapter~\ref{chap:structure} introducing the QRFA model.

The approaches we present in this thesis are designed to be domain independent.
We analyse patterns and design functionality that can be applied in different scenarios where information-seeking dialogues can occur.
Chapter~\ref{chap:structure} validates the hypothesis that information-seeking dialogues across different domains (Open Data search, web search, querying bus schedules or restaurant reservation databases) share common structural properties.
The running example of Open Data search introduced in this chapter, i.e. providing access to publicly available structured data sources, is further developed through Chapters~\ref{chap:structure} and \ref{chap:browsing}.

We foresee that individual tables will be integrated in the future through an annotation process providing links between the datasets on the level of columns, rows and entities they contain~\cite{neumaier2017data,neumaier2019enabling}.
Therefore, our goal is enabling dialogue-based access to knowledge graphs.
In Chapter~\ref{chap:qa} we show how a large-scale knowledge graph, which contains billions of facts~\cite{DBLP:journals/semweb/LehmannIJJKMHMK15}, can be efficiently used to provide answers to complex questions posed in natural language, which is the task relevant for any conversational system that aims to support access to encyclopedic knowledge.
Finally, we consider human conversations, such as Ubuntu trouble-shooting discussions in an on-line chat room (Chapter~\ref{chap:coherence}), to be an important source of real-life samples of information-seeking dialogues that should inspire and inform design of conversational search systems.

\chapter{Functional Structure of a Conversation}\blfootnote{This chapter was published as \citep{DBLP:journals/corr/abs-1812-10720}.}
\label{chap:structure}

    \epigraph{
- Ask me a question…\\
- I really like when you ask the questions.\\
- Tell me something…\\
- I read on Wikipedia that in the stage version of “The Wizard of Oz”, Dorothy’s little dog Toto was replaced by a cow named Imogene. \\ --- from a conversation with Google Assistant, 27 April 2019
    }

We believe that the goal for conversational systems is not to replicate already existing search functionalities enhanced with voice control and speech recognition features but to aim at creating qualitatively different interaction experiences that can resemble negotiation processes that occur between humans in real-life settings.
This process should lead to a better understanding of what the information need is and what information there is to satisfy this need.
The ability to maintain balance in mixed-initiative interactions is crucial for the conversational success and for the usability of a dialogue system~\cite{DBLP:conf/acl/WalkerW90}.
Both participants of an information-seeking dialogue should be able to ask clarifying questions that can help to understand the position and intentions of the conversation partner.
There is still a limited understanding of how mixed-initiative interactions in the context of information-seeking behaviour actually occur~\cite{trippas2018informing,DBLP:journals/jd/Savolainen18}

We propose a novel approach to analyse information-seeking conversations that aim at understanding the structure of the interaction processes and developing a grounded theory, which can help to improve conversational search systems. 
Analyzing an interaction process boils down to discovering patterns in sequences of alternating utterances exchanged between a user and an agent. 
Process mining techniques have been successfully applied to analyze structured event logs, discovering the underlying process models or evaluating whether the observed behavior is in conformance with a known pre-specified process. 
In this chapter, we apply process mining techniques to discover patterns in conversational transcripts and extract a new model of information-seeking dialogues, QRFA, for Query, Request, Feedback, Answer.
Our results are grounded in an empirical evaluation across multiple conversational datasets from different domains, which was never attempted before.
We show that the QRFA model better reflects conversation flows observed in real informa\-tion-seeking conversations than models proposed previously. 
Moreover, QRFA allows us to identify malfunctioning in dialogue system transcripts as deviations from the expected conversation flow described by the model via conformance analysis.

\section{Introduction}
\label{sec:intro}

Interest in information-seeking dialogue systems is growing rapidly, in information retrieval, language technology, and machine learning.
There is, however, a lack of theoretical understanding of the functionality such systems should provide~\cite{trippas2018informing}.
Different information-seeking models of dialogue systems use different terminology as well as different modeling conventions, and conversational datasets are annotated using different annotation schemes~\cite[see, e.g.,][]{trippas2016how,DBLP:journals/dad/WilliamsRH16a}.
These discrepancies hinder direct comparisons and aggregation of the results.
Moreover, the evaluation of conversational datasets has largely been conducted based on manual efforts.
Clearly, it is infeasible to validate models on large datasets without automated techniques.

In this chapter we describe and demonstrate the application of a data-driven approach that can be applied to a large volume of conversational data, to identify patterns in the conversation dynamics. 
It is directly rooted in Conversational Analysis (CA)~\cite{schegloff1968sequencing}, which proposes to analyze regularities such as adjacency pairs and turn-taking in conversational structures, and Speech Act Theory~\cite{searle1969speech,austin1976things} to identify utterances with functions enabled through language (speech acts). 
To this end we leverage state-of-the-art techniques developed in the context of process mining~\cite{Aalst16}, which has traditionally been applied in the context of operational business processes such as logistics and manufacturing, to discover and analyze patterns in sequential data.

Against this background, we create a new annotation framework that is able to generalize across conversational use cases and bridge the terminology gap between diverse theoretical models and annotation schemes of the conversational datasets collected to date. 
We develop and evaluate a new information-seeking model, which we name QRFA, for Query, Request, Feedback, Answer, which shows better performance in comparison with previously proposed models and helps to detect malfunctions from dialogue system transcripts.
It is based on the analysis of 15,931 information-seeking dialogues and evaluated on the task of interaction success prediction in 2,118 held-out dialogues.
We use the COR model~\cite{sitter1992modeling} as our baseline (see Section~\ref{sec:models}) and show in an empirical evaluation that it is not able to adequately reflect the structure of information-seeking dialogues across four publicly available datasets and propose an alternative model.

The QRFA model is derived and evaluated using process mining techniques~\cite{Aalst16}, which makes this approach scalable. 
We view every conversation to be an instance of a general information-seeking process. 
This inclusive perspective helps us to extract and generalize conversation flows across conversations from different domains, such as bus schedules and dataset search.

To the best of our knowledge, we present the first grounded theory of information-seeking dialogues that is empirically derived from a variety of conversational datasets.
Moreover, we describe the methodology we used to develop this theory that can be used to revise and further extend the proposed theory.
We envision that the model and the approach that we describe in this chapter will help not only to better understand the structure of information-seeking dialogues but also to inform the design of conversational search systems, their evaluation frameworks and conversational data sampling strategies.
More concretely, we discovered a set of functional components for a conversational system as different interaction patterns and the distribution over the space of next possible actions.

The remainder of the chapter is organized as follows. 
Section~\ref{sec:background} provides a gentle introduction to process mining. 
In Section~\ref{sec:related_work3} we discuss related work in the context of process mining from conversations. 
Section~\ref{sec:approach3} provides details of our approach to mining processes from conversations. 
In Section~\ref{sec:experiments} we report on the results of applying our conversation mining approach to several conversational datasets and we describe the model we obtained as a result. 
We conclude in Section~\ref{sec:conclusion3}.

\section{Background}
\label{sec:background}

Process mining (PM) has been designed to deal with structured data organized into a process log rather than natural language, such as conversational transcripts. 
However, we view a conversation as a sequence of alternating events between a user and an agent, thus a special type of process, a communication or information exchange process, that can be analyzed using PM by converting conversational transcripts into process logs. 
Basic concepts and techniques from PM, which we adopt in our discourse analysis approach, are described below.

A process is a structure composed of events aligned between each other in time. 
The focus of PM is on extracting and analyzing process models from event logs. 
Each event in the log refers to the execution of an activity in a process instance. 
Additional information such as a reference to a resource (person or device) executing the activity, a time stamp of the event, or data recorded for the event, may be available.

Two major tasks in PM are \emph{process discovery} and \emph{conformance checking}. 
The former is used to extract a process model from an event log, and the latter to verify the model against the event log, i.e., whether the patterns evident from the event log correspond to the structure imposed by the model. 
It is possible to verify conformance against an extracted model as well as against a theoretical (independently constructed) model.

In this work, we adopt state-of-the-art PM techniques to analyze conversational transcripts by extracting process models from publicly available datasets of information-seeking dialogues, and to verify and further extend a theoretical model of information-seeking dialogues based on the empirical evidence from these corpora.

\section{Related Work}
\label{sec:related_work3}

There are relatively few prior publications that demonstrate the benefit of applying process mining (PM) techniques to conversational data.  
\citet{DBLP:journals/internet/CiccioM13} use a corpus of e-mail correspondence to illustrate how the structure of a complex collaborative process can be extracted from message exchanges. 
\citet{wang2015analytical} analyze a sample of discussion threads from an on-line Q\&A forum by applying process mining and network analysis techniques and comparing patterns discovered across different thread categories based on their outcomes (solved, helpful and unhelpful threads).

\citet{DBLP:conf/bpm/RichettiGBS17} analyze the performance of a customer support service team by applying process mining to conversational transcripts that were previously annotated with speech acts using a gazetteer. 
Their results reveal similar structural patterns in the conversational flow of troubleshooting conversations with different durations, i.e., less and more complex cases, which require additional information seeking loops.

To the best of our knowledge, there is no prior work going beyond the individual use cases mining conversations from a specific domain. 
In contrast, we analyze multiple heterogeneous conversational datasets from various domains to be able to draw conclusions on structural similarities as well as differences stemming from variance introduced by labeling approaches and specific characteristics of the underlying communication processes. 

Before applying the proposed approach, utterances have to be annotated with activity labels, such as speech acts~\cite{searle1976classification}. 
The task of utterance classification is orthogonal to our work. 
Dialogue corpora to be used for process mining can be manually annotated by human annotators or automatically by using one of the classification approaches proposed earlier~\cite{stolcke2000dialogue,DBLP:conf/emnlp/CohenCM04,DBLP:conf/emnlp/JeongLL09,DBLP:conf/emnlp/JoYJR17}.

\section{Conversation Mining}
\label{sec:approach3}


We consider every conversation $C$ in a transcript $\mathcal{C}$ to be an instance of the same communication process, the model of which we aim to discover. 
A conversation is represented as a sequence of utterances $C = \langle u_1, u_2, \ldots\rangle$.
An \textit{utterance}, in this case, is defined rather broadly as a text span within a conversation transcript attributed to one of the conversation participants and explicitly specified during the annotation process (utterance labeling step).
We denote the set of all utterances as $U = \{ u_1, u_2, \ldots\}$. 
Our approach to conversational modeling consists of three steps: (1)~utterance labeling, (2)~model discovery, and (3)~conformance checking, which is useful for model validation and error detection in conversation transcripts.



\subsection{Utterance labeling}

A utterance $u_i$ in a conversation $C$ can be mapped to multiple labels $l^1_i,\ldots,l^n_i$, each belonging to pre-defined label sets $L^1,\ldots,L^n$, respectively.
The label sets, not necessarily disjoint, may correspond to different annotation schemes.
We denote the general multi-labeling of utterances as a mapping function $\hat{\lambda} : U \to L_1 \times \cdots \times L_n$.
It can stem from manual annotations of different human analysts, or categories returned by multiple machine-learned classifiers.
For the sake of readability, we assume in the remainder of this section that all annotations share a single set of labels $L$, thus the mapping function used henceforth is reduced to $\lambda : U \to L$.

\subsection{Extracting the model of the conversation flow}

We apply a process discovery approach to collect patterns of the conversation structure from  transcripts. 
The goal of process discovery is to extract a model that is representative of empirically observed behavior
stored in an \emph{event log}~\cite{Aalst16}.
Event logs can be abstracted as sets of sequences (\emph{traces}), where each element in the sequence (the \emph{event}) is labeled with an activity (\emph{event class}) plus optional \emph{attributes}.
We reduce a conversation transcript $\mathcal{C}$ along with its labeling to an event log, as follows:
a conversation $C$ is a trace, an utterance $u$ is an event, and the label of $u$, $\lambda(u)=l$, is the event class.

There is a wide variety of algorithms that can be applied for process discovery.
Imperative workflow mining algorithms, such as the seminal $\alpha$-algorithm~\cite{Aalst16} or the more recent Inductive Miner (IM)~\cite{DBLP:conf/bpm/LeemansFA14}, extract procedural process models that depict the possible process executions, in the form of, e.g., a Petri net~\cite{Aalst/JCSC1998:ApplicationofPetriNetsToWorkflowManagement}.
Other approaches, such as frequent episode mining~\cite{DBLP:conf/simpda/LeemansA14} or declarative constraints mining~\cite{DiCiccio.Mecella/CIDM2013:TwoStepFast,DBLP:journals/tmis/CiccioM15},
extract local patterns and aggregate relations between activities.
%
%
%
One such relation is the \emph{succession} between two activities, denoting that the second one occurs eventually after the first one. 
In our context, succession between $l_i$ and $l_j$ holds true in a sub-sequence $\langle u_i, \ldots, u_j\rangle$, with $i < j$, if $u_j \mapsto l_j$ and $u_i \mapsto l_i$. 
The frequency of such patterns observed across conversations can indicate dependencies between the utterance labels $l_i$ and $l_j$. Those dependencies can be used to construct a model describing a frequent behavior (model discovery) as well as to detect outliers breaking the expected sequence (error analysis), upon the setting of thresholds for minimum frequency.
The discovery algorithm described in \cite{DBLP:journals/tmis/CiccioM15} requires a linear pass through each sequence to count, for every label $l \in L$,
\begin{inparaenum}[(1)]
	\item its number of occurrences per sequence, and 
	\item the distance at which other labels occur in the same sequence in terms of number of utterances in-between.
\end{inparaenum}

\subsection{Conformance analysis and model validation}

The goal of the model validation step is conformance checking, i.e., to assess to which extent the patterns evident from the transcript fit the structure imposed by the model.
We use it here to also evaluate the predictive power of the model, i.e., the ability of the model to generalize to unseen instances of the conversation process. 
A good model should fit the transcripts but not overfit it.
The model to be validated against can be the one previously extracted from transcripts, or a theoretical model, i.e., an independently constructed one. 
In the former case, we employ standard cross-validation techniques by creating a test split separate from the development set that was used to construct the model.

Model quality can be estimated with respect to a conversation transcript. Likewise, the quality of the conversation can be estimated with respect to a pre-defined model. 
In other words, discrepancies between a conversation model and a transcript indicate either inadequacy of the model or errors (undesired behavior or recording malfunctions) in the conversation. 

To compute fitness, i.e., the ability of the model to replay the event log, we consider the measure first introduced in \cite{AalstAD12WIDM}, based on the concept of alignment.
Alignments keep consistent the replay of the whole sequence and the state of the process by adding so-called non-synchronous moves if needed.
The rationale is, the more non-synchronous moves are needed, the lower the fitness is.
Thus, a penalty is applied by means of a cost function on non-synchronous moves.
For every sequence fitness is computed as the complement to 1 of the total cost of the optimal alignment, divided by the cost of the worst-case alignment.
Log fitness is calculated by averaging the sequence fitness values over all sequences in a log.

\section{QRFA Model}
\label{sec:experiments}

In this section, we apply the approach to extraction and validation of a conversational model proposed in Section~\ref{sec:approach} to develop a new model (QRFA). 
We (1)~collect \textit{datasets} of publicly available corpora with information-seeking dialogue transcripts; (2)~analyze and link their annotation schemes to each other and to the COR model (Figure~\ref{fig:model_sitter}); (3)~analyze the conversation flows in the datasets; and (4)~\textit{evaluate} QRFA and compare the results with COR as a baseline.

\subsection{Conversational datasets}
We used publicly available datasets of information-seeking dialogues that are annotated with utterance-level labels (see the dataset statistics in Table~\ref{datasets-table}).

\begin{table}[h]
\centering
	\BottomFloatBoxes
		\ttabbox{%
			\begin{small}
				\begin{tabular}{l rrrr }
					\toprule
					\bf Dataset & \bf Dialogues & \bf Utterances & \bf Labels &  \\ \midrule
					SCS         & 39            &            101 &         13 &  \\
					ODE         & 26            &            417 &         20 &  \\
					DSTC1       & 15,866        &        732,841 &         37 &  \\
					DSTC2       & 2,118         &         40,854 &         21 &  \\ \bottomrule
				\end{tabular}
			\end{small}
		}{%
			\caption{Dataset statistics.}%
			\label{datasets-table}%
		}
		\hspace*{-.5em}
		\ttabbox{%
			\centering
			\begin{small}
				\begin{tabular}{l @{\hspace{.5em}}ll >{\hspace{1em}}l @{\hspace{.5em}}ll }
					\toprule
					& \multicolumn{2}{c}{Proactive}  &  &  \multicolumn{2}{c}{Reactive}   \\
					\cmidrule{2-3} \cmidrule{5-6}
					\textit{User} & \textbf{Q}uery   & Information &  & \textbf{F}eedback & Positive    \\
					&                  & Prompt      &  &                   & Negative    \\
					\textit{Agent}                              & \textbf{R}equest & Offer       &  & \textbf{A}nswer   & Results     \\
					&                  & Understand  &  &                   & Backchannel \\
					&      \multicolumn{2}{c}{}      &  &                   & Empty       \\ \bottomrule
				\end{tabular} 
			\end{small}%
		}{%
			\caption{New functional annotation schema for information-seeking conversation utterances.} %
			\label{schema-table} %
		}
\end{table}

\textbf{Spoken Conversational Search.} This dataset\footnote{\url{https://github.com/JTrippas/Spoken-Conversational-Search}}~\cite[\textit{SCS},][]{trippas2016how} contains human-human conversations collected in a controlled laboratory study with 30 participants.
The task was designed to follow the setup, in which one of the conversation participants takes over the role of the information Seeker and another of the Intermediary between the Seeker and the search engine.
It is the same dataset that was used to develop the SCS model illustrated in Figure~\ref{fig:model_trippas}. 
All dialogues in the dataset are very short and contain at most three turns, with one label per utterance. 
The efficiency of the interaction and the user satisfaction from the interaction are not clear.


\textbf{Open Data Exploration.} This dataset\footnote{\url{https://github.com/svakulenk0/ODExploration_data}} (\textit{ODE}) was collected in a laboratory study with 26 participants and a setup similar to the SCS but with the task formulated in the context of conversational browsing, in which the Seeker does not communicate an explicit information request. 
The goal of the Intermediary is to iteratively introduce and actively engage the Seeker with the content of the information source. 
All dialogues in this dataset contain one label per utterance. 
The majority of the conversation transcripts (92\%) exhibit successful interaction behavior leading to a positive outcome, such as satisfied information need and positive user feedback (only 2 interactions were unsuccessful), and can be considered as samples of effective information-seeking strategies.

\textbf{Dialog State Tracking Challenge.} These datasets%
\footnote{\url{https://www.microsoft.com/en-us/research/event/dialog-state-tracking-challenge}}~\cite[\textit{DSTC1} and \textit{DSTC2}, ][]{DBLP:journals/dad/WilliamsRH16a} provide annotated human-computer dialogue transcripts from an already implemented dialogue system for querying bus schedules and a restaurant database. 
The transcripts may contain more than one label per utterance, which is different from the previous two datasets. 
The efficiency of the interaction and user satisfaction from the interaction with the agent are not clear.

\subsection{QRFA model components}
\label{sec:qrfa_labels}

Since all datasets and the theoretical model that we consider use different annotation schemes, we devise a single schema to be able to aggregate and compare conversation traces.
To the best of our knowledge, no such single schema that is able to unify annotations across a diverse set of information-seeking conversation use cases has been proposed and evaluated before.
Our schema is organized hierarchically into two layers of abstraction to provide a more simple and general as well as more fine-grained views on the conversation components.

First, we separate utterances into four basic classes: two for User (Query and Feedback) and two for Agent (Request and Answer).
This distinction is motivated by the role an utterance plays in a conversation.
Some of the utterances explicitly require a response, such as a question or a request, while others constitute a response to the previous utterance, such as an answer.
Such a distinction is reminiscent of the Forward and Backward Communicative Functions that are foundational for the DAMSL annotation scheme~\cite{core1997coding}. 
The labels also reflect the roles partners take in a conversation. 
The role of the Agent is to provide Answers to User's Queries. During the conversation the Agent may Request additional information from the User and the Agent may provide Feedback to the Agent's actions. 

The initial set of four labels (QRFA) are further subdivided to provide a more fine-grained level of detail.
See Table~\ref{schema-table} and the descriptions below:

\textbf{Q}uery provides context (or input) for \textit{Information} search (question answering), as the default functionality provided by the agent (e.g., ``Where does ECIR take place this year?''), but can also \textit{Prompt} the agent to perform actions, such as cancel the previous query or request assistance, e.g., ``What options are available?''

\textbf{R}equest is a pro-active utterance from the agent, when there is a need for additional information (Feedback) from the user. It was the only class that caused disagreement between the annotators, when trying to subdivide it into two groups of requests: the ones that contain an \textit{Offer}, such as an offer to help the User or presenting the options available (e.g., ``I can group the datasets by organization or format''), and the ones whose main goal is to \textit{Understand} the user need, such as requests to repeat or rephrase the Query (e.g., ``Sorry I am a bit confused; please tell me again what you are looking for''). 

\textbf{F}eedback from the user can be subdivided by sentiment into \textit{Positive}, such as accept or confirm, and \textit{Negative}, such as reject or be discontented.
 
\textbf{A}nswer corresponds to the response of the agent, which may contain one of the following: (1)~\textit{Results}, such as a search engine result page (SERP) or a link to a dataset, (2)~\textit{Backchannel} response to maintain contact with the User, such as a promise or a confirmation (e.g., ``One moment, I'll look it up.''), and (3)~\textit{Empty} result set (e.g., ``I am sorry but there is no other Indian restaurant in the moderate price range'').

Two authors of the paper on which this chapter is based independently aligned the annotation schemes of the datasets and the COR model to match the single schema with an inter-annotator agreement of 94\%.
We found the first more abstract level of annotation sufficient for our experiments to make the conversation models easier to interpret. 
The complete table containing alignments across the schemes is made available to the community to enable reproducibility and encourage future work in this direction (Tables~\ref{table:schemas1} and \ref{table:schemas2}).\footnote{\url{https://github.com/svakulenk0/conversation_mining/blob/master/annotations/alignments_new.xls}}

The sample conversation introduced in Chapter~\ref{chap:background} will receive the following annotations in accordance with the QRFA schema

\textbf{Q} How many courses are there at WU? \\
\textbf{R} For which semester? \\
\textbf{Q} I want to have the latest data possible. \\
\textbf{A} There were 2,576 courses offered at WU in the SS17. \\
\textbf{F} Thank you!

\subsection{QRFA model dynamics}

\begin{figure}[tb]
    \centering
    \includegraphics[width=0.5\textwidth]{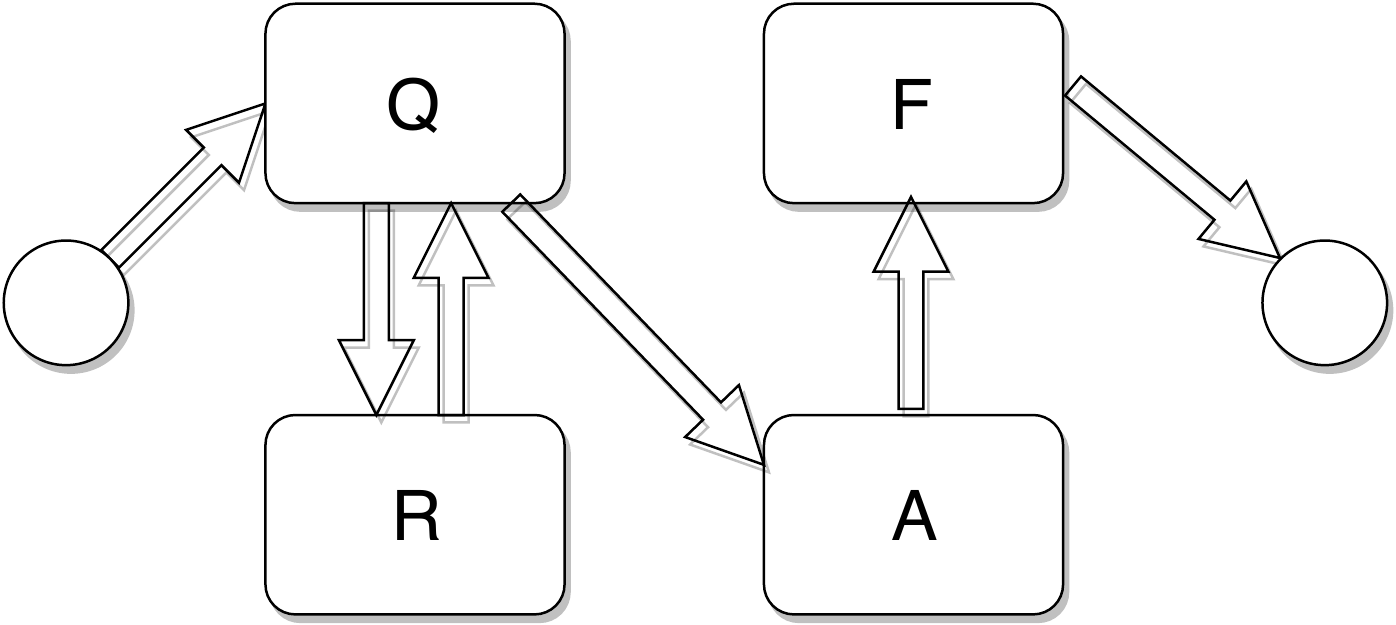} %
    \caption{A process model extracted from a sample conversation.} %
    \label{fig:sample_trace}
\end{figure}

The next step of conversation mining is building a process model given the dialogue annotations.
For the running example introduced in Section~\ref{sec:qrfa_labels} the process model corresponds to the sequence of labels that models a single trace, i.e. an instance of the information-seeking process: $\textbf{START} \rightarrow \textbf{Q} \rightarrow \textbf{R} \rightarrow \textbf{Q} \rightarrow \textbf{A} \rightarrow \textbf{F} \rightarrow \textbf{END}$.
For brevity we collapse all duplicate labels into a single node, i.e. one node per label, and obtain a directed cyclic graph (see Figure~\ref{fig:sample_trace}).

\begin{figure}[tb]
    \centering
    \includegraphics[trim={0 6.5cm 0 0},clip,width=0.5\textwidth]{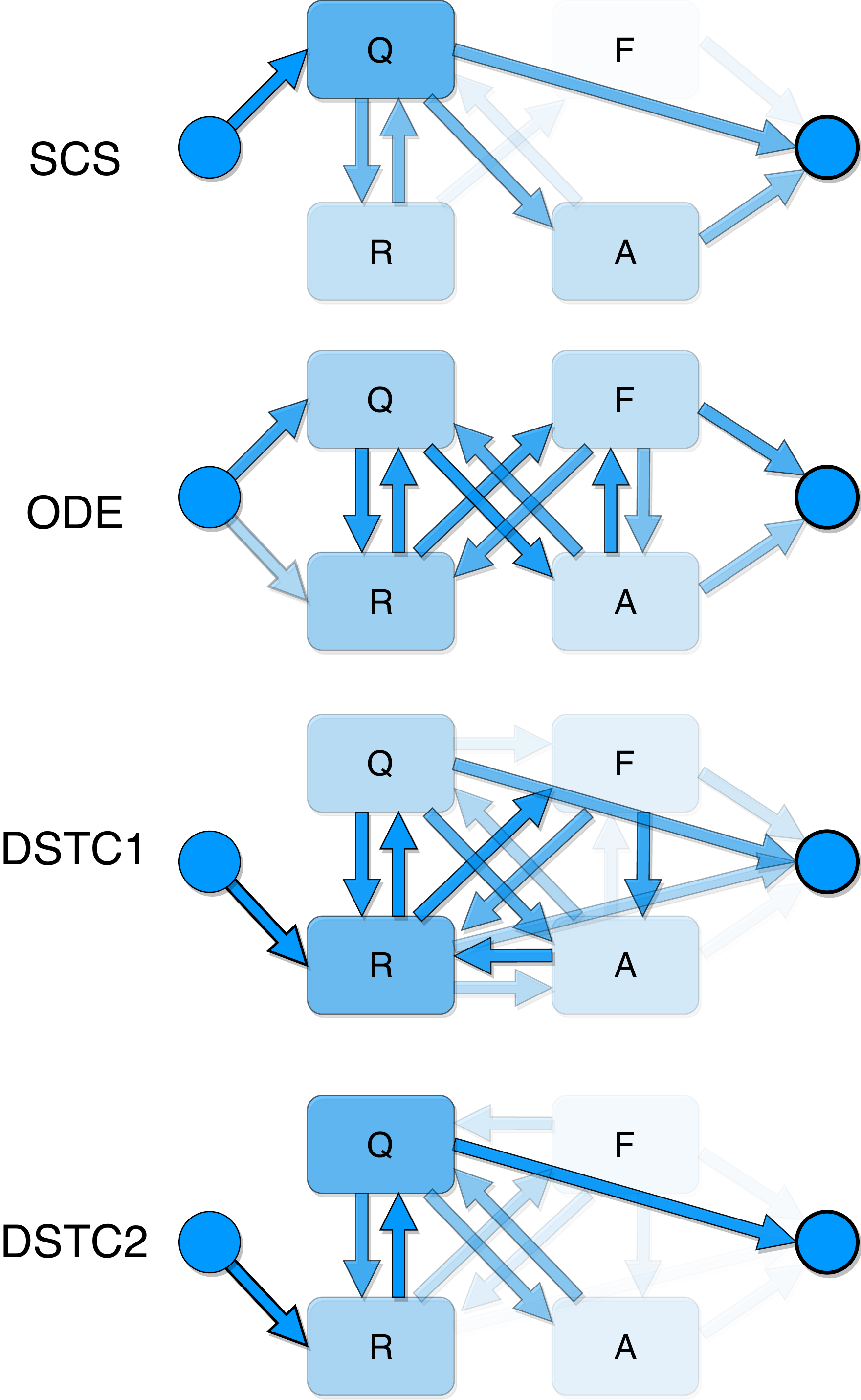} %
    \caption{Conversation flows in the SCS, ODE and DSTC1 datasets. Color intensity indicates frequency.} %
    \label{fig:conversation_flows}
\end{figure}

We used the ProM Episode Miner plug-in~\cite{DBLP:conf/simpda/LeemansA14} and a declarative process mining tool, MINER\-ful%
\footnote{\url{https://github.com/cdc08x/MINERful}}~\cite{DiCiccio.Mecella/CIDM2013:TwoStepFast,DBLP:journals/tmis/CiccioM15}, to discover frequent sequence patterns in the conversation transcripts.
Figure~\ref{fig:conversation_flows} illustrates the conversation flows in each of the three datasets used for model discovery (one of the datasets, DSTC2, is held out for model evaluation).
Color intensity (opacity) indicates the frequency of the observed sequences between the pairs of utterances within the respective dataset (the frequency counts for all transitions across all the datasets are available on-line\footnote{\url{https://github.com/svakulenk0/conversation_mining/tree/master/results/}}).
An empirically derived information-seeking conversation model would be the sum of the models extracted from the three conversation transcripts.

\begin{figure}[tb]
    \centering
	\includegraphics[width=0.6\textwidth, trim= -5mm 0 0 0]{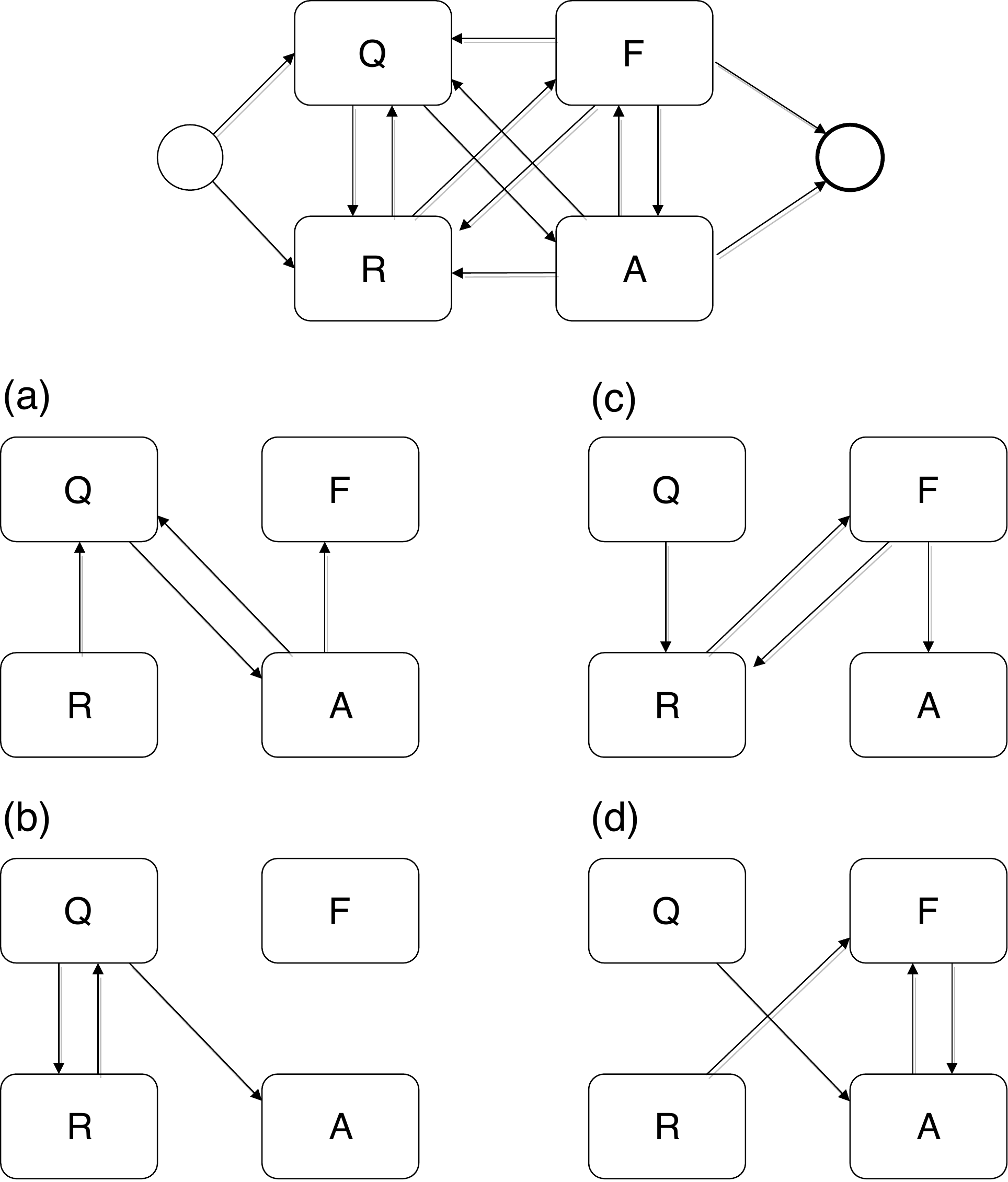} %
	\caption{The QRFA model for conversational search composed of the ``four virtuous cycles of information-seeking'': (a) question answering loop, (b) query refinement loop, (c) offer refinement loop, (d) answer refinement loop.} %
	\label{fig:cycles} %
\end{figure}

However, an empirically derived model guarantees neither correctness nor optimality since the transcripts (training data) may contain errors, i.e., negative patterns.
Instead of blindly relying on the empirical ``as is'' model, we analyze and revise it (re-sample) to formulate our theoretical model of a successful information-seeking conversation (Figure~\ref{fig:cycles}).
For example, many conversations in the SCS and DSTC1 datasets are terminated right after the User Query (Q$\rightarrow$END pattern) for an unknown reason, which we consider to be undesirable behavior: the User's question is left unanswered by the Agent.
Therefore, we discard this transition from our prescriptive model, which specifies how a well-structured conversation ``should be'' (Figure~\ref{fig:cycles}).
Analogously to discarding implausible transitions, the power of the theoretical modeling lies in the ability to incorporate transitions are still considered legitimate from a theoretical point of view even though they were not observed in the training examples.
Incompleteness in empirically derived models may stem from assumptions already built into the systems by their designers, when analyzing dialogue system logs, or also differences in the annotation guidelines, e.g., one label per utterance constraint.
In our case, we noticed that adding the FQ transition, which was completely absent from our training examples, will make the model symmetric.
The symmetry along the horizontal axes reflects the distribution of the transitions between the two dialogue partners.
Hence, the FQ transition mirrors the AR transition, which is already present in our transcripts, but on the User side.
The semantics of an FQ transition is that the User can first give feedback to the Agent and then follow up with another question.
\citet[Figure 1, Example 1]{trippas2018informing} empirically show that utterances in information-seeking dialogues tend to contain multiple moves, i.e., can be annotated with multiple labels.

The final shape of a successful information-seeking conversation according to our model is illustrated in Figure~\ref{fig:cycles}.
To analyze this model in more detail, we decompose it into a set of connected components, each containing one of the cycles from the original model. 
We refer to them as ``four virtuous cycles of information-seeking,''\footnote{A virtuous cycle refers to complex chains of events that reinforce themselves through a feedback loop. A virtuous circle has favorable results, while a vicious circle has detrimental results.} representing the possible User-Agent exchanges (feedback loops) in the context of: (a)~question answering, (b)~query refinement, (c)~offer refinement, and (d)~answer refinement. 
To verify that the loops actually occur and estimate their frequencies, we mined up to 4-label sequences from the transcripts using the Episode Miner plug-in.

\afterpage{\begin{landscape}
\begin{table}[]
\begin{tabular}{|l|ll|ll|}
\hline
\multicolumn{1}{|c|}{} & \multicolumn{4}{c|}{\cellcolor[HTML]{FFFC9E}User} \\
\multicolumn{1}{|c|}{} & \multicolumn{2}{c}{\cellcolor[HTML]{FFFC9E}\textbf{Q}uery} & \multicolumn{2}{c|}{\cellcolor[HTML]{FFFC9E}\textbf{F}eedback} \\
\multicolumn{1}{|c|}{\multirow{-3}{*}{Models}} & \multicolumn{1}{c}{\cellcolor[HTML]{FFFC9E}\textbf{Information}} & \multicolumn{1}{c}{\cellcolor[HTML]{FFFC9E}\textbf{Prompt}} & \multicolumn{1}{c}{\cellcolor[HTML]{FFFC9E}\textbf{Positive}} & \multicolumn{1}{c|}{\cellcolor[HTML]{FFFC9E}\textbf{Negative}} \\
\hline
COR & request & withdraw & accept & reject \\
10 labels &  &  & be contented & be discontented \\
\hline
CS~\cite{DBLP:conf/chiir/RadlinskiC17} & rt &  & rr &  \\
10 labels &  &  & rp &  \\
 &  &  & rnp &  \\
 &  &  &  & rc \\
\hline
SCS & Initial information request & SERP information request & Confirms &  \\
13 labels & Intent clarification &  &  &  \\
 & Query repeat &  &  &  \\
 & Query embellishment &  &  &  \\
\hline
ODE & set(keywords) & question(data) & confirm() & reject() \\
17 labels &  & more() & success() &  \\
 &  & prompt(link) &  &  \\
 &  & verify() &  &  \\
\hline
DSTC1 & inform & restart & affirm & negate \\
32 labels & nextbus & repeat &  &  \\
 & prevbus & tellchoices &  &  \\
 &  & goback &  &  \\
\hline
DSTC2 & inform &  & ack & negate \\
22 labels & request &  & affirm &  \\
 & confirm &  & thankyou &  \\
 & deny &  &  &  \\
 & repeat &  &  &  \\
 & reqalts &  &  & \\
 \hline
\end{tabular}
{
\caption{Schema alignments (Part 1).}%
\label{table:schemas1}%
}
\end{table}
\begin{table}[]
\centering\small
\begin{tabular}{|l|ll|lll|}
 \hline
\multicolumn{1}{|c|}{} & \multicolumn{5}{c|}{\cellcolor[HTML]{DAE8FC}Agent} \\
\multicolumn{1}{|c|}{} & \multicolumn{2}{c}{\cellcolor[HTML]{DAE8FC}\textbf{R}equest} & \multicolumn{3}{c|}{\cellcolor[HTML]{DAE8FC}\textbf{A}nswer} \\
\multicolumn{1}{|c|}{\multirow{-3}{*}{Models}} & \multicolumn{1}{c}{\cellcolor[HTML]{DAE8FC}\textbf{Offer}} & \multicolumn{1}{c}{\cellcolor[HTML]{DAE8FC}\textbf{Understand}} & \multicolumn{1}{c}{\cellcolor[HTML]{DAE8FC}\textbf{Results}} & \multicolumn{1}{c}{\cellcolor[HTML]{DAE8FC}\textbf{Backchannel}} & \multicolumn{1}{c|}{\cellcolor[HTML]{DAE8FC}\textbf{Empty}} \\
 \hline
COR & offer &  & assert & promise & reject \\
10 labels & withdraw &  &  &  &  \\
 \hline
CS~\cite{DBLP:conf/chiir/RadlinskiC17} & a0 &  & a1i &  &  \\
10 labels & a1p &  & a2+i &  &  \\
 & a2+p &  &  &  &  \\
 &  &  &  &  &  \\
  \hline
SCS & Query refine- & Asks to & SERP w/o modification & Confirms &  \\
13 labels & ment offer & repeat & SERP with modification &  &  \\
 &  &  & SERP with modification + Suggestion &  &  \\
 &  &  & Scanning document theme &  &  \\
  \hline
ODE & list(keywords) & prompt(keywords) & bool(data) & confirm() &  \\
17 labels & prompt(link) & verify() & count(data) & success() &  \\
 &  &  & top(keywords) &  &  \\
 &  &  & link(dataset) &  &  \\
  \hline
DSTC1 & request &  & schedule & ack & sorry \\
32 labels & open-request & are-you-there & morebuses & hold-on & canthelp.cant\_find\_stop \\
 & example & bebrief &  & impl-conf & canthelp.from\_equals\_to \\
 &  & expl-conf &  &  & canthelp.no\_buses\_at\_time \\
 &  & please-repeat &  &  & canthelp.no\_connection \\
 &  & please-rephrase &  &  & canthelp.nonextbus \\
 &  &  &  &  & canthelp.route\_doesnt\_run \\
 &  &  &  &  & canthelp.system\_error \\
 &  &  &  &  & canthelp.uncovered\_route \\
 &  &  &  &  & canthelp.uncovered\_stop \\
  \hline
DSTC2 & request &  & inform & impl-conf & canthelp \\
22 labels & welcomemsg & expl-conf & offer &  & canthelp.exception \\
 & reqmore & repeat &  &  &  \\
 & select & confirm-domain &  &  & \\
  \hline
\end{tabular}
{
\caption{Schema alignments (Part 2).}%
\label{table:schemas2}%
}
\end{table}
\end{landscape}}

\subsection{QRFA model evaluation}
Our evaluation of the QRFA model is twofold. 
Firstly, we measure model fitness with respect to the conversational datasets including a held-out dataset (DSTC2), which was not used during model development, to demonstrate the ability of the model to fit well across all available datasets and also generalize to unseen data. 
Secondly, we hypothesize that deviations from the conversation flow captured in the QRFA model signal anomalies, i.e., undesired conversation turns. Therefore, we also compare the model's performance on the task of error detection in conversational transcripts with human judgments of the conversation success.

\textbf{Fitness and generalization.}
To analyze the model fit with respect to the actual data we applied the conformance checking technique proposed by \citet{AdriansyahDA11EDOC}, available as a ProM plug-in under the name ``Replay a Log on Petri Net for Conformance Analysis.'' 
To this end, we translated the COR and QRFA models into the Petri net notation and ran a conformance analysis for each model on every dataset.
Exit and Restart activities are part of the ``syntactic sugar'' added for the Petri net notation and we set them to \textit{invisible} in order to avoid counting them, when assigning the costs during analysis.
Table~\ref{table:evaluation} (top) contains the fitness measures of the COR and QRFA models for all the datasets. 
We use the default uniform cost function that assigns a cost of~1 to every non-synchronous move. 
Fitness is computed separately for each sequence (dialogue) as a proportion of the events that are in alignment with the model specification.
For the sample conversation introduced in the end of Section~\ref{sec:qrfa_labels} the fitness measure of the QRFA model is 1 (6/6) since all 6 events (transitions) within the dialogue sequence $\textbf{START} \rightarrow \textbf{Q} \rightarrow \textbf{R} \rightarrow \textbf{Q} \rightarrow \textbf{A} \rightarrow \textbf{F} \rightarrow \textbf{END}$ can be completely replayed (simulated) on this model as a finite-state automaton (Figure~\ref{fig:cycles}).
Assume that one of the events in the sequence deviates from the model specification, e.g., $\textbf{START} \rightarrow \textbf{Q} \rightarrow \textbf{R} \rightarrow \textbf{Q} \rightarrow \textbf{A} \rightarrow \textit{Q} \rightarrow \textbf{END}$.
In this case the fitness of the QRFA model with respect to this dialogue sequence is 0.83 (5/6) since one out of 6 events in the sequence ($\textit{Q} \rightarrow \textbf{END}$) does not conform to the model specification.

We measure generalization as the fitness of a model on the sequences that were not used to develop the model.
The ability of the model to generalize to a different held-out dataset is significant ($0.99$ on average). 
This result demonstrates the out-of-sample generalizability of the model, which is a more challenging task than testing the model on the held-out (test) splits from the same datasets (label frequency distributions) used for the model development.
Remarkably, the baseline COR model managed to fully fit only a single conversation across all four datasets.
This comparison clearly shows the greater flexibility that the QRFA model provides, which in turn indicates the requirement for information-seeking dialogue systems to be able to operate in four different IR modes (Figure~\ref{fig:conversation_flows}) and seamlessly switch between them when appropriate.

\textbf{Conversation success and error detection.}
Since only one of the datasets, namely ODE, was annotated with a success score, we add manual annotations for the rest of the datasets (2 annotators, inter-annotator agreement: 0.85).
We produced annotations for 89 dialogues in total, for the full SCS dataset and a random sample for each of the DSTC datasets.\footnote{\url{https://github.com/svakulenk0/conversation\_mining/tree/master/annotations}}
Criteria for the success of a conversational interaction are defined in terms of informational outcomes, i.e., the search results were obtained and the information need was satisfied, as well as emotional outcomes, i.e., whether the interaction was pleasant and efficient.

\begin{table}[t]
\begin{floatrow}
\ttabbox{%
\begin{adjustbox}{width=.98\textwidth}%
\begin{tabular}{lllllllllllllll}
\toprule
\multicolumn{1}{l}{Dataset} & \multicolumn{3}{l}{SCS} & \multicolumn{3}{l}{ODE} & \multicolumn{3}{l}{DSTC1} & \multicolumn{3}{l}{DSTC2} & \multicolumn{2}{l}{Average} 
\\ \cmidrule(r){1-1}\cmidrule(r){2-4}\cmidrule(r){5-7}\cmidrule(r){8-10}\cmidrule(r){11-13}\cmidrule{14-15}
Metric/Model                           & COR      & QRFA     & GS       & COR      & QRFA     & GS       & COR      & QRFA      & GS        & COR      & QRFA      & GS        & COR             & QRFA             \\ \midrule
Average/case                           & 0.58        & 0.89     &          & 0.74        & 1        &          & 0.66        & 0.96      &           & 0.7         & 0.99      &           & 0.67               & 0.96             \\ 
Max.                                   & 0.8         & 1        &          & 1           & 1        &          & 1           & 1         &           & 0.91        & 1         &           & 0.93               & 1                \\ 
Min.                                   & 0.4         & 0.8      &          & 0.6         & 1        &          & 0           & 0         &           & 0.53        & 0.8       &           & 0.38               & 0.65             \\ 
Std. Deviation                         & 0.17        & 0.1      &          & 0.09        & 0        &          & 0.08        & 0.05      &           & 0.05        & 0.02      &           & 0.10               & 0.04             \\ \midrule
Cases with value 1                   & 0           & 0.46     & 0.37     & 0.04        & 1        & 0.92     & 0           & 0.14      & 0.07      & 0           & 0.83      & 0.79      & 0.01               & 0.61             \\ 
Error detection Precision                   &            & 1     &      &         & 1        &      &            &  1    &       &           & 0.67       &       &              &       0.92         \\ 
Error detection Recall                     &            & 0.78     &      &         & 0       &      &            &  0.83    &       &           & 0.57       &       &              &       0.55         \\ 
\bottomrule
\end{tabular}
\end{adjustbox}
}{%
\caption{Evaluation results of QRFA and COR models of information-seeking dialogues on the conversational datasets in terms of model fitness/generalization (top) and error detection abilities (bottom). The gold standard (GS) column refers to the manual annotations of the conversational datasets with the conversation success score (inter-annotator agreement: 0.85).}%
\label{table:evaluation}%
}
\end{floatrow}
\end{table}

Results of the conversation success prediction task are summarized in Table~\ref{table:evaluation} (bottom); QRFA correlates well with human judgments of conversation success based on the model fitness obtained via conformance checking (Cases with value 1).
We also took a closer look at the cases annotated as unsuccessful in terms of fitness to the QRFA model and reported Precision/Recall metrics for the conversation failure detection task.
For example, the model predicted all conversations in the ODE corpus as success (100\% success rate) and overlooked 8\% that actually failed, hence the Recall for conversation failure detection is 0 in this case. 

Table~\ref{table:evaluation} shows that half of the errors affecting conversation success are due to a violation of structural requirements formulated via the QRFA model.
The model overestimates the success rate of a dialog agent since only syntactic information in some cases is not enough to evaluate the overall performance, such as the quality of the answer obtained.
However, it shows very promising results, clearly indicating the faulty cases, such as the situations when the user's query was left unanswered by the agent (SCS and DSTC1).
Our evaluation shows that the QRFA model reflects the patterns of successful information-seeking conversations and the deviations from its shape likely indicate flaws in the conversation flow. 
These results are demonstrated across four conversational datasets from different domains. 
In particular, then, QRFA does not overfit the errors from the datasets used for development and it generalizes to the held-out dataset.

We conclude that the QFRA model satisfies the four quality criteria for a process model defined by \citet{Aalst16}: (1)~fitness -- it fits across four conversational datasets without overfitting, which allows it to successfully detect deviations (errors) in the information-seeking process (Table~\ref{table:evaluation}); (2)~precision -- all types of  interaction described by the model are observed in the conversation transcripts; (3)~generalization -- the model is able to describe the structure and deviations in previously unseen conversations; and (4)~simplicity -- it contains a minimal number of elements necessary to describe the conversation dynamics in information-seeking dialogues.

\section{Conclusion}
\label{sec:conclusion3}

We have proposed an annotation schema and a theoretical model of information-seeking dialogues grounded in empirical evidence from several public conversational datasets.
Our annotation schema resembles the approach used in DAMSL, where utterances are classified into Forward and Backward Communicative Function, but adopts labels to our information-seeking setting, where roles are more distinct due to information asymmetry between participants.
The patterns that we have discovered extend and correct the assumptions built into the COR model and also incorporate frameworks previously proposed within the information retrieval community.

Our empirical evaluation indicates that, however simple, the QRFA model still provides a better fit than the most comprehensive model proposed previously by explaining the conversational flow in the available information-seeking conversational datasets. 
Moreover, we have described an efficient way to provide sanity checking diagnostics of a dialogue system using process mining techniques (conformance checking) and have shown how the QRFA model helps to evaluate the performance of an existing dialogue system from its transcripts. 

In future work, we plan to evaluate the QRFA model against new conversational datasets and further extend it to a finer granularity level if required.
Our experiments so far have utilized hand-labeled conversation transcripts.
Introducing automatically generated labels may propagate errors into the model extraction phase.
Nevertheless, discovering patterns in raw conversational data that is automatically tagged with semantic labels is an exciting research direction~\cite{DBLP:conf/semweb/VakulenkoRCSP18}.
In addition, the predictions of the QRFA model may be an informative signal for evaluating or training reinforcement learning-based dialogue systems~\cite{li-dialogue-2019}.

Wide adoption of information-seeking dialogue systems will lead to a massive increase in conversational data, which can potentially be used for improving dialogue systems. 
We believe that QRFA and similar models will become important for informing the design of dialogue systems, motivating the collection of new information-seeking conversational data, specifying the functional requirements the systems should satisfy, and providing means for their evaluation.

The analysis we performed in this chapter is specifically focused on the functional structure of an information-seeking dialogue.
The discovered patterns reflect regularities that are common for this type of dialogues in general, regardless of the domain.
The QRFA model provides a level of abstraction that does not take into account semantic relations between concepts mentioned in a dialogue.
We define the semantic structure of a conversation as another important dimension of the communication process, which is orthogonal to its functional structure.
In Chapter~\ref{chap:coherence} we describe how the structure of the background knowledge is reflected in a conversation and has an effect on its coherence.

\chapter{Semantic Structure of a Conversation}\blfootnote{This chapter was published as \citep{DBLP:conf/semweb/VakulenkoRCSP18}.}
\label{chap:coherence}

\epigraph{
- What is the weather in Heidelberg?\\
- Today is 90\% chance of rain...\\
- Tell me about the castle.\\
- The Castle is a 1926 novel by Franz Kafka… \\ --- from a conversation with Apple Siri, 10 May 2019
    }

Human conversations tend to be concise and leave out a lot of detail by relying on shared background knowledge(indirect illocution)~\cite{searle1985expression,DBLP:conf/emnlp/MogheABK18,grewendorf2012speech,stalnaker1970}.
It is still possible to correctly interpret ambiguous statements with implicit relations based on the context due to the assumption postulated in Grice's maxim of relevance (relation) that all conversation participants under the cooperative principle should contribute to the ongoing conversation topic~\cite{grice1975logic}.
In other words, the focus of a conversation is not expected to constantly jump from one topic to another but gradually evolve by smoothly transitioning between related concepts within the semantic space.
This transition happens in parallel to the dynamics of the functional transitions that we analysed in Chapter~\ref{chap:structure}, such as question-answer pairs.
We attempt to measure the rhythm of a ``healthy'' conversation by traversing a knowledge graph and identify situations in which an ``offbeat'' can signal anomalies or a rapid topic switch.
The implications are manifold: on the one hand, this measure can inform a response selection procedure of which concepts should be considered relevant based on the conversation context; on the other hand, anomalies discovered during a conversation can indicate a knowledge gap, as an inability to correctly interpret the conversation partner, that needs to be resolved.

Thus, we introduce the task of measuring semantic (in)coherence in a conversation with respect to background knowledge, which relies on the identification of semantic relations between concepts introduced during a conversation. 
We propose and evaluate graph-based and machine learning-based approaches for measuring semantic coherence using knowledge graphs, their vector space embeddings and word embedding models, as sources of background knowledge. 
We demonstrate how these approaches are able to uncover different coherence patterns in conversations on the Ubuntu Dialogue Corpus.

\section{Introduction}

Conversational interfaces are seeing a rapid growth in interest.
Conversational systems need to be able to model the structure and semantics of a human conversation in order to provide intelligent responses. 
The requirement conversations be coherent is meant to improve the probability distribution over possible dialogue states and candidate responses.

A conversation is an information exchange between two or more participants.\footnote{We use the terms ``dialogue'' and ``conversation'' interchangeably, while ``dialogue'' refers specifically to a two-party conversation.} 
An essential property of a conversation is its \textit{coherence}; \citet{de1981textlinguistics}
describe it as a ``continuity of senses.''
Coherence constitutes the outcome of a cognitive process, and is, therefore, an inherently subjective measure.
It is always relative to the background knowledge of participants in the conversation and depends on their interpretation of utterances.
Coherence reflects the ability of an observer to perceive meaningful relations between the concepts and to be critical of the new relations being introduced. 
Meaning emerges through the interaction of the knowledge presented in the conversation with the observer's stored knowledge of the world~\cite{petofi1974semantics}. 
In other words, a conversation has to be assigned an interpretation, which depends on the knowledge available to the agent.

In this chapter we focus on analyzing semantic relations that hold within dialogues, i.e., relations that hold between the concepts (entities) mentioned in the course of the same dialogue.
We call this type of relation \emph{semantic coherence}. 
We focus on semantic relations but ignore other linguistic signals that make a text coherent from a grammatical point of view. 
A classic example to illustrate the difference is due to~\citet{reason:Chomsky57a}: ``Colorless green ideas sleep furiously'' -- a syntactically well formed English sentence that is semantically incoherent.

Our hypothesis is that, apart from word embeddings, recognizing concepts in the text of a conversation and determining their semantic closeness in a background knowledge graph can be used as a measure for coherence.
To this end, we propose and evaluate several approaches to measure semantic coherence in dialogues using different sources of background knowledge: both text corpora and knowledge graphs.
The contributions that we make in this chapter are threefold: 
(1)~we introduce a dialogue graph representation, which captures relations within the dialogue corpus by linking them through the semantic relations available from the background knowledge;
(2)~we formulate the semantic coherence measuring task as a binary classification task, discriminating between real dialogues and generated adversary samples,\footnote{As there is no standard corpus available for this task, we test against 5 ways to generate artificial negative samples.} and 
(3)~we investigate the performance of state-of-the-art and novel algorithms on this task: top-$k$ shortest path induced subgraphs and convolutional neural networks trained using vector embeddings.

The main challenge in applying structural knowledge to natural language understanding becomes apparent when we do not just try to differentiate between genuine conversations and completely random ones, but create adversarial examples as conversations that have similar characteristics compared to the positive examples from the dataset.
Then, the results achieved using word embeddings are usually best and suggest that knowledge graph (KG) embeddings would potentially be an efficient way to harness the structure of entity relations.
However, KG embedding-based models rely on entity linking being correct and cannot easily recover from errors made at the entity linking stage compared to other graph-based approaches that we use in our experiments.

\added{Our goal is to examine the relation between the structure stored in a background knowledge source, such as a knowledge graph or word vectors, and the structure of a conversation. To the best of our knowledge, the only available resource for real examples of incoherent conversations is the datasets that accompany the Dialogue System Technology Challenge (DSTC) Dialogue Breakdown Detection shared task~\cite{DBLP:conf/lrec/HigashinakaFKI16}. However, the aforementioned dialogues are examples of chit-chat conversations that usually do not contain any entity mentions that we could use to link to an external knowledge graph, such as DBpedia. Ubuntu dialogues, in contrast, do contain many software-related terminology that can be associated with knowledge graph entities. Since the Ubuntu dialogues dataset does not have any coherence or dialogue break-down annotations, we had to resort to develop strategies. The generated dialogues can exhibit different degrees of incoherence: from uniformly random entity sets to topic shifts. The latter are also likely to occur within a real conversation setting, when the topic of a conversation is abruptly changed by one of the conversation partners. The benefit of generating such samples is that we know exactly when the topic change occurs. We believe that these automated techniques of generating negative samples from unlabeled data -- in the light of non-availability of human-annotated datasets -- are a contribution as such. There are likely to occur many other types of incoherence in dialogues. However, the study of this phenomenon requires a systematic collection and analysis of empirical data, which is a promising direction for future work but is well beyond the scope of this chapter.}

\section{Related Work}
\label{section:related4}

Previous work on discourse analysis demonstrates good results in recognizing discourse structure based on lexical cohesion for specific tasks such as topic segmentation in multi-party conversations~\cite{DBLP:conf/acl/GalleyMFJ03}. 
Term frequency distribution on its own already provides a strong signal for topic drift.
A more sophisticated approach to assess text coherence is based on the entity grid representation~\cite{DBLP:journals/coling/BarzilayL08}, which represents a text as a matrix that captures occurrences of entities (columns) across sentences (rows) and indicates the role entity plays in the sentence (subject, object, or other). 
This approach relies on a syntactic (dependency) parser to annotate the entity roles and is, therefore, also targeted at measuring lexical cohesion rather than semantic relations between concepts.
The de facto standard testbed for discourse coherence models is the information (sentence) ordering task~\citep{DBLP:conf/acl/Lapata03}; 
it was recently extended to a convolutional neural network-based model for coherence scoring~\cite{DBLP:conf/acl/NguyenJ17}. 
The best results to date were demonstrated by incorporating a fraction of semantic information from an external knowledge source (entity types classification) into the original entity grid model~\citep{DBLP:conf/acl/ElsnerC11a}. 
\citet{DBLP:conf/cikm/CuiLZZ17} push the state-of-the-art on the sentence ordering task by incorporating word embeddings at the input layer of a convolutional neural network instead of the entity grid.
In summary, background knowledge has been found to be able to provide a strong signal for measuring coherence in discourse.
In contrast to previous research focused on measuring coherence in a monologue, we consider the task of evaluating coherence in a written dialogue setting by analyzing the largest multi-turn dialogue corpus available to date, the Ubuntu Dialogue Corpus~\cite{DBLP:conf/sigdial/LowePSP15}.

Research in dialogue systems focuses on developing models able to generate or select from candidate utterances, based on previous interactions.
\citet{DBLP:journals/dad/LowePSCLP17} evaluated several baseline models on the Ubuntu Dialogue Corpus for the next utterance classification task.
Their error analysis suggests that the models can benefit from an external knowledge of the Ubuntu domain, which could provide the missing semantic links between the concepts mentioned in the course of the conversation.
These results motivated us to consider evaluating whether relations accumulated in large knowledge graphs could provide missing semantics to make sense of a conversation.
Our work seeks to discover the potential and limitations of KGs to support natural language \emph{understanding} beyond single search queries or factoid question answering towards a holistic interactive experience, which recognizes and supports the natural (coherent) flow of a conversation.


\section{Measuring Semantic Coherence}
\label{section:approach}

In this section, we describe several approaches to modeling a conversation and measuring its coherence.
We use dialogues, i.e., a two-party conversation to illustrate our approaches.
Our approaches could also be applied to multi-party conversations.

We propose to measure dialogue coherence with a numeric score that indicates more coherent parts of a conversation and provides a signal for topic drift.
Our approach is based on the assumption that naturally occurring human dialogues, on average, exhibit more coherence than their random permutations.


\subsection{Dialogue graph}


We model a dialogue as a graph $D$, which contains 4 types of nodes ${P, U, W, C}$ and edges $E$ between them.
$P$ refers to the set of conversation participants, $U$ -- the set of utterances, $W$ -- the set of words and $C$ -- the set of concepts. 

The words $w$ in a conversation are grouped into utterances 
$\forall w \in W, \exists (u, w) \in E $ such that $u \in U$,\footnote{\label{fn:order}For simplicity, we ignore the role of word order; it can be re-constructed from the order within the conversation $T$ if needed, see below.} 
which belong to one of the conversation participants $ \forall u \in U \exists(p, u) \in E$ such that $p \in P $. 
Every utterance can belong to only one of the participants, while the same words can be re-used in different utterances by the same or different participants.
Words may refer to concepts from the background knowledge $(w, c) \in E$, where $w \in W, c \in C$. 
Several words may refer to a single concept, while the same concept may be represented by different sets of words.
The sequence in which words appear in a conversation is given by the consecutive set of edges 
$T = \{(w_1, w_2), (w_2, w_3), 
\ldots \} $ such that  $T \subset E$, 
indicating the dialogue flow.

The first three types of nodes $P$, $U$, and $W$ together with their relations are available from the dialogue transcript itself, while the set of concepts $C$ and relations between them constitute the semantic representation (meaning) of a dialogue.
The meaning is not directly observable, but is constructed by an observer (one of the dialogue participants or a third party) based on the available background knowledge.
The background knowledge supplies additional links, which we refer to as \textit{semantic relations}. They link words to concepts they may refer to: $(w, c)$ (see footnote~\ref{fn:order}) and different concepts to each other $(c_i, c_j)$. These external relations provide the missing links between words, which explain and justify their co-occurence.
The absence of such links gives an important signal to the observer, and may indicate a topic switch or discourse incoherence.
However, some of the valid links may also be missing from the background knowledge.

\begin{figure}[t]
\centering
\includegraphics[width=\textwidth]{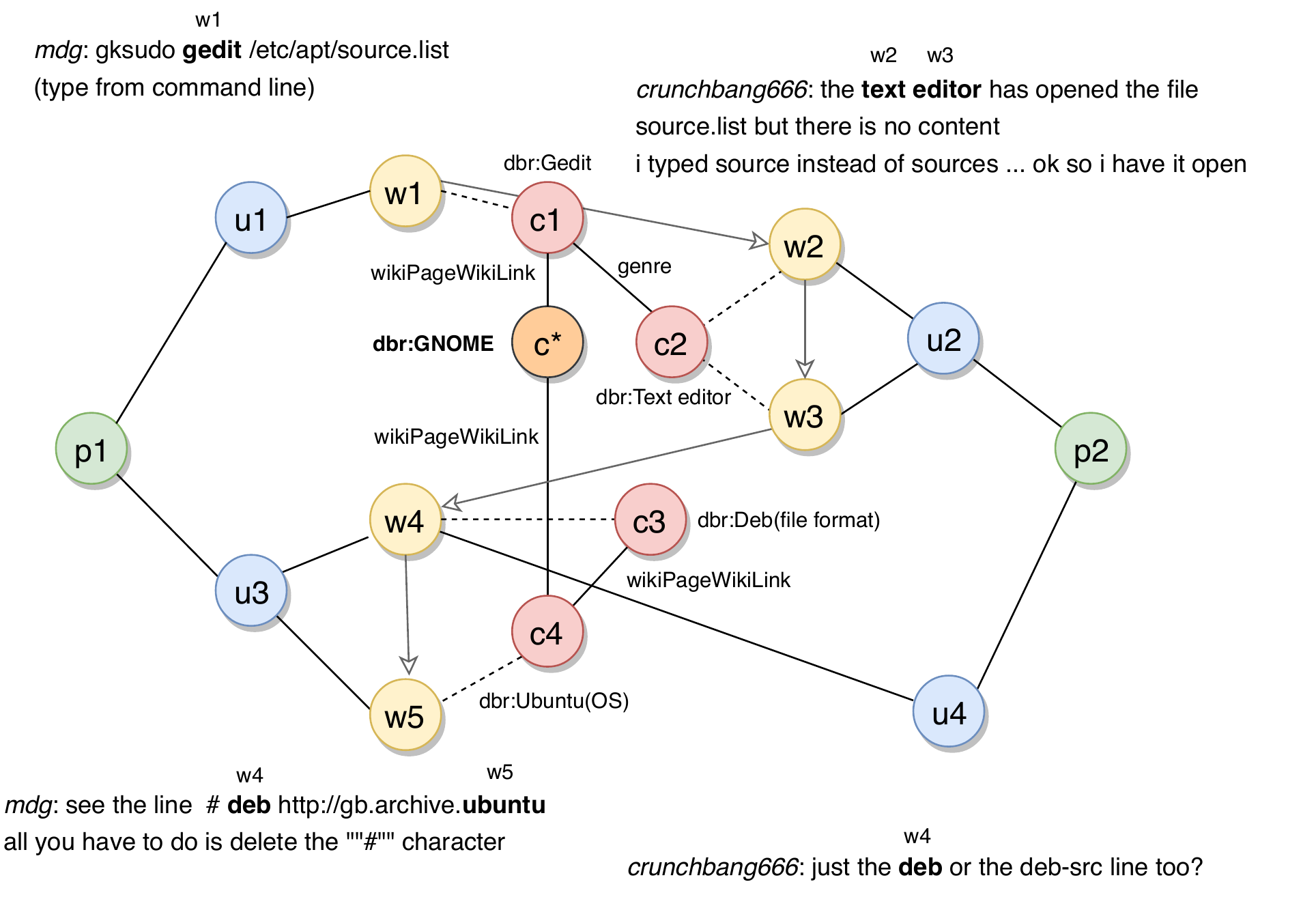}
\caption{Dialogue graph example along with the annotated dialogue. We focus specifically on the layer of concepts in the middle $[c_1, \ldots, c_4]$  attempting to bridge the semantic gap in the lexicon of a conversation using available knowledge models: word embeddings and a knowledge graph.}
\label{fig:graph}
\end{figure}
An example dialogue graph is illustrated in Fig.~\ref{fig:graph}.
The dialogue consists of four utterances represented by  nodes $u1$--$u4$. In the graph we also illustrate a subgraph extracted from the background knowledge, which links the concepts $c1$ \texttt{dbr:Gedit} and $c4$ \texttt{dbr:Ubuntu(OS)} to the concept $c*$ \texttt{dbr:GNOME}, which was not mentioned in the conversation explicitly. This link represents semantic relation between the dialogue turns: ($u1, u2$) and ($u3, u4$), indicating semantic coherence in the dialogue flow. In this example, the semantic relation extracted from the background knowledge corresponds to the shortest path of length 2, i.e., the distance between the concepts mentioned in the dialogue was two relations introducing one external concept from the background knowledge. $c*$ can consist of more than one entity, but encompass a whole subgraph summarizing various relations, which hold between entities, and are represented via alternative paths between them in a knowledge graph. In the next section, we describe our approach to empirically learn semantic relations that are characteristic for human dialogues, using different sources of background knowledge and different knowledge representation models.



\subsection{Semantic relations}

We collect semantic relations between concepts referenced in a dialogue from our background knowledge.
We consider two common sources of background knowledge: (1)~unstructured data: word co-occurrence statistics from text corpora; (2)~semi-structur\-ed data: entities and relations from knowledge graphs.
In order to be able to use a KG as a source of background knowledge we need to perform an entity linking step, which maps words to semantic concepts $(w, c)$, where concepts refer to entities stored in KG.
We consider two  approaches to retrieve relations between the entities mentioned in a dialogue, namely vector space embeddings and subgraph induction via the top-k shortest paths algorithm.

\subsubsection{Embeddings}

Embeddings are generated using the distributional hypothesis by representing an item via its context, i.e., its position and relations it holds with respect to other items.
Embeddings are multi-dimensional vectors (of a fixed size), which encode the distributional information of an item (a word in the a or a node in a graph), i.e., its position and relations to other items in the same space.
This is achieved by computing vector representations towards an optimality criteria defined with a certain output function, which depends on the embedding vectors being trained. Thus, embeddings efficiently encode (compress) an original sparse representation of the relations (e.g., an adjacency matrix) for each of the items. It provides an easy and fast way to access this information (relationship structure). Following this approach, every concept $c_i$ in our dialogue graph~\ref{fig:graph} is assigned to an n-dimensional vector, which encodes its location in the semantic space, and loses all the edges, which explicitly specified its relations to other concepts in the space.

We consider two types of embeddings to represent concepts mentioned in a dialogue, one for each of our background knowledge sources: word embeddings trained on a text corpus, and entity embeddings trained on a KG. For word embeddings, we use word2vec~\cite{DBLP:conf/nips/MikolovSCCD13}, in particular the skip-gram variant, which aims to create embeddings such that they are useful for predicting words which are in the neighborhood of a given word.
GloVe~\cite{pennington2014glove} is a word embedding method, with the explicit goal of embedding analogies between entities.
This method does not work directly on the text corpus, but rather on co-occurrence counts which are derived from the original corpus.

For graph embeddings, we use two methods that can be scaled to large graphs, such as DBpedia and Wikidata: biased RDF2Vec~\cite{DBLP:conf/wims/CochezRPP17} (using random walks) and Global RDF Vector Space Embeddings~\cite{DBLP:conf/semweb/CochezRPP17};
we refer to the latter ones as \textit{KGlove} embeddings.
RDF2Vec is based on word2vec. It works by first generating random walks on the graph, where the edges have received weights which influence the probability of following these edges.
During the walk, a sentence is generated consisting of the identifiers occurring on the nodes and edges traversed.
For each entity in the graph, many walks are performed and hence a large text is generated. This text is then used for training word2vec.
KGlove is based on GloVe, but instead of counting the co-occurrence counts from text, they are computed from the graph using personalized PageRank scores starting from each node or entity in the graph.
These counts (i.e., probabilities) are then used as the input to an optimization problem that aims to encode analogies by creating embedding vectors corresponding to the co-occurrences.

\subsubsection{Subgraph induction}

An embedding (usually) carries a single representation for an item (word or entity), which is designed to capture all relations the item has regardless of the task or the context in which the item occurs.
For example, an embedding representation may neglect some of the infrequent relations, which can become more relevant than others depending on the situation (context).
In order to contrast the embedding-based approach, we also implement a more traditional graph-based approach to represent entity relations in a KG.
Given a sequence of entities, as they appear in a dialogue, i.e., $[c_1, c_2 \ldots c_n]$, we extract relations, as top-k shortest paths, between every entity $c_i$ and all the entities that were mentioned in the same dialogue before $c_i$, i.e., $(c_1, c_i), (c_2, c_i), \ldots, (c_{i-1}, c_i)$. 

For the top-k shortest path computation, we apply an approach based on bidirectional breadth-first search~\cite{SavenkovMUP17} using the space-efficient binary {Header, Dictionary, Triples} (HDT) encoding~\cite{fernandez2013} of the KG.
This approach maps entities discussed in the dialogue to KG concepts, and then interprets paths  between concepts in the KG as semantic relations between the respective entities.
Many such relations are never mentioned in the conversation and only become explicit through the path enumeration over the KG. By increasing the number of desired shortest paths $k$ and the maximum path length $\ell$, one can discover more relations, including those that might be omitted or obscured in the entity embedding representation in the case of a random walk or frequency-based embedding algorithms.
An obvious downside of this increase in recall is reduced efficiency.

\subsection{Dialogue classification}
We measure semantic coherence by casting the task into a classification problem. The score produced by the classifier 
corresponds to our measure of semantic coherence. 

Since human dialogues are expected to exhibit a certain degree of incoherence due to topic drift and since relations are missing from our background knowledge, we cannot assume every concept in our dialogue dataset to be coherent with respect to the other concepts in the same dialogue.
However, it is reasonable to assume that on average a reasonably large set of concepts extracted from a human dialogue exhibits a higher degree of coherence than a randomly generated one.
We build upon this assumption and cast the task of measuring semantic coherence as a binary classification task, in which real dialogues have to be distinguished from corrupted (incoherent) dialogues. 
We consider positive and negative examples for whole conversations, represented as a sequence of words or entities, which constitute the input for the binary classifier. 
Effectively, these examples provide a supervision signal for measuring and aggregating distances between words/concepts by learning the weights for the neural network classifier.

\subsubsection{Negative sampling}
\label{section:sampling}

To produce negative (adversarial) examples for the binary classification task we propose five sampling strategies:

\begin{itemize}[nosep]
\item RUf: {Random uniform}. For every positive example we choose a sequence of entities (or words for training on word embeddings) of the same size from the vocabulary uniformly at random; so, we double the size of the dataset effectively by supplementing it with completely randomly generated (i.e., presumably incoherent) counterexamples.

\item SqD: {Sequence disorder}. Randomly permute the original sequence, which is similar in spirit to the sentence ordering task for evaluating discourse coherence~\cite{DBLP:conf/acl/Lapata03}. The key difference is that we rearrange the order of words (entities), which may also occur within the same sentence (utterance), rather than permuting whole sentences. 

\item VoD: {Vocabulary distribution}. For every positive example choose a sequence of entities of the same length from the vocabulary using the same frequency distribution as in the original corpus; so, VoD is very similar to RuF, but tries to emulate ``structure'' to some extent by choosing similar term frequencies.

\item VSp: {Vertical split}. Create a negative example by permuting two positive examples replacing utterances of one of the conversation participants with utterances of a participant from a different conversation.

\item HSp: {Horizontal split}. Create a negative example by permuting two positive examples merging the first half of one conversation with the second half of a different conversation.
\end{itemize}

\added{Data augmentation is a powerful technique that helps to take advantage of a vast unlabeled data source without requiring expensive manual annotation efforts for model training and evaluation. For example, application of different perturbation techniques to naturally occurring data, such as rotations and clipping, proved useful for image recognition. This approach does not guarantee resemblance to any naturally occurring data points but is designed to better explore the boundaries of the manifold on which the natural data points are located, i.e., study their properties in terms of similarities and differences to artificially constructed samples.}

\added{In this chapter we apply the same idea to study the properties of human dialogues. The most straight-forward and realistic approach to generating negative samples of incoherent conversations is appending another dialog chosen at random to the end of each dialog (``horizontal split''). This kind of perturbation constitutes a controlled ``topic drift,'' when the subject of conversation shifts to another subject in the middle of a conversation. Sampling perturbations (``random distribution'' and ``vocabulary distribution'') were designed to provide a lower bound and evaluate the data distribution by measuring how likely a randomly generated dialog is to appear coherent to the model. The last type of negative samples presented to the model (``vertical split'') evaluates the ability of the model to pick on the asymmetric property of a conversation, i.e., utterance distribution between the conversation participants.}

\subsubsection{Convolutional neural network}

To solve the binary classification task we train a classifier using a convolutional neural network architecture, which is applied to sequences of words and entities to distinguish irregular semantic drift, which was deliberately injected into conversations, from smooth drift which occur within real conversations.

It is a standard architecture previously employed for a variety of natural language tasks, such as text classification~\cite{DBLP:journals/corr/Kim14f}. The network consists of (1) an input layer, which appends the pre-trained embeddings for each of the word (entity) from the dialogue sequence; (2) a convolutional layer, which consist of filters (arrays of trainable weights) sliding over and learning predictive local patterns in the previous layer of the input embeddings; (3) a max pooling layer, which aggregates the features learned by the neighboring filters; (4) the hidden layer, a fully connected layer, which allows combining features from all the dimensions with a non-linear function; and (5) the output layer is a fully connected layer, which aggregates the scores to make the final prediction.
See also Section~\ref{sec:implementation4} for details.

\section{Evaluation Setup}
\label{section:evaluation4}

The source code of our implementation and evaluation procedures is publicly accessible.\footnote{\url{https://github.com/vendi12/semantic\_coherence}} We also release our dataset used in the evaluation, which contains dialogue annotations with DBpedia entities and shortest paths, for reproducibility and further references.\footnote{\url{https://github.com/vendi12/semantic_coherence/tree/master/data}}

\subsection{Dataset}

\subsubsection{Dialogues}
Our experiments were performed on a sample of dialogues from the Ubuntu Dialogue Corpus\footnote{\url{https://github.com/rkadlec/ubuntu-ranking-dataset-creator}}~\cite{DBLP:conf/sigdial/LowePSP15}, which contains 1,852,869 dialogues in total, with one dialogue per file in TSV format, and is the largest conversational dataset to date. 
There are multiple challenges related to using this corpus, however. 
The dialogues were automatically extracted from a public chat using several heuristics selecting two user handles and segmenting based on the timestamps. 
The dialogues cannot be considered as perfectly coherent since some of the related utterances are missing from the dialogues; there can be several different topics discussed within the same conversation and the asynchronous nature of on-line communication often results in semantic mismatch in the dialogue sequence. 
While we cannot guarantee local coherence of the real dialogues, we expect them to be on average more coherent, when comparing to the dialogues randomly generated by sampling entities (words) from the vocabulary or merging entities (words) from different dialogues, which we refer to as negative samples, or adversaries, in our binary classification task.

We proceed by annotating a sample of 291,848 dialogues from the Ubuntu Dialogue Corpus with the DBpedia entities using the DBpedia Spotlight public web service\footnote{\url{http://model.dbpedia-spotlight.org/en/annotate}}~\cite{DBLP:conf/i-semantics/DaiberJHM13}. 
The input to the entity linking API is the text for each utterance in a conversation. 
Next, we considered only the dialogues where both participants contribute at least 3 new entities each, i.e., every dialogue in our dataset contains minimum 6 entities shared between the dialogue partners. 
The threshold for entities per conversation was chosen to ensure there is enough semantic information for measuring coherence. 
This way, we end up with a sample of 45,510 dialogues, which we regard as true positive examples of coherent dialogue. It contains 17,802 distinct entities and 21,832 distinct words that refer to these. 
The maximum size of a dialogue in this dataset is 115 entities or 128 words referring to them. We shuffled the dialogues and selected 5,000 dialogues for our test set. 
While this procedure means we cannot test our approach on short conversations, with fewer entities, we consider 45K dialogues to be a representative dataset for evaluating our approach.

The negative samples for both training and test set were generated using five different sampling strategies described in Section~\ref{section:sampling}. Each development set consists of 81,020 samples (50\% positive and 50\% negative). 
We further split it into a training and validation set: 64,816 and 16,204 (20\%) samples, respectively. 
Our test set comprises the remaining 5,000 positive examples, and 5,000 generated negative samples.

\subsubsection{Knowledge models}

We compared the performance on our task across two types of embeddings models trained on two different knowledge source types: GloVe~\cite{pennington2014glove} and Word2Vec~\cite{DBLP:conf/nips/MikolovSCCD13} for the word embeddings, and biased RDF2vec~\cite{DBLP:conf/wims/CochezRPP17} and KGloVe~\cite{DBLP:conf/semweb/CochezRPP17} for the knowledge graph entity embeddings.

We utilise two publicly available word embedding models: GloVe embeddings pre-trained on the Common Crawl corpus (2.2M words, 300 dimensions)\footnote{\url{https://nlp.stanford.edu/projects/glove/}} and Word2Vec model trained on the Google News corpus (3M words, 300 dimensions).\footnote{\url{https://code.google.com/archive/p/word2vec/}} 1,578 words from our dialogues (7\%) were not found in the GloVe embeddings dataset and received a zero vector in our embedding layer. Thus, GloVe embeddings cover 20,254 words from our vocabulary (93\%). Word2Vec embeddings cover only 73\% of our vocabulary.

For RDF2Vec and KGloVe we used publicly available pre-trained global embeddings of the DBpedia entities (see~\cite{DBLP:conf/wims/CochezRPP17} and~\cite{DBLP:conf/semweb/CochezRPP17}, respectively). 
For KGlove we used all different embeddings, while for RDF2Vec we experimented with the embeddings that gave the best performance in \cite{DBLP:conf/wims/CochezRPP17}.
KGlove embeddings cover 17,258 entities from our vocabulary (97\%), while
Rdf2Vec provides 62--77\% due to different importance sampling strategies of the embedding approaches.

The shortest paths used were extracted from dumps of DBpedia (April 2016, 1.04 billion triples) and Wikidata (March 2017, 2.26 billion triples).\footnote{\url{http://www.rdfhdt.org/datasets/}}

\subsection{Implementation}
\label{sec:implementation4}
Our neural network model is implemented using the Keras library with a TensorFlow backend. 
The one-dimensional (temporal) convolutional layer contains 250 filters of size 3 and stride (step) 1. The max pooling layer is global, the hidden layer is set to 250 dimensions.
There are two activation layers with rectified linear unit (ReLU) after the convolutional and the hidden layers to capture also non-linear dependencies between input and output, and two dropout layers with rate 0.2 after the embeddings and hidden layers to avoid overfitting.
The last ReLU activation is projected onto a single-unit output layer with a sigmoid activation function to obtain a coherence score on the interval between 0 and 1.

The network is trained using the Adam optimizer with the default parameters~\cite{DBLP:journals/corr/KingmaB14} to minimize the binary cross-entropy loss between the predicted and correct value.
All models were trained for 10 epochs in batches of 128 samples and early stopping after 5 epochs if there is no improvement in accuracy on the validation set.

To compute the shortest paths we merged the dumps of DBpedia and Wikidata into a single 36GB binary file in  HDT format~\cite{fernandez2013} (DBpedia+Wikidata HDT), with an additional 21GB index on the subject and the object components of triples.
We set the parameters of the algorithm in our experimental evaluation as follows: $k$ for the number of shortest paths to be retrieved from the graph to 5, the maximum length $\ell$ of a path to 9 edges (relations) and a timeout terminating the query after 2 seconds to cope with the scalability issues of the algorithm.
Our top-k shortest paths algorithm implementation is available via a SPARQL endpoint\footnote{\url{http://wikidata.communidata.at}} using the syntax shown in Fig.~\ref{fig:k-shortest}.

\begin{figure}[ht]
\begin{verbatim}[fontsize=\small]
 PREFIX ppf: <java:at.ac.wu.arqext.path.>
 PREFIX dbr: <http://dbpedia.org/resource/>                
 SELECT * WHERE { 
 ?X ppf:topk ("--source" dbr:Directory_service 
                         dbr:Gnome dbr:GNOME 
                         dbr:Desktop_environment 
              "--target" dbr:Desktop_computer 
              "--k" 5 "--maxlength" 9 "--timeout" 2000) }\end{verbatim}
\caption{\label{fig:k-shortest} k-shortest path query (cf.~\cite{SavenkovMUP17} to extract relevant connections between entities from the knowledge graph}
\end{figure}
 The function \texttt{at.ac.wu.arqext.path.topk} is a user defined extension available as a Jena ARQ extension.\footnote{\url{https://bitbucket.org/vadim\_savenkov/topk-pfn}}
 
 \section{Evaluation Results}
\label{section:results}

Table~\ref{tbl:context} reports the most common entities and relations, which while not being mentioned in the course of a dialogue, were on the shortest paths (in the KG) between other entities that were explicitly mentioned in the dialogue, i.e., which constitute an implicit dialogue context. 
While Dbpedia Spotlight dereferenced ``Ubuntu'' mentions to the concept related to philosophy rather than to the popular software distribution, the graph-based approach succeeds in recovering the correct meaning of the word by extracting this concept from the shortest paths that lie between the other entities mentioned in dialogues. Almost all relations obtained from the KG correspond to the links between the corresponding Wikipedia web pages (wikiPageWikiLink).

\begin{table}[h]
\begin{floatrow}
\centering
\ttabbox{%
\begin{adjustbox}{width=.98\textwidth}%
\begin{tabular}{clrlrlr}
\toprule
Top & \multicolumn{2}{c}{Mentioned entities} & \multicolumn{2}{c}{Context entities} & \multicolumn{2}{c}{Relations}     \\
\cmidrule(r){2-3}
\cmidrule(r){4-5}
\cmidrule{6-7}
\#  & Label                     & Count      & Label                        & Count & Label                     & Count \\
\midrule
1                       & Ubuntu(philosophy)      & 1605                      & Ubuntu(OS) & 1058                      & wikiPageWikiLink & 51014                     \\
2                       & Sudo                      & 708                       & Linux                       & 725                       & gold/hypernym             & 319                       \\
3                       & Booting                   & 676                       & Microsoft\_Windows          & 208                       & ontology/genre            & 178                       \\
4                       & APT(Debian)             & 405                       & FreeBSD                     & 175                       & operatingSystem  & 140                       \\
5                       & Live\_CD                  & 314                       & Smartphone                  & 171                       & rdf-schema\#seeAlso       & 116        \\              
\bottomrule
\end{tabular}
\end{adjustbox}
}{%
\caption{The top 5 most common entities and relations in the Ubuntu Dialogue dataset: mentioned entities -- from linking dialogue utterances to DBpedia entities via Dbpedia Spotlight Web service; context entities and relations -- from the shortest paths between the mentioned entities in DBpedia.}
\label{tbl:context}
}
\end{floatrow}
\end{table}

\subsection{Semantic distance}

\begin{figure}[!t]
\centering
\includegraphics[width=0.76\textwidth]{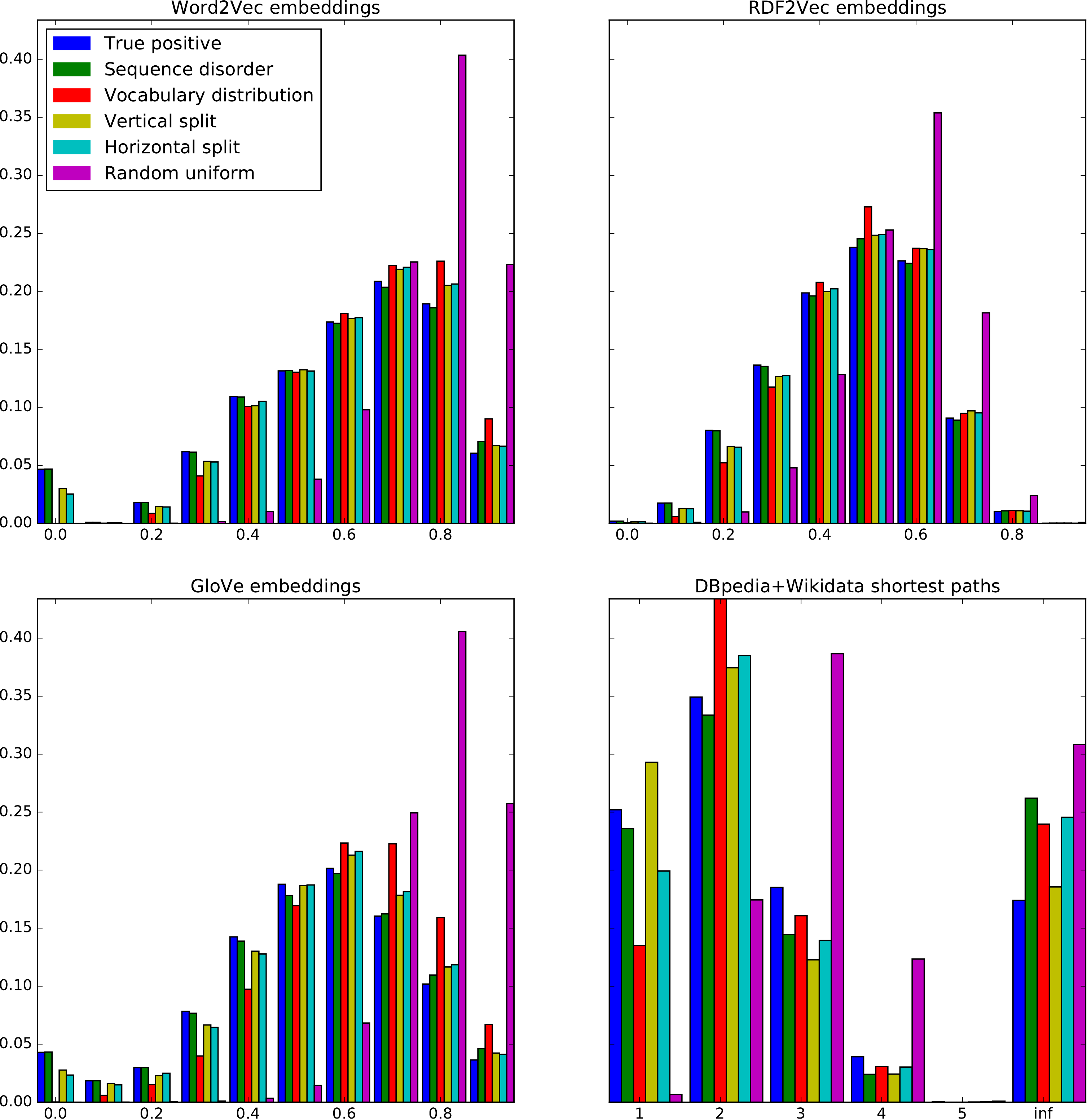}
\caption{Distribution of cosine distances for different data splits using Word2Vec and GloVe word embeddings (left), and RDF2Vec KG embeddings (top right), compared with the distribution of shortest path lengths in DBpedia+Wikidata KG (bottom right). Words in real dialogues (True positive) are more related than frequent domain words (Vocabulary distribution), and much more than a random sample (Random uniform).}
\label{fig:sem_dist_distrs}
\end{figure}

The length of the shortest path (number of edges, i.e., relations on the path) is a standard measure used to estimate semantic (dis)similarity between entities in a knowledge graph~\cite{DBLP:journals/corr/abs-1203-1889}. We observe how it correlates with a standard measure to estimate similarity between vectors in a vector space, \textit{cosine distance}, defined as: 
$1 - \cos(x, y) = 1 -\frac {x y^\intercal}{|| x||  || y||}$ 
Fig.~\ref{fig:sem_dist_distrs} showcases different perspectives on semantic similarity (coherence) between the entities in real and generated dialogues as observed in different semantic spaces (w.r.t. the knowledge models), alignments and differences between them. The barplots reflect the distributions of the semantic distances between entities in dialogues. The semantic distances are measured using cosine distances between vectors in the vector space for word (Word2Vec and GloVe) and KG (RDF2Vec) embeddings, and in terms of the shortest path lengths in the DBpedia+Wikidata KG. We observe that the real dialogues (True positive) tend to have smaller distances between entities: 1--2 hops or at most 0.3 cosine distance, while randomly generated sequences are skewed further off. Embeddings produce much more fine-grained (continuous) representation of semantic distances in comparison with the shortest path length metric. Distributions produced by different word embeddings are very similar in shape, while the one from KG embeddings is steeper and skewed more to the center, there are only a few entities further than 0.7, while this is the top for the random distances in word embeddings.

We also discover the bottleneck of our shortest path algorithm at length 5. Since the set of relevant entities for which the paths are computed grows proportionally to the dialogue length, depending on the degree of the node the number of expanded nodes quickly reaches the limit on the memory size.  In our case, the algorithm retrieved the paths of length at most 5 due to the 2-second timeout, while the parameter for the maximum length of the path $\ell$ was set to 9.

\subsection{Classification results}
Our evaluation results from training a neural network on the task of measuring (in)cohe\-rence in dialogues are listed in Table~\ref{tbl:results}. It summarizes the outcomes of models trained on different embeddings using different types of adversarial samples (negative sampling strategies are described in Section~\ref{section:sampling}). For the KG embeddings, we report only the approaches that performed best across different test splits.\footnote{The full result table is available on-line: \url{https://github.com/vendi12/semantic_coherence/blob/master/results/results.xls}}

\begin{table}[!t]
\centering
\ttabbox{%
\begin{adjustbox}{width=.98\textwidth}%
\begin{tabular}{p{2cm}lllllllllllll}
\toprule
 &  & \multicolumn{12}{c}{Accuracy}                                                                                      \\ 
\cmidrule{3-14}
                                  &  Data &  & \multicolumn{10}{c}{TNeg}                                           &  \\ 
    \cmidrule{4-13}
Embeddings & split & TPos & RUf  & Avg  & VoD  & Avg  & SqD  & Avg  & VSp  & Avg  & HSp  & Avg  & Avg  \\ 
    \midrule
Word2Vec                          & RUf                         & 0.99                  & 0.99 & \hl{\textbf{0.99}} & 0.02 & 0.50 & 0.02 & 0.50 & 0.01 & 0.50 & 0.01 & 0.50 & 0.60                 \\
                                  & VoD                         & 0.89                  & 0.62 & 0.75 & 0.90 & \hl{0.89} & 0.53 & 0.71 & 0.18 & 0.54 & 0.20 & 0.54 & \textbf{0.69}                 \\
                                  & SqD                         & 0.75                  & 0.65 & 0.70 & 0.88 & 0.81 & 0.81 & \hl{0.78} & 0.27 & 0.51 & 0.29 & 0.52 & 0.66                 \\
                                  & VSp                         & 0.59                  & 0.50 & 0.55 & 0.82 & 0.71 & 0.41 & 0.50 & 0.59 & \hl{0.59} & 0.61 & 0.60 & 0.59                 \\
                                  & HSp                         & 0.62                  & 0.39 & 0.50 & 0.71 & 0.66 & 0.38 & 0.50 & 0.55 & 0.58 & 0.63 & \hl{0.63} & 0.58                 \\ 
\midrule
GloVe                       & RUf                         & 0.99                  & 0.99 & \hl{\textbf{0.99}} & 0.00 & 0.50 & 0.01 & 0.50 & 0.00 & 0.50 & 0.00 & 0.50 & 0.60                 \\
                                  & VoD                         & 0.93                  & 0.38 & 0.66 & 0.93 & \hl{\textbf{0.93}} & 0.39 & 0.66 & 0.19 & 0.56 & 0.08 & 0.51 & 0.66                 \\
                                  & SqD                         & 0.76                  & 0.71 & 0.73 & 0.91 & 0.84 & 0.82 & \hl{\textbf{0.79}} & 0.16 & 0.46 & 0.15 & 0.45 & 0.66                 \\
                                  & VSp                         & 0.60                  & 0.25 & 0.42 & \underline{0.92} & 0.76 & 0.43 & 0.51 & 0.65 & \hl{0.62} & 0.66 & 0.63 & 0.59                 \\
                                  & HSp                         & 0.71                  & 0.34 & 0.52 & 0.81 & 0.76 & 0.30 & 0.50 & 0.55 & \textbf{0.63} & 0.66 & \hl{\textbf{0.68}} & 0.62                 \\ 
\midrule
rdf2vec PRS            & RUf                         & 0.98                  & 0.99 & \hl{\textbf{0.99}} & 0.02 & 0.50 & 0.02 & 0.50 & 0.02 & 0.50 & 0.01 & 0.50 & 0.60                 \\
                                  & VoD                         & 0.79                  & 0.68 & 0.73 & 0.83 & \hl{0.81} & 0.34 & 0.57 & 0.36 & 0.57 & 0.35 & 0.57 & 0.65                 \\
                                  & SqD                         & 0.59                  & 0.48 & 0.54 & 0.72 & 0.66 & 0.67 & \hl{0.63} & 0.43 & 0.51 & 0.40 & 0.50 & 0.56                 \\
rdf2vec PR                 & HSp                         & 0.57                  & 0.59 & 0.58 & 0.72 & 0.64 & 0.43 & 0.50 & 0.59 & 0.58 & 0.67 & \hl{0.62} & 0.58                 \\ 
\midrule
KGloVe Uni        & RUf                         & 0.92                  & 0.97 & \hl{0.94} & 0.11 & 0.51 & 0.09 & 0.50 & 0.08 & 0.50 & 0.07 & 0.50 & 0.59                 \\
                                  & VoD                         & 0.54                  & 0.88 & 0.71 & 0.73 & \hl{0.64} & 0.61 & 0.58 & 0.51 & 0.52 & 0.52 & 0.53 & 0.60                 \\
                                  & SqD                         & 0.55                  & 0.62 & 0.58 & 0.64 & 0.59 & 0.63 & \hl{0.59} & 0.47 & 0.51 & 0.45 & 0.50 & 0.56                 \\
KGloVe PrO & HSp                         & 0.31                  & 0.81 & 0.56 & 0.75 & 0.53 & 0.69 & 0.50 & 0.77 & 0.54 & 0.70 & \hl{0.51} & 0.53                 \\
KGloVe PR     & HSp                         & 0.47                  & 0.69 & 0.58 & 0.61 & 0.54 & 0.54 & 0.50 & 0.57 & 0.52 & 0.65 & \hl{0.56} & 0.54     \\ 
\bottomrule
\end{tabular}
\end{adjustbox}
}{%
\caption{Accuracy on the test set across different embedding and sampling approaches. 
The table shows for 7 different embedding strategies (4 types), how the embedding performs when trained with data from different generated adversarial examples.
For example, the underlined value in the table (0.92), means that GloVe word embeddings, when trained with genuine and Vertical split (VSp) adversarial examples, is able to correctly find 92\% of the Vocabulary distribution (VoD) adversarial examples in the test set.
In the same row, in the TPos column, it can be seen that 60\% of the genuine messages were correctly identified. Hence, this results in an average accuracy of 0.76.
In blue highlight, we indicate the results where the adversarial examples for training the model where of the same type as for testing the model.
In bold, we indicate the best result for each adversarial example type.
	Abbreviations: TPos -- True Positive, TNeg -- True Negative, RUf -- Random uniform, VoD -- Vocabulary distribution, SqD -- Sequence disorder, VSp -- Vertical split, HSp -- Horizontal split, Avg -- Average, PRS -- PageRank Split, PR -- PageRank, Uni -- Uniform, PrO -- Predicate Object.}
\label{tbl:results}
}
\end{table}

From the results we observe that the easiest task was to distinguish real dialogues from randomly generated sequences.
When the model was trained with randomly generate dialogues, accuracies often reach close to 100\%.
However, this same model performs poorly when used for any other type of non-genuine messages we created. 
In the best case (KGloVe Uni), still only 10\% of messages randomly sampled from the vocabulary distribution were correctly detected.
This indicates that there is a need to experiment with the other types as well.
We also observe that the models that are trained with specific adversarial examples are best in separating that type.
However, even when the model is not explicitly trained to recognize a specific type of dialogue, but instead trained on other types of adversarial examples, it is sometimes still able to classify  messages correctly. 
This happens, for example, in the case of KGloVe Uniform where the adversarial messages are sampled from the Vocabulary distribution and the model is still able to detect around 70\% of randomly generated messages.

The dialogues generated by permuting the sequence of entities (words) in the original dialogues (the sequence ordering task) were harder to distinguish (The best performing model resulted in an accuracy of 0.79). 
Finally, the hardest task was to discriminate the adversarial examples generated by merging two different dialogues together (vertical and horizontal splits).
This was expected as these dialogues have short sequences of genuine dialogue inside, making them hard to classify.

The best performance across all test settings was achieved using the word embeddings models, especially GloVe performed well. 
KG embeddings, while performing reasonably well on the easier tasks (RUf and VoD), fell short to distinguish more subtle changes in semantic coherence.
For the KG embedding weighting approaches, we noticed that the ones which performed well in earlier work, also worked better in this task.
In particular, it was noticed that the weighting biased by PageRank computed on the Wikipedia links graph results in better results in machine learning tasks.

As discussed in Section \ref{section:evaluation4}, RDF2vec has fewer entity embeddings than KGloVe, when trained from the same original graph (DBpedia). KGloVe will provide an embedding, even when not much is known about a specific entity.
In case of a node that does not have any edges, KGloVe will assign a random vector to it.
In contrast, RDF2Vec will prune infrequent nodes.
Another problem that affects KG embeddings are incorrectly recognized entities.
There is no linking required for needed word embeddings since it represents different meanings of the word in a single vector.



\begin{figure}[!t]
\centering

\includegraphics[width=0.92\textwidth]{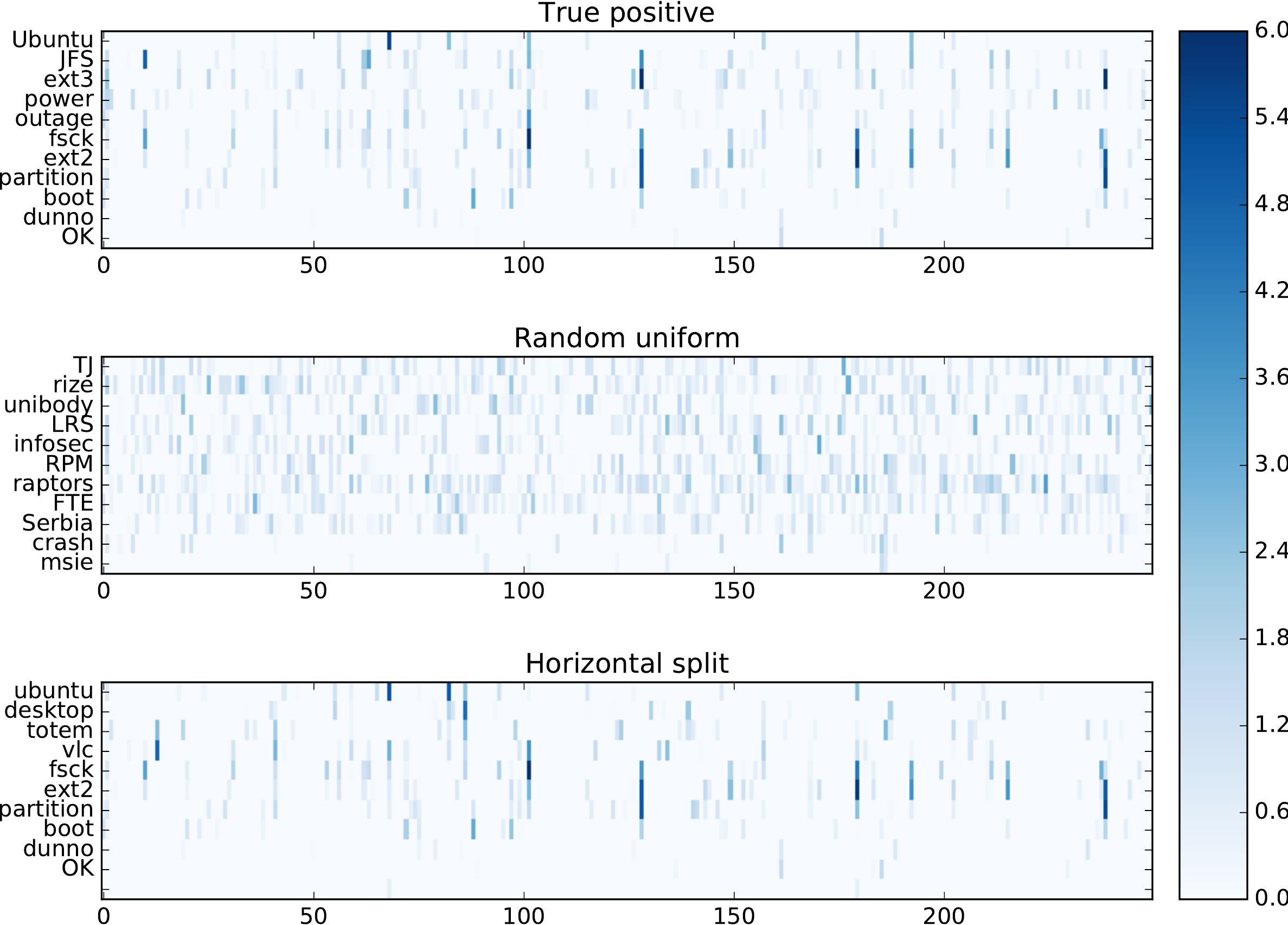}
\caption{Heatmap of the activations on the output of the word embeddings layer. Notice the vertical-bar pattern indicating a stronger semantic relation between the words in a real dialogue (top) in comparison with a random word sequence (middle). The topic drift effect can be observed when two different dialogues are concatenated (horizontal split -- bottom): the bars at the top are shifted in comparison with the bars in the second half of the conversation, comparing to the coherence patterns observed in the real dialogue (top).}

\label{fig:heatmap}
\end{figure}

Overall, we want to be able not only to tell to which degree a dialogue is (in)coherent but also to identify the regions in the dialogue where coherence was disrupted, or to partition the dialogue into coherent segments indicating the shifts between different topics. 
Visualization of the activations in the output of the convolutional layer of the Glove word embeddings-based model exhibits distinct vertical activation patterns, which can be interpreted as traces of local coherence the model is able to recognize (See Fig.~\ref{fig:heatmap}).

\section{Conclusion}
\label{section:conclusion4}

We considered the task of measuring semantic coherence of a conversation, which introduces an important and challenging problem that requires processing vast amounts of heterogeneous knowledge sources to infer implicit relations between the utterances, i.e., bridging the semantic gap in understanding natural language.
We proposed and evaluated several approaches to this problem using alternative sources of background knowledge, such as structured (knowledge graph) and unstructured (text corpora) knowledge representations. 
These approaches detect semantic drift in conversations by measuring  coherence with respect to the background knowledge. 
Our models were trained for dialogues but the approach does not restrict the number of conversation participants. 
The model's performance depends to a large extent on the choice of background knowledge source, with respect to the conversation domain. 
The conversation needs to contain a sufficient number of recognized entities to signal its position within the semantic space.
Using entity annotations in the processing mining approach for analysing and evaluating conversation success introduced in Chapter~\ref{chap:structure} also constitutes a promising direction for the future work. 

\added{The theory of indirect illocutionary acts introduced by Searle (1979) suggests that conversation participants are able to infer the intention of a speaker even when it is not expressed directly. This is achieved by applying reasoning based on a shared background knowledge. We took this idea a step further by hypothesizing that there is a direct relation between an utterance and a response to it that can be tracked using the background knowledge.}

\added{In our experiments we track latent relationships between utterances adjacent in a conversation by projecting them upon a large knowledge graph. Every response in a conversation is likely to have a relation to the previous utterance that can be described as a set of top-$k$ shortest paths between the first utterance and the next responses. The idea is that the latent relationships are represented and thereby can be made explicit through the background knowledge graph. The subgraph extracted as a result of the traversal between these sets of concepts may reflect the reasoning process behind the response explicitly surfacing the indirect illocution “in-between the lines” of a conversation.}

Our results indicate promising directions as well as challenges in applying structural knowledge to analyse natural language.
We show that the use of word embeddings in text classification is superior to some existing knowledge graph embeddings.
This is an important insight, advancing research by uncovering limitations of state-of-the-art knowledge graph embeddings and indicating directions for improvements.

Knowledge graph embeddings constitute a potentially powerful method to efficiently harness entity relations for tasks that require estimates of semantic similarity.
However, their use relies on the correctness of the entity linking performance. 
Errors made at this stage in the pipe-line approach do propagate into the classification results, but we noticed that they are rather consistent, which partially mitigates the problem.
Our experiments showed that graph-based approaches are more robust to errors in entity linking than knowledge graph embeddings, which is an important insight for future work.

We follow up on these findings in Chapter~\ref{chap:qa} and show how to make question answering over knowledge graphs more robust to uncertainty in entity linking and relation detection.
Our approach integrates word embeddings to interpret a question and find parts of the knowledge graph that can be relevant to this question.
It is able to consider several alternative interpretations in parallel and aggregate their scores via message-passing directly on the graph structure.

\chapter{Question Answering}\blfootnote{This chapter was published as \citep{DBLP:conf/cikm/VakulenkoGPRC19}.}
\label{chap:qa}

\epigraph{  When it is not in our power to determine what is true, we ought to act according to what is most probable. \\ --- Descartes, Discourse on Method, 1637}

Natural language is an inherently ambiguous communication tool.
Therefore, any attempt at interpreting a natural language utterance requires a probabilistic model capable of reasoning under uncertainty.
Moreover, natural language understanding and interpretation is often context-dependent and require access to background knowledge that provides a set of relevant concepts and relations between them.
This chapter introduces an approach that estimates and incrementally aggregates confidence scores over the structure of a knowledge graph in the context of a natural language understanding task.
Our approach mitigates the issue previously discussed in Chapter~\ref{chap:coherence} when all alternative interpretations were discarded at an early linking and disambiguation stage.
Instead, we show how to maintain and deliver a large pool of candidate interpretations ranked by the corresponding confidence scores in a scalable manner.
We demonstrate the performance of the proposed approach in the context of a question answering task, which is one of the core AI tasks and provides sufficient evaluation criteria.

More specifically, we focus on the task of question answering over knowledge graphs (KGQA), which has evolved from simple single-fact questions to complex questions that require graph traversal and aggregation.
In this chapter we propose a novel approach for complex KGQA that uses unsupervised message passing, which propagates confidence scores obtained by parsing an input question and matching terms in the knowledge graph to a set of possible answers.
First, we identify entity, relationship, and class names mentioned in a natural language question, and map these to their counterparts in the graph.
Then, the confidence scores of these mappings propagate through the graph structure to locate the answer entities.
Finally, these are aggregated depending on the identified question type.
We demonstrate that this approach can be efficiently implemented as a series of sparse matrix multiplications mimicking joins over small local subgraphs.
Our evaluation results show that the proposed approach outperforms the state-of-the-art on the LC-QuAD benchmark.
Moreover, we are able to show that the performance of the approach depends only on the quality of the question interpretation results, i.e., given a correct relevance score distribution, our approach always produces a correct answer ranking. 
Our error analysis reveals correct answers missing from the benchmark dataset and inconsistencies in the DBpedia knowledge graph.
Finally, we provide a comprehensive evaluation of the proposed approach accompanied with an ablation study and an error analysis, which showcase the pitfalls for each of the question answering components in more detail.

\section{Introduction}
The amount of data shared on the Web grows every day~\cite{gandomi2015beyond}.
Information retrieval systems are very efficient but they are limited in terms of the representation power for the underlying data structure that relies on an index for a single database table, i.e., a homogeneous collection of textual documents that share the same set of attributes, e.g., web pages or news articles~\cite{manning2010introduction}.
Knowledge graphs (KGs), i.e., graph-structured knowledge bases, such as DBpedia~\cite{DBLP:journals/semweb/LehmannIJJKMHMK15} or Wikidata~\cite{DBLP:journals/cacm/VrandecicK14}, can interlink datasets with completely different schemas~\cite{DBLP:journals/dagstuhl-reports/BonattiDPP18}.
Moreover, SPARQL is a very expressive query language that allows us to retrieve data from a KG that matches specified graph patterns~\cite{sparql}.
Query formulation in SPARQL is not easy in practice since it requires knowledge of which datasets to access, their vocabulary and structure~\cite{DBLP:journals/internet/FreitasCOO12}.
Natural language interfaces can mitigate these issues, making data access more intuitive and also available for the majority of lay users~\cite{hendrix1982natural,DBLP:conf/semweb/KaufmannB07}.
One of the core functionalities for this kind of interfaces is question answering (QA), which goes beyond keyword or boolean queries, but also does not require knowledge of a specialised query language~\cite{DBLP:conf/rweb/UngerFC14}.

QA systems have been evolving since the early 1960s with efforts in the database community to support natural language queries by translating them into structured queries~\citep[see, e.g.,][]{green-automatic-1963,woods-lunar-1977,bronnenberg-question-1980}.
Whereas a lot of recent work has considered answering questions using unstructured text corpora~\cite{DBLP:conf/emnlp/RajpurkarZLL16} or images~\cite{DBLP:conf/cvpr/GoyalKSBP17}, we consider the task of answering questions using  information stored in KGs.
KGs are an important information source that provide a convenient intermediate representation that can integrate information from different sources and different modalities, such as images and text~\cite{DBLP:conf/www/FariaUSMF18}.
The resulting models are at the same time abstract, compact, and interpretable~\cite{wilcke2017knowledge}.

Question answering over knowledge graphs (KGQA) requires matching an input question to a subgraph, in the simplest case matching a single labeled edge (\emph{triple}) in the KG, a task also called \emph{simple} question answering~\cite{DBLP:journals/corr/BordesUCW15}.
The task of \emph{complex} question answering is defined in contrast to simple KGQA and requires matching more than one triple in the KG~\cite{DBLP:conf/semweb/TrivediMDL17}.
Previously proposed approaches to complex KGQA formulate it as a subgraph matching task~\cite{DBLP:conf/coling/BaoDYZZ16,maheshwari2018learning,DBLP:conf/coling/SorokinG18}, which is an NP-hard problem (by reduction to the subgraph isomorphism problem)~\cite{DBLP:conf/sigmod/ZouHWYHZ14}, or attempt to translate a natural language question into template-based SPARQL queries to retrieve the answer from the KG~\cite{diefenbach2018towards}, which requires a large number of candidate templates~\cite{DBLP:journals/corr/abs-1809-10044}.

We propose an approach to complex KGQA, called \OurApproach{}, based on an unsupervised message-passing algorithm, which allows for efficient reasoning under uncertainty using text similarity and the graph structure.
The results of our experimental evaluation demonstrate that \OurApproach{} is able to manage uncertainties in interpreting natural language questions, overcoming inconsistencies in a KG and incompleteness in the training data, conditions that restrict applications of alternative supervised approaches.

A core aspect of \OurApproach{} is in disentangling reasoning from the question interpretation process.
We show that uncertainty in reasoning stems from the question interpretation phase alone, meaning that under correct question interpretations \OurApproach{} will always rank the correct answers at the top.
\OurApproach{} is designed to accommodate uncertainty inherent in perception and interpretation processes via confidence scores that reflect natural language ambiguity, which, just like for humans, depends on the ability to interpret terms correctly.
These ranked confidence values are then aggregated through our message-passing in a well-defined manner, which allows us to simultaneously consider multiple alternative interpretations of the seed terms, favoring the most likely interpretation in terms of the question context and relations modeled within the KG.
Rather than iterating over all possible orderings, we show how to evaluate multiple alternative question interpretations in parallel via efficient matrix operations.

Another assumption of \OurApproach{} that proves useful in practice is to deliberately disregard subject-object order, i.e., edge directions in a knowledge graph, thereby treating the graph as undirected.
Due to relation sparsity, this model relaxation turns out to be sufficient for most of the questions in the benchmark dataset.
We also demonstrate that due to insufficient relation coverage of the benchmark dataset any assumption on the correct order of the triples in the KG is prone to overfitting.
More than one question-answer example per relation is required to learn and evaluate a supervised model that predicts relation directionality.

Our evaluation on LC-QuAD\footnote{\url{http://lc-quad.sda.tech}}~\cite{DBLP:conf/semweb/TrivediMDL17}, a recent large-scale benchmark for complex KGQA, shows that \OurApproach{} significantly outperforms the state-of-the-art, without the need to translate a natural language question into a formal query language such as SPARQL.
We also show that \OurApproach{} is interpretable in terms of activation paths, and simple, effective and efficient at the same time.
Moreover, our error analysis demonstrates limitations of the LC-QuAD benchmark, which was constructed using local graph patterns.

The rest of the chapter is organized as follows.
Section~\ref{sec:related_work} summarizes the state of the art in KGQA.
Section \ref{sec:approach} presents our approach, \OurApproach{}, with particular attention to the \emph{question interpretation} and \emph{answer inference} phases.
In Section~\ref{sec:evaluation}, we describe the setup for the experimental evaluation of \OurApproach{} on the LC-QuAD dataset. We also provide a detailed ablation, scalability and error analysis in Section~\ref{sec:results}. Finally, Section \ref{sec:conclusion} concludes and lists future work.

\section{Related Work}
\label{sec:related_work}
The most commonly used KGQA benchmark is the SimpleQuestions \cite{DBLP:journals/corr/BordesUCW15} dataset, which contains questions that require identifying a single triple to retrieve the correct answers.
Recent results~\cite{DBLP:conf/emnlp/PetrochukZ18} show that most of these simple questions can be solved using a standard neural network architecture.
This architecture consists of two components: (1) a conditional random fields (CRF) tagger with GloVe word embeddings for subject recognition given the text of the question, and (2) a bidirectional LSTM with FastText word embeddings for relation classification given the text of the question and the subject from the previous component.
Approaches to simple KGQA cannot easily be adapted to solving complex questions, since they rely heavily on the assumption that each question refers to only one entity and one relation in the KG, which is no longer the case for complex questions.
Moreover, complex KGQA also requires matching more complex graph patterns beyond a single triple.

Since developing KGQA systems requires solving several tasks, namely entity, relation and class linking, and afterwards query building, they are often implemented as independent components and arranged into a single pipeline~\cite{DBLP:conf/semweb/DubeyBCL18}.
Frameworks such as QALL-ME~\cite{DBLP:journals/ws/FerrandezSKDFNITONMG11}, OKBQA~\cite{DBLP:conf/semweb/KimUNFHKCKUKC17} and Frankenstein~\cite{DBLP:conf/esws/SinghBRS18}, allow one to share and reuse those components as a collaborative effort.
For example, Frankenstein includes 29 components that can be combined and interchanged~\cite{DBLP:conf/www/SinghRBSLUVKP0V18}.
However, the distribution of the number of components designed for each task is very unbalanced.
Most of the components in Frankenstein support entity and relation linking, 18 and 5 components respectively, while only two components perform query building~\cite{DBLP:journals/corr/abs-1809-10044}.

There is a lack of diversity in approaches that are being considered for retrieving answers from a KG.
OKBQA and Frankenstein both propose to translate natural language questions to SPARQL queries and then use existing query processing mechanism to retrieve answers.\footnote{\url{http://doc.okbqa.org/query-generation-module/v1/}}
We show that using matrix algebra approaches is more efficient in case of natural language processing than traditional SPARQL-based approaches since they are optimized for parallel computation, thereby allowing us to explore multiple alternative question interpretations at the same time~\cite{DBLP:conf/hpec/KepnerABBFGHKLM16,jamour2018demonstration}.

Query building approaches involve query generation and ranking steps~\cite{maheshwari2018learning,DBLP:conf/esws/ZafarNL18}.
These approaches essentially consider KGQA as a subgraph matching task~\cite{DBLP:conf/coling/BaoDYZZ16,maheshwari2018learning,DBLP:conf/coling/SorokinG18}, which is an NP-hard problem (by reduction to the subgraph isomorphism problem)~\cite{DBLP:conf/sigmod/ZouHWYHZ14}.
In practice, \citet{DBLP:journals/corr/abs-1809-10044} report that the question building components of Frankenstein fail to process 46\% questions from a subset of LC-QuAD due to the large number of triple patterns.
The reason is that most approaches to query generations are template-based~\cite{diefenbach2018towards} and complex questions require a large number of candidate templates~\cite{DBLP:journals/corr/abs-1809-10044}.
For example, WDAqua~\cite{diefenbach2018towards} generates 395  SPARQL queries as possible interpretations for the question ``Give me philosophers born in Saint Etienne.''

In summary, we identify the query building component as the main bottleneck for the development of KGQA systems and propose \OurApproach{} as an alternative to the query building approach.
We also observe that the pipeline paradigm is inefficient since it requires KG access first for disambiguation and then for query building again.
\OurApproach{} accesses the KG only to aggregate the confidence scores via graph traversal after question parsing and shallow linking that matches an input question to labels of nodes and edges in the KG.

The work most similar to ours is the spreading activation model of Treo~\cite{DBLP:journals/dke/FreitasOOSC13}, which is also a no-SPARQL approach based on graph traversal that propagates relatedness scores for ranking nodes with a cut-off threshold.
Treo relies on POS tags, the Stanford dependency parser, Wikipedia links and TF/IDF vectors for computing semantic relatedness scores between a question and terms in the KG.
Despite good performance on the QALD 2011 dataset, the main limitation of Treo is an average query execution time of 203s~\cite{DBLP:journals/dke/FreitasOOSC13}.
In this chapter we show how to scale this kind of approach to large KGs and complement it with machine learning approaches for question parsing and word embeddings for semantic expansion.

Our approach overcomes the limitations of the previously proposed graph-based approach in terms of efficiency and scalability, which we demonstrate on a compelling benchmark.
We evaluate \OurApproach{} on LC-QuAD~\cite{DBLP:conf/semweb/TrivediMDL17}, which is the largest dataset used for benchmarking complex KGQA.
WDAqua is our baseline approach, which is the state-of-the-art in KGQA as the winner of the most recent Scalable Question Answering Challenge (SQA2018)~\cite{DBLP:conf/esws/NapolitanoUN18}.
Our evaluation results demonstrate improvements in precision and recall, while twice reducing average execution time over the SPARQL-based WDAqua, which is also orders of magnitude faster than the results reported for the previous graph-based approach Treo.

There is other work on KGQA that uses embedding queries into a vector space~\cite{DBLP:conf/nips/HamiltonBZJL18,DBLP:conf/semweb/WangWLCZQ18}.
The benefit of our graph-based approach is in preserving the original structure of the KG that can be used for both executing precise formal queries and answering ambiguous natural language questions at the same time.
The graph structure also makes the results traceable and, therefore, interpretable in terms of relevant paths and subgraphs in comparison with vector space operations.

\OurApproach{} uses message passing, a family of approaches that were initially developed in the context of probabilistic graphical models~\cite{pearl1988probabilistic,koller2009probabilistic}.
Recently, graph neural networks trained to learn patterns of message passing have been shown to be effective on a variety of tasks~\cite{DBLP:conf/icml/GilmerSRVD17,DBLP:journals/corr/abs-1806-01261}, including KG completion~\cite{DBLP:conf/esws/SchlichtkrullKB18}.
We show that our unsupervised approach to message passing performs well on the complex question answering task and helps to overcome sampling biases in the training data, which supervised approaches are prone to.

\section{Approach}
\label{sec:approach}
\OurApproach{}, our KGQA approach, consists of two consecutive phases: (1) question interpretation, and (2) answer inference.
In the question interpretation phase we identify the sets of entities and predicates that we consider relevant for answering the input question along with the corresponding confidence scores.
In the second phase these confidence scores are propagated and aggregated directly over the structure of the KG, to provide a confidence distribution over the set of possible answers.

Our notion of a KG is inspired by common concepts from the Resource Description Framework (RDF)~\cite{rdf-primer}, a standard representation used in many large-scale knowledge graphs, e.g., DBpedia and Wikidata:

\begin{definition}
\label{def:kg}
We define a (knowledge) \emph{graph} $K=\langle E,G,P\rangle$ as a tuple that contains sets of \emph{entities} $E$ (nodes) and \emph{properties} $P$, both represented by Unique Resource Identifiers (URIs), and a set of directed labeled edges $\langle e_i, p, e_j\rangle\in G$, where $e_i, e_j \in E$ and $p \in P$.
\end{definition}

\noindent%
The set of edges $G$ in a KG can be viewed as a (blank-node-free) RDF graph, with subject-predicate-object triples $\langle e_i, p, e_j\rangle$.
In analogy with RDFS, we refer to a subset of entities $C \subseteq E$ appearing as objects of the special property \texttt{\small rdf:type} as \emph{Classes}.
We also refer to classes, entities and properties collectively as \textit{terms}.
We ignore RDF literals, except for \texttt{\small rdfs:labels} that are used for matching questions to terms in KG.

The task of \emph{question answering over a knowledge graph} (KGQA) is: given a natural language question $Q$ and a knowledge graph $K$, produce the correct answer $A$, which is either a subset of entities in the KG $A \subseteq E$ or a result of a computation performed on this subset, such as the number of entities in this subset (\smalltt{COUNT}) or an assertion (\smalltt{ASK}).
These types of questions are the most frequent in existing KGQA benchmarks~\cite{DBLP:journals/corr/BordesUCW15,DBLP:conf/semweb/TrivediMDL17,DBLP:conf/semweb/UsbeckGN018}.
In the first phase \OurApproach{} maps a natural language question $Q$ to a structured model $q$, which the answer inference algorithm will operate on then. 

\subsection{Question interpretation}

To produce a question model $q$ we follow two steps: (1) \textit{parse}, which extracts references (entity, predicate and class mentions) from the natural language question and identifies the question type; and (2) \textit{match}, which assigns each of the extracted references to a ranked list of candidate entities, predicates and classes in the KG.

\begin{figure*}[!t]
\centering
\includegraphics[width=\textwidth]{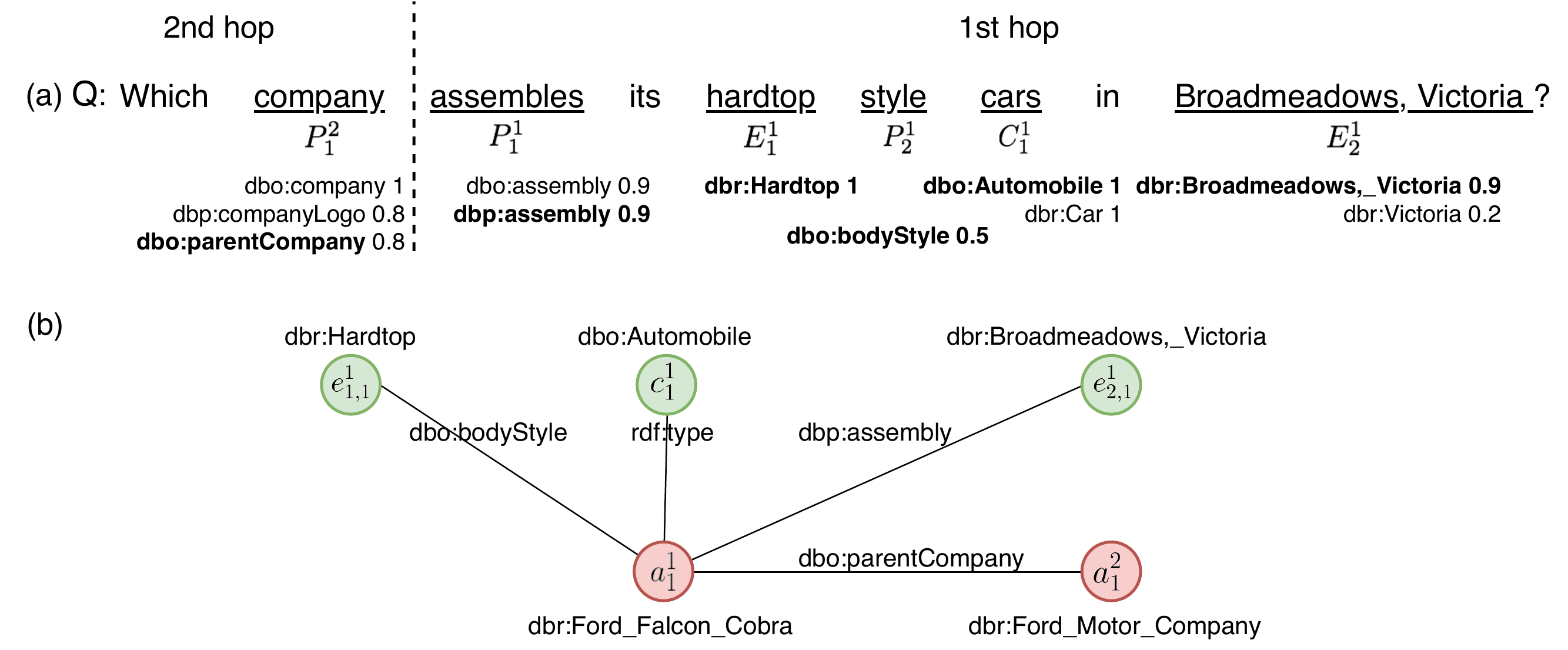}
\caption{(a) A sample question $Q$ highlighting different components of the question interpretation model: references and matched URIs with the corresponding confidence scores, along with (b) the illustration of a sample KG subgraph relevant to this question. The URIs in bold are the correct matches corresponding to the KG subgraph.}
\label{fig:qi}
\end{figure*}

Effectively, a complex question requires answering several sub-questions, which may depend on or support each other. A dependence relation between the sub-questions means that an answer $A^1$ to one of the questions is required to produce the answer $A^2$ for the other question: $A^2=f(A^1,K)$. 
We call such complex questions \textit{compound} questions and match the sequence in which these questions should be answered to \textit{hops} (in the context of this chapter, one-variable graph patterns) in the KG.
Consider the sample compound question in Figure~\ref{fig:qi}, which consists of two hops: (1) find the car types assembled in Broadmeadows Victoria, which have a hardtop style, (2) find the company, which produces these car types. There is an intermediate answer (the car types with the specified properties), which is required to arrive at the final answer (the company).

Accordingly, we define (compound) questions as follows:

\begin{definition}\label{def:question}
A \emph{question model} is a tuple $q=\langle t_q, Seq_q \rangle$, where $t_q \in T$ is a question type required to answer the question $Q$, and $Seq_q = (\langle E^i, P^i, C^i\rangle)_{i=1}^h$ is a sequence of $h$ hops over the KG, $E^i$ is a set of entity references, $P^i$ -- a set of property references, $C^i$ -- a set of class references relevant for the \textit{i}-hop in the graph, and $T$ -- a set of question types, such as $\{\smalltt{SELECT}, \smalltt{ASK}, \smalltt{COUNT}\}$.
\end{definition}

\noindent%
Hence, the question in Figure~\ref{fig:qi} can be modeled as: $\langle \smalltt{SELECT}, (\langle E^1={}$\{``hardtop'', ``Broadmeadows, Victoria''\}, $P^1={}$\{``assembles'', ``style''\}, $C^1={}$\{``cars''\}$\rangle, \langle E^2=\emptyset, P^2={}$\{``company''\}, $C^2=\emptyset\rangle) \rangle$, where $E^i$, $P^i$, $C^i$ refer to the entities, predicates and classes in hop $i$.

Further, we describe how the question model $q$ is produced by parsing the input question $Q$, after which we match references in $q$ to entities and predicates in the graph $K$.

\paragraph{\bf Parsing.}
Given a natural language question $Q$, the goal is to classify its type $t_q$ and parse it into a sequence $Seq_q$ of reference sets according to Definition~\ref{def:question}.
Question type detection is implemented as a supervised classification model trained on a dataset of annotated question-type pairs that learns to assign an input question to one of the predefined 
types $t_q \in T$.

We model reference (mention) extraction $Seq_q$ as a sequence labeling task~\cite{DBLP:conf/icml/LaffertyMP01}, in which a question is represented as a sequence of tokens (words or characters).
Then, a supervised machine learning model is trained on an annotated dataset to assign labels to tokens, which we use to extract references to entities, predicates and classes.
Moreover, we define the set of labels to group entities, properties and classes referenced in the question into $h$ hops.

\paragraph{\bf Matching.}
Next, the question model (Definition~\ref{def:question}) is further updated with an \emph{interpreted question model} $I(q) = (t_q, SEQ_q)$
in which each component of $Seq_q$ is represented by sets of pairs from $(E \cup P \cup C) \times [0, 1]$ obtained by matching the references to concrete terms in $K$ (by their URIs) as follows: for each entity (or property, class, resp.) reference in $Seq_q$, we retrieve a ranked list of most similar entities from the KG along with the matching confidence score.

Fig.~\ref{fig:qi} also shows the result of this matching step on our example.
For instance, the property references for the first hop are replaced by the set of candidate URIs: $P^1 = \{P_1^1, P_2^1 \} \in SEQ_q$ within $I(q)$, where $P_1^1 = \{ \texttt{(\small  dbo:assembly}, 0.9)$, $(\texttt{\small dbp:assembly}, 0.9)\}$, $P_2^1 = \{ (\texttt{\small dbo:bodyStyle}, 0.5)\}$.

\subsection{Answer inference}

Our answer inference approach 
iteratively traverses and aggregates confidence scores across the graph based on the initial assignment from $I(q)$.
An answer set $A^i$, i.e., a set of entities along with their confidence scores $E \times [0, 1]$, is produced after each hop $i$ and used as part of the input to the next hop $i+1$, along with the 
terms matched for this hop in $I(q)$, i.e., $SEQ_q(i+1) = \langle E^{i+1}, P^{i+1}, C^{i+1}\rangle$.
The entity set $A^h$ produced after the last hop $h$ can be further transformed to produce the final answer: $A_q=f_{t_q}(A^h)$ via an aggregation function $f_{t_q} \in F$ from a predefined set of available aggregation functions $F$ defined for each of the question types $t_q \in T$.
We compute the answer set $A^i$ for each hop inductively in two steps: (1) \emph{subgraph extraction} and (2)~\emph{message passing}.

\paragraph{\bf Subgraph extraction.} 
This step refers to the retrieval of relevant triples from the KG that form a subgraph.
Thus, the URIs of the matched entities and predicates in the query are used as seeds to retrieve the triples in the KG that contain at least one entity (in subject or object position), and one predicate from the corresponding reference sets.
Therefore, the extracted subgraph will contain $n$ entities, which include all entities from $E^{i}$ and the entities adjacent to them through properties from $P^{i}$.

The subgraph is represented as a set of $k$ adjacency matrices with $n$ entities in the subgraph: $\mathbb{S}^{k\times n\times n}$, where $k$ is the total number of matched property URIs.
There is a separate $n\times n$ matrix for each of the $k$ properties used as seeds, where $\mathbb{S}_{pij} = 1$ if there is an edge labeled $p$ between the entities $i$ and $j$ , and 0 otherwise.
All adjacency matrices are symmetric, because $I(q)$ does not model edge directionality, i.e., it treats $K$ as undirected.
Diagonal entries are assigned 0 to ignore self loops.

\begin{algorithm}[tb]
\caption{Message passing for KGQA}
\label{alg:mp}
\begin{flushleft}
\textbf{Input:} adjacency matrices of the subgraph $\mathbb{S}^{k\times n\times n}$,\\ entity $\mathbb{E}^{l\times n}$ and property reference activations $\mathbb{P}^{m\times k}$\\
\noindent\textbf{Output:} answer activations vector $A \in \mathbb{R}^n$\\ 
\end{flushleft}
\begin{algorithmic}[1]
\STATE  $W^{n}, N_P^{n}, \mathbb{Y_E}^{l\times n} = \emptyset$
\FOR{$P_j \in \mathbb{P}^{m\times k}, j \in \{1, ..., m\}$}
    \STATE {$\mathbb{S}_j = \bigoplus_{i=1}^{k} P_j \otimes \mathbb{S}$} \commentA{property update}
    \STATE $\mathbb{Y} = \mathbb{E} \oplus\otimes \mathbb{S}_j$ \commentA{entity update}
    \STATE {$W = W + \bigoplus_{i=1}^{l} \mathbb{Y}_{ij}$} \commentA{sum of all activations}
    \STATE {${N}_{P_j} = \sum_{i=1}^{l} 1$ if $\mathbb{Y}_{ij} > 0$ else $0$} \STATE $\mathbb{Y_E} = \mathbb{Y_E} \oplus \mathbb{Y}$ \commentA{activation sums per entity}
\ENDFOR
\STATE $W = 2\cdot W / (l + m) $ \commentA{activation fraction}
\STATE $N_E = \sum_{i=1}^{l} 1$ if $\mathbb{Y_E}_{ij} > 0$ else $0$ \STATE \textbf{return} $A = (W \oplus N_E \oplus N_P) / (l + m + 1)$
\end{algorithmic}
\end{algorithm}

\begin{figure*}[!t]
\centering
\includegraphics[width=\textwidth]{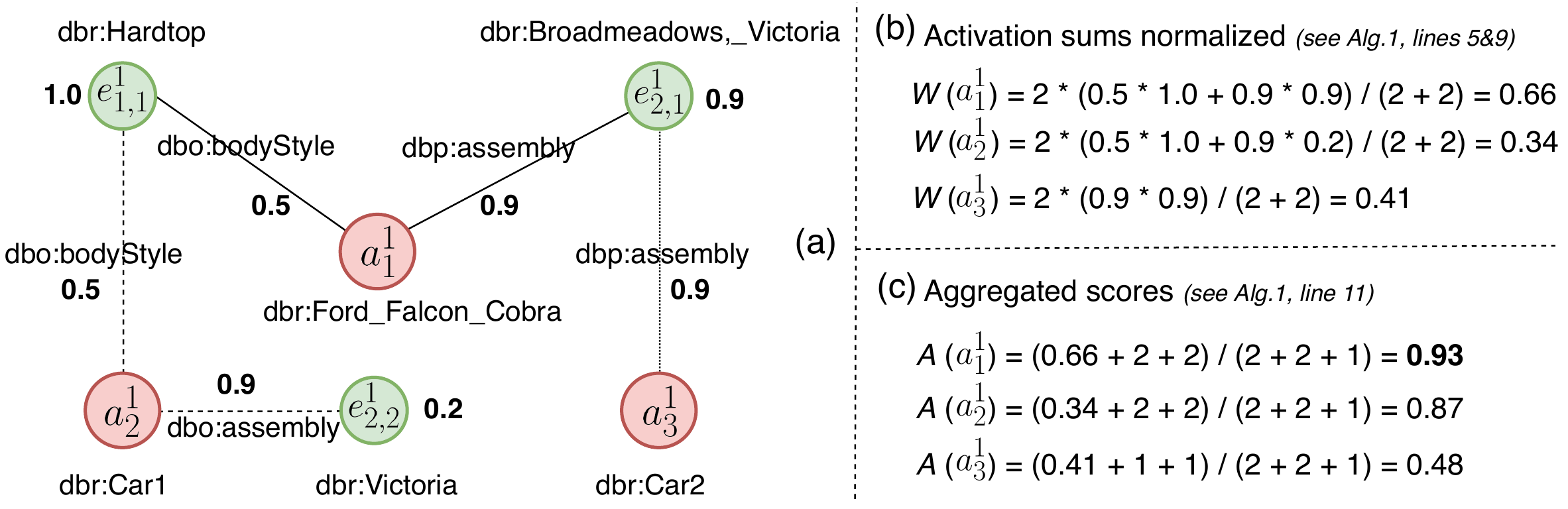}
 \caption{
(a) A sample subgraph with three entities as candidate answers, (b) their scores after predicate and entity propagation, and (c) the final aggregated score.
 }
\label{fig:agg}
\end{figure*}

\paragraph{\bf Message passing.} The second step of the answer inference phase involves message passing,\footnote{The pseudocode of the message passing algorithm is presented in Algorithm~\ref{alg:mp}.} i.e., propagation of the confidence scores from the entities $E^i$ and predicates $P^i$, matched in the question interpretation phase, to adjacent entities in the extracted subgraph.
This process is performed in three steps, (1)~\emph{property update}, (2)~\emph{entity update}, and (3)~\emph{score aggregation}.
Algorithm~\ref{alg:mp} summarizes this process, detailed as follows.

For each of $m$ property references $P_j \in \mathbb{P}^{m\times k}, j \in \{1, \ldots, m\}$ where $m = |P^i|$, we
\begin{enumerate}[nosep,leftmargin=*]
\item \emph{select} the subset of adjacency matrices from $\mathbb{S}^{k\times n\times n}$ for the property URIs if $p_{ij} > 0$, where $p_{ij} \in P_j$, and propagate the confidence scores to the edges of the corresponding adjacency matrices via element-wise multiplication.
Then, all adjacency matrices are combined into a single adjacency matrix $\mathbb{S}_j^{n\times n}$, which contains all of their edges with the sum of confidence scores if edges overlap (\emph{property update}: line 3, Algorithm~\ref{alg:mp}).

\item \emph{perform} the main message-passing step via the sum-product update, in which the confidence scores from $l$ entity references, where $l = |E^i|$, are passed over to the adjacent entities via all edges in $\mathbb{S}_j^{n\times n}$ (\emph{entity update}: line 4, Algorithm~\ref{alg:mp}).

\item \emph{aggregate} the confidence scores for all $n$ entities in the subgraph into a single vector $A$ by combining the sum of all confidence scores with the number of entity and predicate reference sets, which received non-zero confidence score.
The intuition behind this \emph{score aggregation} formula (line 11,  Algorithm~\ref{alg:mp}) is that the answers that received confidence from the majority of entity and predicate references in the question should be preferred.
The computation of the answer scores for our running example is illustrated in Fig.~\ref{fig:agg}.
\end{enumerate}

\noindent%
The minimal confidence for the candidate answer is regulated by a threshold to exclude partial and low-confidence matches.
Finally, we also have an option to filter answers by considering only those entities in the answer set $A^i$ that have one of the classes in $C^i$.

The same procedure is repeated for each hop in the sequence using the corresponding URI activations for entities, properties and classes modeled in $SEQ_q(i)=\langle E^i, P^i, C^i\rangle$ and augmented with the intermediate answers produced for the previous hop $A^{i-1}$.
Lastly, the answer to the question $A_q$ is produced based on the entity set $A^h$, which is either returned `as is' or put through an aggregation function $f_{t_q}$ conditioned on the question type $t_q$.

\begin{figure*}[!t]
\centering
\includegraphics[width=0.8\textwidth]{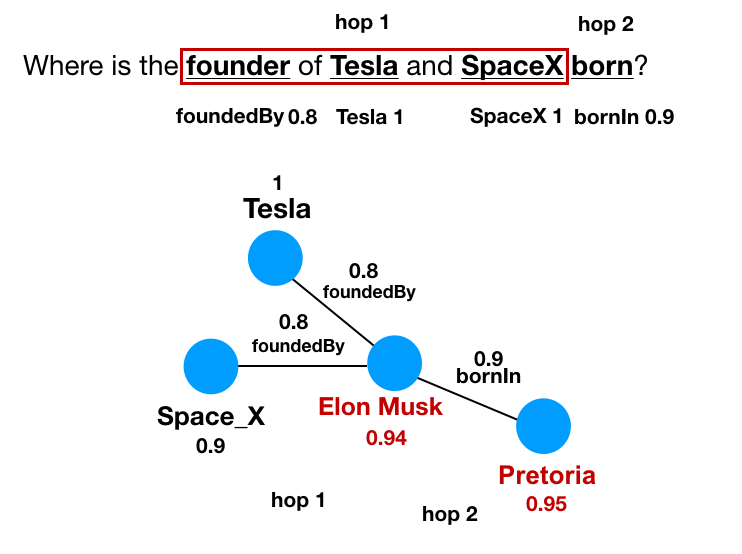}
 \caption{
A sample subgraph illustrating the message passing algorithm.
 }
\label{fig:sample}
\end{figure*}

\added{Our message passing algorithm effectively defines a breadth-first search procedure through a knowledge graph.
A simple example of message passing for complex question answering over a knowledge graph is given in Figure~\ref{fig:sample}.
The initial weights of the nodes originate from the question interpretation phase, which includes parsing and matching of the question text to the entities and relations stored in knowledge graph. The matching function also assigned scores to each of the mention spans parsed from the question text. In our example, ``founder’’ was matched to the relation with the label ``foundedBy’’ with the confidence score 0.8, the mention span ``Tesla’’ to the entity with the label ``Tesla’’ with the score 1, etc.
In the first hop the activation scores from the two nodes ``Tesla'' and ``SpaceX'' are being propagated to the adjacent node ``Elon Musk''.
The basic operator in message passing is implemented via the sum-product rule that can be efficiently performed using matrix multiplications: $1 \times 0.8 + 1 \times 0.9 = 1.52$.
Next, the score is normalised to 1 by dividing by the number of triples: $1.52 / 2 = 0.76$.
Finally, the score is further adjusted to account for the fraction of the matched entities and properties: $(0.76 + 2 + 1) / (1 + 2 + 1) = 0.94$.
Exactly the same procedure is repeated for the second hop. The node in the knowledge graph corresponding to the entity labeled ``Elon Musk’’, which was obtained as an intermediate answer from the previous hop with the confidence score of 0.94, and the relation parsed from the question text as the relation to be expanded during the second hop and matched to the relation with the label ``bornIn’’ with the confidence score of 0.9 are to be considered in the second hop: $(0.94 \times 0.9 + 1 + 1) / (1 + 1 + 1) = 0.95$.
Thus, the message passing procedure iteratively explores the structure of the knowledge graph and aggregates the scores assigned to the entities and properties in the graph based on the question interpretation procedure.
}

\section{Evaluation Setup}
\label{sec:evaluation}

We evaluate \OurApproach{}, our KGQA approach, on the LC-QuAD dataset of complex questions constructed from the DBpedia KG~\cite{DBLP:conf/semweb/TrivediMDL17}.
First, we report the evaluation results of the end-to-end approach, which incorporates our message-passing algorithm in addition to the initial question interpretation (question parser and matching functions).
Second, we analyze the fraction and sources of errors produced by different KGQA components, which provides a comprehensive perspective on the limitations of the current state-of-the-art for KGQA, the complexity of the task, and limitations of the benchmark.
Our implementation and evaluation scripts are open-sourced.\footnote{\url{https://github.com/svakulenk0/KBQA}}

\paragraph{\bf Baseline.}
We use WDAqua~\cite{diefenbach2018towards} as our baseline; to the best of our knowledge, the results produced by WDAqua are the only published results on the end-to-end question answering task for the LC-QuAD benchmark to date.
It is a rule-based framework that integrates several KGs in different languages and relies on a handful of SPARQL query patterns to generate SPARQL queries and rank them as likely question interpretations.
We rely on the evaluation results reported by the authors~\cite{diefenbach2018towards}.
WDAqua results were produced for the full LC-QuAD dataset, while other datasets were used for tuning the approach.

\paragraph{\bf Metrics.}
We follow the standard evaluation metrics for the end-to-end KGQA task, i.e., we report precision (P) and recall (R) macro-averaged over all questions in the dataset, and then use them to compute the F-measure (F). 
Following the evaluation setup of the QALD-9 challenge~\cite{DBLP:conf/semweb/UsbeckGN018} we assign both precision and recall equal to 0 for every question in the following cases: (1) for \smalltt{SELECT} questions, no answer (empty answer set) is returned, while there is an answer (non-empty answer set) in the ground truth annotations; (2) for \smalltt{COUNT} or \smalltt{ASK} questions, an answer differs from the ground truth; (3) for all questions, the predicted answer type differs from the ground truth.
In the ablation study, we also analyze the fraction of questions with errors for each of the components separately, where an error is a not exact match with the ground-truth answer.

\paragraph{\bf Hardware.}
\added{The computational setup in our experiments is comparable to the setup used for evaluating WDAqua~\cite{diefenbach2018towards}. That is, all experiments were performed on a single commodity server equipped with Intel
Core i7-8700K CPU 3.70GHz, RAM 16 GB and 240 GB
SSD, which is comparable to the setup reported for the competing approach: Intel Xeon E5-2667 CPU 3.2GHz, RAM 16 GB and 500 GB SSD~\cite{diefenbach2018towards}. All message-passing operations are being performed on a CPU since they require sparse matrix multiplications.}

\subsection{The LC-QuAD dataset}
\label{sec:dataset}
The LC-QuAD dataset\footnote{\url{https://github.com/AskNowQA/LC-QuAD}}~\cite{DBLP:conf/semweb/TrivediMDL17} contains 5K question-query pairs that have correct answers in the DBpedia KG (2016-04 version).
The questions were generated using a set of SPARQL templates by seeding them with DBpedia entities and relations, and then paraphrased by human annotators.
All queries are of the form \smalltt{ASK}, \smalltt{SELECT}, and \smalltt{COUNT}, fit to subgraphs with diameter of at most 2-hops, contain 1--3 entities and 1--3 properties.

We used the train and test splits provided with the dataset (Table~\ref{tab:dataset}).
Two queries with no answers in the graph were excluded.
All questions are also annotated with ground-truth reference spans\footnote{\url{https://github.com/AskNowQA/EARL}} to evaluate performance of entity linking and relation detection~\cite{DBLP:conf/semweb/DubeyBCL18}.

\begin{table}
\centering
\ttabbox{%
\begin{tabular}{lrrr}
\toprule
 & \multicolumn{3}{c}{Questions} \\
\cmidrule{2-4}
Split & All & Complex & Compound  \\
\midrule
all  & 4,998 (100\%)       &     3,911 (78\%)       & 1,982 (40\%)           \\
train  & 3,999 \phantom{0}(80\%)       &      3,131 (78\%)           & 1,599 (40\%)          \\
test   & 999 \phantom{0}(20\%)        &       780 (78\%)          & 383 (38\%)         \\
\bottomrule
\end{tabular}
}{
\caption{Dataset statistics: number of questions across the train and test splits; number of complex questions that reference more than one triple; number of complex questions that require two hops in the graph through an intermediate answer-entity.}
\label{tab:dataset}
}
\end{table}

\subsection{Implementation details}
\label{sec:implementation5}

Our implementation uses the English subset of the official DBpedia 2016-04 dump losslessly compressed into a single HDT file\footnote{\url{http://fragments.dbpedia.org/hdt/dbpedia2016-04en.hdt}}~\cite{fernandez2013}.
HDT is a state-of-the-art compressed RDF self-index, which scales linearly with the size of the graph and is, therefore, applicable to very large graphs in practice.
This KG contains 1B triples, more than 26M entities (\emph{dbpedia.org} namespace only) and 68,687 predicates.
Access to the KG for subgraph extraction and class constraint look-ups is implemented via the Python HDT API.\footnote{\url{https://github.com/Callidon/pyHDT}}
This API builds an additional index \cite{martinez2012exchange} to speed up all look-ups, and consumes the HDT mapped in disk, with ${\sim}$3\% memory footprint.\footnote{Overall, DBpedia 2016-04 takes 18GB in disk, and 0.5GB in main memory.}

Our end-to-end KGQA solution integrates several components that can be trained and evaluated independently.
The pipeline includes two supervised neural networks for (1)~question type detection and (2)~reference extraction; and unsupervised functions for (3)~entity and (4)~predicate matching, and (5)~message passing.

\paragraph{\bf Parsing.}
Question type detection is implemented as a bi-LSTM neural-network classifier trained on pairs of question and type.
We use another biLSTM+CRF neural network for extracting references to entities, classes and predicates for at most two hops using the set of six labels: \{``E1'', ``P1'', ``C1'', ``E2'', ``P2'', ``C2''\}.
Both classifiers use GloVe word embeddings pre-trained on the Common Crawl corpus with 840B tokens and 300 dimensions~\cite{pennington2014glove}. 

\paragraph{\bf Matching.}
The labels of all entities and predicates in the KG (\texttt{\small rdfs:label} links) are indexed into two separate catalogs and also embedded into two separate vector spaces using the English FastText model trained on Wikipedia~\cite{bojanowski2017enriching}.
We use two ranking functions for matching and assigning the corresponding confidence scores: index-based for entities and embedding-based for predicates.
The index-based ranking function uses 
BM25~\cite{manning2010introduction} to calculate 
confidence scores
for the top-500 matches on the combination of n-grams and Snowball stems.\footnote{\url{https://www.elastic.co/guide/en/elasticsearch/reference/current}}
Embeddings-based confidence scores are computed using the Magnitude library\footnote{\url{https://github.com/plasticityai/magnitude}}~\cite{patel2018magnitude} for the top-50 nearest neighbors in the FastText embedding space.

Many entity references in the LC-QuAD questions can be handled using simple string similarity matching techniques; e.g., ``companies'' can be mapped to ``dbo:Comp\-any''.
We built an ElasticSearch (Lucene) index to efficiently retrieve such entity candidates via string similarity to their labels.
The entity labels were automatically generated from entity URIs by stripping the domain part of the URI and lower-casing, e.g., entity ``http://dbpedia.org/ontology/Company'' received the label ``company'' to better match question words.
LC-QuAD questions also contain more complex paraphrases of the entity URIs that require semantic similarity computation beyond fuzzy string matching, such as ``movie'' refers to ``dbo:Film'', ``stockholder'' to ``dbp:owner'' or ``has kids'' to ``dbo:child''.
We embeded entity and predicate labels with FastText~\cite{bojanowski2017enriching} to detect semantic similarities beyond string matching.

Index-based retrieval scales much better than nearest neighbour computation in the embedding space, which is a crucial requirement for the 26M entity catalog.
In our experiments, syntactic similarity was sufficient for entity matching in most of the cases, while property matching required capturing more semantic variations and greatly benefited from using pre-trained embeddings.

\section{Evaluation Results}
\label{sec:results}

Table~\ref{tab:results1} shows the performance of \OurApproach{} on the KGQA task in comparison with the results previously reported by  \citet{diefenbach2018towards}.

Our results demonstrate a noticeable improvement in recall (we were able to retrieve answers to 50\% of the benchmark questions), while maintaining a comparable precision score.
For the most recent QALD challenge the guidelines were updated to penalize systems that miss the correct answers, i.e., that are low in recall, which gives a clear signal of its importance for this task~\cite{DBLP:conf/semweb/UsbeckGN018}.
While it is often trivial for users to filter out a small number of incorrect answers that stem from interpretation ambiguity, it is much harder for users to recover missing correct answers.
Indeed, we showed that \OurApproach{} is able to identify correct answers that were missing even from the benchmark dataset since they were overlooked by the benchmark authors due to sampling bias.

\begin{table}
\centering
\ttabbox{%
\begin{tabular}{lcccc}
\toprule
\bf Approach & \bf P & \bf R & \bf F & \bf Runtime \\
\midrule
WDAqua & 0.22\smash{\rlap{*}} & 0.38 & 0.28    & 1.50 s/q \\
\OurApproach{} (our approach)      & 0.25 & 0.50 & 0.33    &   0.72 s/q  \\
\bottomrule
\end{tabular}
}{
\caption{Evaluation results. (*) P of the WDAqua baseline is estimated from the reported precision of 0.59 for answered questions only. Runtime is reported in seconds per question as an average across all questions in the dataset. The distribution of runtimes for \OurApproach{} is Min: 0.01, Median: 0.67 Mean: 0.72, Max: 13.83}
\label{tab:results1}
}
\end{table}

\subsection{Ablation study} 
Table~\ref{tab:results2} summarizes the results of our ablation study for different setups.
We report the fraction of all questions that have at least one answer that deviates from the ground truth (\emph{Total} column), questions with missing term matches (\emph{No match}) and other errors. 
Revised errors is the subset of other errors that were considered as true errors in the manual error analysis.

\begin{figure}[!t]
\centering
\includegraphics[width=0.5\textwidth]{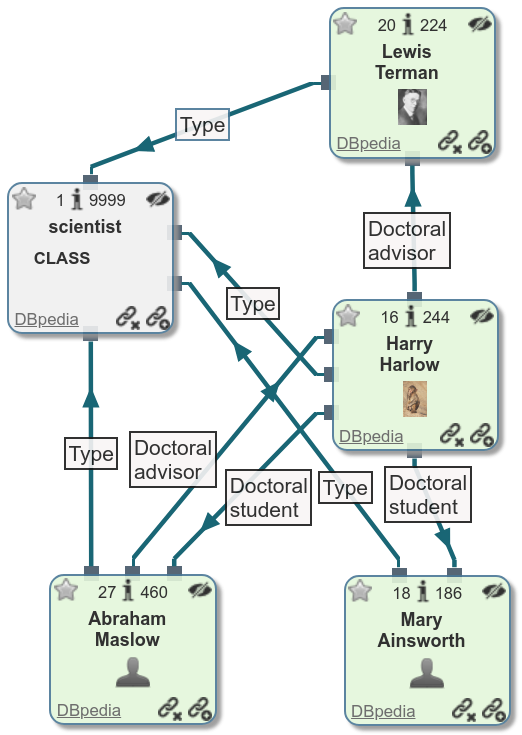}
\vspace{0.2cm}
\caption{\emph{Directed relation} example ({\tt\small dbp:doctoralStudents} and {\tt\small dbp:doctoralAdvisor} hierarchy) that requires modeling directionality of the relation. LC-QuAD question \#3267: ``Name the scientist whose supervisor also supervised Mary Ainsworth?'' (correct answer: Abraham Maslow) can be easily confused with a question: ``Name the scientist who supervised also the supervisor of Mary Ainsworth?'' (correct answer: Lewis Terman).
The LC-QuAD benchmark is not suitable for evaluating directionality interpretations, since only 35 questions (3.5\%) of the LC-QuAD test split use relations of this type, which explains high performance results of \OurApproach{} that treats all relation as undirected.}
\label{fig:supervision}
\end{figure}

\begin{figure}[!t]
\centering
\includegraphics[width=0.6\textwidth]{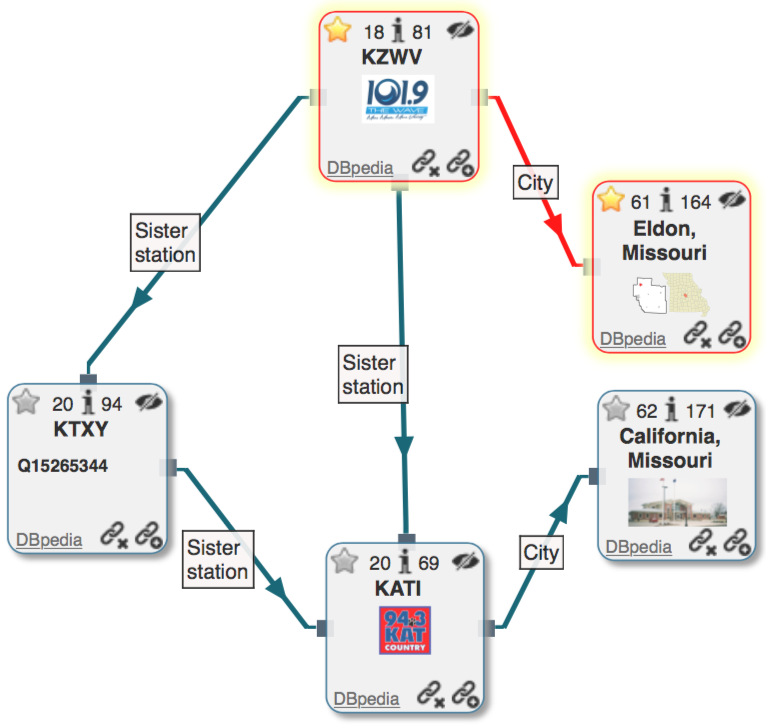}
\caption{\emph{Undirected relation} example ({\tt\small dbo:sisterStation}) that reflects bi-directional association between the adjacent entities (Missouri radio stations). LC-QuAD question \#4486: ``In which city is the sister station of KTXY located?'' (correct answer: {\tt\small dbr:California,Missouri}, {\tt\small dbr:Missouri}; missing answer: {\tt\small dbr:Eldon,Missouri}). DBpedia does not model bi-directional relations and the relation direction is selected at random in these cases. LC-QuAD does not reflect bi-directionality either by picking only one of the directions as the correct one and rejecting correct solutions ({\tt\small dbr:KZWY} $\rightarrow$ {\tt\small dbr:Eldon,Missouri}). \OurApproach{} was able to retrieve this false negative sample due to the default undirectionality assumption built into the question interpretation model.}
\label{fig:sister}
\end{figure}

Firstly, we make sure that the relaxations in our question interpretation model hold true for the majority of questions in the benchmark dataset (95\%)
by feeding all ground truth entity, class and property URIs to the answer inference module (Setup 1 in Table~\ref{tab:results2}).
We found that only 53 test questions (5\%) require one to model the exact order of entities in the triple, i.e., subject and object positions.
These questions explicitly refer to a hierarchy of entity relations, such as {\tt\small dbp:doctoralStudents} and {\tt\small dbp:doctoralAdvisor} (see Figure~\ref{fig:supervision}\footnote{The numbers on top of each entity show its number of predicates and triples.}$^,$\footnote{All sample graph visualizations illustrating different error types discovered LC-QuAD dataset in Figures~\ref{fig:supervision}, \ref{fig:sister}, \ref{fig:rome} were generated using the LODmilla web tool, \url{http://lodmilla.sztaki.hu}~\cite{sztaki8012}, with data from the DBpedia KG.}), and their directionality has to be interpreted to correctly answer such questions.
We also recovered a set of correct answers missing from the benchmark for  relations that are symmetric by nature, but were considered only in one direction by the benchmark, e.g., {\tt\small dbo:related}, {\tt\small dbo:associatedBand}, and {\tt\small dbo:sisterStation} (see Figure~\ref{fig:sister}).

These results indicate that a more complex question model attempting to reflect structural semantics of a question in terms of the expected edges and their directions (parse graph or lambda calculus) is likely to fall short when trained on this dataset: 53 sample questions are insufficient to train a reliable supervised model that can recognize relation directions from text, which explains poor results of a slot-matching model for subgraph ranking reported on this dataset~\cite{maheshwari2018learning}.
\added{Given the parse of an input question as a query graph, the best slot-matching approach achieves F1 0.7 for the LC-QuAD dataset, which is well below the performance of the message-passing algorithm when given the correct question parse (F1 0.98 in Table~\ref{tab:results2})}

There were only 8 errors (1\%) due to the wrong question type detected caused by misspelled or grammatically incorrect questions (row 2 in Table~\ref{tab:results2}).
Next, we experimented with removing class constraints and found out that although they generally help to filter out incorrect answers (row 3) our matching function missed many correct classes even using the ground-truth spans from the benchmark annotations (row 4).

The last four evaluation setups (5--8) in Table~\ref{tab:results2} show the errors from parsing and matching reference spans to entities and predicates in the KG.
Most errors were due to missing term matches (10--34\% of questions), which indicates that the parsing and matching functions constitute the bottleneck in our end-to-end KGQA.
Even with the ground-truth span annotations for predicate references the performance is below 0.6 (34\% of questions), which indicates that relation detection is much harder than the entity linking task, which is in line with results reported by \citet{DBLP:conf/semweb/DubeyBCL18} and \citet{DBLP:journals/corr/abs-1809-10044}.

\begin{landscape}
\begin{table}[]
\vspace{0.2cm}
 \setlength{\tabcolsep}{5pt}
\centering\small
\ttabbox{%
\begin{tabular}{ll|cccc|ccc|c|cc}
\hline
 &    \multirow{3}{*}{\bf Setup}              & \multicolumn{4}{c|}{\multirow{2}{*}{\bf Question interpretation}} &  \multirow{3}{*}{\bf P}  & \multirow{3}{*}{\bf R} & \multirow{3}{*}{\bf F}& \multicolumn{3}{c}{\bf Questions with errors}       \\
\cline{10-12}
   &    &  &  & &    &     &   &    & \multirow{2}{*}{Total}           & \multicolumn{2}{c}{New errors} \\
   \cline{3-6}\cline{11-12}
   &    & Q. type & Entity & Property & Class   &   &   &  &        & No match & Other$\rightarrow$Revised  \\
   
\hline
1   & Question model*     & \multicolumn{4}{c|}{\cellcolor{lightgray}GT}            & 0.97 & 0.99 & 0.98 & \phantom{0}9\%             & --        & ~9\%   $\rightarrow$ ~5\%     \\
\hline
2   & Question type  & PR & \multicolumn{3}{c|}{\cellcolor{lightgray} GT}    & 0.96 & 0.98 & 0.97 & 10\%             & --        & ~1\%   $\rightarrow$ ~1\%     \\
\hline
3   & Ignore classes  & \multicolumn{3}{c|}{\cellcolor{lightgray}GT}&    None      & 0.94 & 0.99 & 0.96 & 14\%             & --        & ~5\%   $\rightarrow$ ~3\%     \\
4   & Classes GT span$^+$  & \multicolumn{3}{c|}{\cellcolor{lightgray}GT}&    GT span$^+$     & 0.89 & 0.92 & 0.90 & 17\%             & \phantom{0}8\%      &  --    \\
\hline
5   & Entities GT span$^+$  & {\cellcolor{lightgray}GT}  &  GT span$^+$ & \multicolumn{2}{c|}{\cellcolor{lightgray} GT} & 0.85 & 0.88 & 0.86 & 20\%            & 10\%     & ~1\%   $\rightarrow$ ~1\%     \\
6   & Entities PR & {\cellcolor{lightgray}GT}& PR& \multicolumn{2}{c|}{\cellcolor{lightgray} GT} &  0.64 & 0.74 & 0.69 & 46\%            & 27\%     & 10\%  $\rightarrow$ ~5\%     \\
\hline
7   & Predicates GT span$^+$  & \multicolumn{2}{c}{\cellcolor{lightgray} GT} & GT span$^+$ & {\cellcolor{lightgray}GT}& 0.56 & 0.59 & 0.57 & 48\%            & 34\%     & ~5\%   $\rightarrow$ ~3\%     \\
8   & Predicates PR    & \multicolumn{2}{c}{\cellcolor{lightgray} GT} & PR & {\cellcolor{lightgray}GT} & 0.36 & 0.53 & 0.43 & 74\%            & 34\%     & 31\%  $\rightarrow$ 19\%    \\
\hline
\end{tabular}
}{
\caption{Ablation study results. (*) Question model results set the optimal performance for our approach assuming that the question interpretation is perfectly aligned with the ground-truth annotations. We then estimate additional (new) errors produced by each of the KGQA components separately. The experiments marked with {\bf GT} use the term URIs and question types extracted from the ground truth queries. {\bf GT span$^+$} uses spans from the ground-truth annotations and then corrects the distribution of the matched entities/properties to mimic correct question interpretation with a low-confidence tail with alternative matches. {\bf PR} (Parsed Results) stands for predictions made by question parsing and matching models (see Section \ref{sec:implementation5}).}
\label{tab:results2}
}
\end{table}
\end{landscape}

The experiments marked \textbf{GT span+} were performed by matching terms to the KG using the ground-truth span annotations, then down-scaling the confidence scores for all matches and setting the confidence score of the match used in the ground-truth query to the maximum confidence score of 1.
In this setup, all correct answers according to the benchmark were ranked at the top, which demonstrates the correctness of the message passing and score aggregation algorithm.

\subsection{Scalability analysis}

\begin{figure}[!t]
\centering
\includegraphics[width=0.7\textwidth]{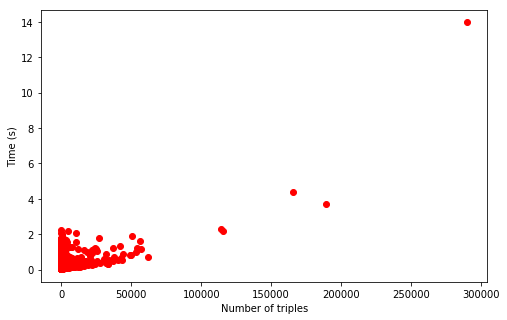}
\caption{Processing times per question from the LC-QuAD test split (Min: 0.01s Median: 0.68s Mean: 0.72s Max: 13.97s).}
\label{fig:scalability}
\end{figure}

As we reported in Table~\ref{tab:results1}, \OurApproach{} is twice as fast as  the WDAqua baseline using a comparable hardware configuration. 
Figure~\ref{fig:scalability} shows the distribution of processing times and the number of examined triples per question from the LC-QuAD test split. The results are in line with the expected fast retrieval of HDT~\cite{fernandez2013}, which scales linearly with the size of the graph.
Most of the questions are processed within 2 seconds (with a median and mean around 0.7s), even those  examining more than 50K triples.
Note that only 10 questions took more than 2 seconds to process and 3 of them took more than 3 seconds. These outliers end up examining a large number of alternative interpretations (up to 300K triples), which could be prevented by setting a tighter threshold. Finally, it is worth mentioning that some questions end up with no results (i.e., 0 triples accessed), but they can take up to 2 seconds for parsing and matching. 

\subsection{Error analysis}

We sampled questions with errors (P $< 1$ or R $< 1$) for each of the setups and performed an error analysis for a total of 206 questions.
The most noticeable result was that half of the errors were due to the incompleteness of the benchmark dataset and inconsistencies in the KG (column \emph{Revised} in Table~\ref{tab:results2}).
Since the benchmark provides only a single SPARQL query per question that contains a single URI for each entity, predicate and class, all alternative though correct matches are missing, e.g., the gold-truth query using {\tt\small dbp:writer} will miss {\tt\small dbo:writer}, or match all {\tt\small dbo:languages} triples but not {\tt\small dbo:language}, etc.


\begin{figure}[!t]
\centering
\includegraphics[width=0.6\textwidth]{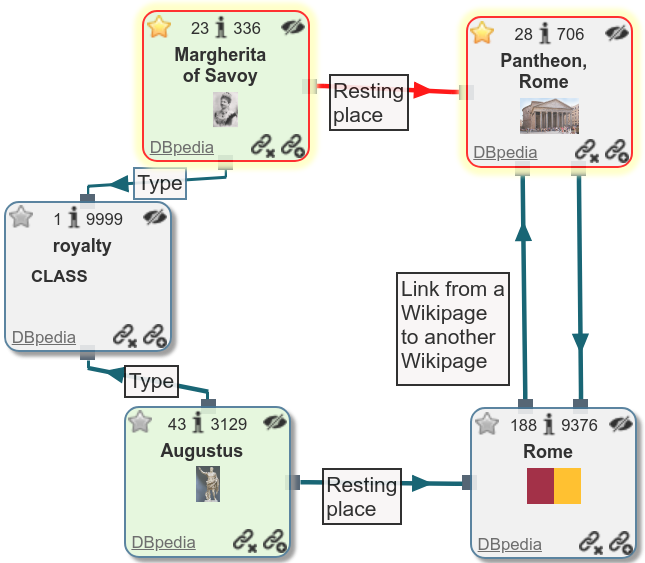}
\vspace{0.2cm}
\caption{\emph{Alternative entity} example that demonstrates a missing answer when only a single correct entity URI is considered ({\tt\small dbr:Rome} and not {\tt\small dbr:Pantheon,Rome}).
LC-QuAD question \#261: ``Give me a count of royalties buried in Rome?'' (correct answer: {\tt\small dbr:Augustus}; missing answer: {\tt\small dbr:Margherita\_of\_Savoy}). \OurApproach{} was able to retrieve this false negative sample due to the string matching function and retaining a list of alternative URIs per entity mention.}
\label{fig:rome}
\end{figure}

\OurApproach{} was able to recover many such cases to produce additional correct answers using: (1) missing or alternative class URIs, e.g., {\tt\small dbr:Fire\_Phone} was missing from the answers for technological products manufactured by Foxconn since it was annotated as a {\tt\small device}, and not as an {\tt\small information appliance}; (2) related or alternative entity URIs, e.g., the set of royalties buried in {\tt\small dbo:Rome} should also include those buried in {\tt\small dbr:PantheonRome} (see Figure~\ref{fig:rome}); (3) alternative properties, e.g., {\tt\small dbo:hometown} as {\tt\small dbo:birthPlace}. 



We discovered alternative answers due to the \textit{majority vote} effect, when many entities with low confidence help boost a single answer.
Majority voting can produce a best-effort guess based on the data in the KG even if the correct terms are missing from the KG or could not be recovered by the matching function, e.g., \emph{``In which time zone is Pong Pha?''} -- even if \emph{Pong Pha} is not in the KG many other locations with similar names are likely to be located in the same geographic area.

Overall, our evaluation results indicate that the answer set of the LC-QuAD benchmark can be used only as a seed to estimate recall but does not provide us with a reliable metric for precision.
Therefore, attempts to further improve performance on such a dataset can lead to learning the biases embedded in the construction of the dataset, e.g., the set of relations and their directions.
We showed that \OurApproach{} is able to mitigate this pitfall by resorting to 
unsupervised message passing that collects answers from all local subgraphs, which contain terms matching the input question, in parallel.

\section{Conclusion}
\label{sec:conclusion}

We have proposed \OurApproach{}, a novel approach for complex KGQA using message passing, which sets the new state-of-the-art results on the LC-QuAD benchmark for complex question answering.
In the experimental evaluation we showed that \OurApproach{} is scalable and can be successfully applied to very large KGs, such as DBpedia, which is one of the biggest cross-domain KGs.
\OurApproach{} does not require supervision in the answer inference phase, which helps to avoid the pitfalls of overfitting the training data and helps to discover correct answers missing from the benchmark due to the limitations of  its construction.
Moreover, the answer inference process can be explained by the extracted subgraph and the confidence score distribution.
\OurApproach{} requires only a handful of hyper-parameters to model confidence thresholds in order to stepwise filter partial results and trade off recall for precision.

\OurApproach{} is built on the basic assumption of considering edges as undirected in the graph, which proved reasonable and effective in our experiments.
The error analysis revealed that, in fact, symmetric edges were often missing in the KG, i.e., the decision on the order of entities in KG triples is made arbitrarily and is not duplicated in the reverse order.
However, there is also a (small) subset of relations, e.g., hierarchy relations, for which relation direction is essential.

\OurApproach{} is not without limitations. 
It is designed to handle questions where the answer is a subset of entities or an aggregate based on this subset, e.g., questions for which the expected answer is a subset of properties in the graph, are currently out of scope.


Question answering over KGs is a hard task due (1) ambiguities stemming from question interpretation, (2) inconsistencies in knowledge graphs, and (3) challenges in constructing a reliable benchmark, which motivate the development of robust methods able to cope with uncertainties and also provide assistance to end-users in interpreting the provenance and confidence of the answers.
\added{Also, complex question answering over KGs is a relatively new task and LC-QuAD is the first dataset dedicated to training and evaluating performance of this specific task. At this stage, to the best of our knowledge, WDAqua is the only end-to-end system that claims to support complex questions and was evaluated on the LC-QuAD dataset.}

\added{The lack of other baselines in this area is partially due to the fact that development and evaluation of a large-scale question answering system is a very challenging task that requires interaction between different components used for question answering: entity matching, relation matching and answer retrieval.
There are several on-going community initiatives designed to consolidate the efforts required to set up and maintain adequate benchmarks~\cite{DBLP:conf/esws/UsbeckNHKRN17,DBLP:conf/esws/SinghBRS18,DBLP:journals/semweb/UsbeckRHCHNDU19}. We are currently working on integrating QAmp (our approach) with the GERBIL QA benchmarking framework~\cite{DBLP:journals/semweb/UsbeckRHCHNDU19}, which will not only make evaluation against existing but also future KGQA approaches easier.
Such frameworks are also important for enabling interchange and evaluation of the individual components.}


\added{Evolution and versioning of knowledge graphs is an important and interesting topic which is being discussed in the semantic web community~\cite{DBLP:conf/esws/2017mepdaw,DBLP:journals/semweb/FernandezUPK19,DBLP:conf/www/X19o,DBLP:conf/esws/FernandezMPR18}.
Approaches to take versions and updates into account for query answering of structured (SPARQL) queries exist but require additional annotations on a temporal dimension for the validity of the triples in the graph.
Along the same lines we assume it is possible to extend the proposed approach to cope with versioned (temporally annotated) graphs by introducing an additional constraint that will either pre-filter relevant triples based on the question interpretation or generating answers within associated validity intervals.
In the latter case the complete answer to our sample question would be ``Ford until October 2016''.
Ideally, such approaches should also include provenance information, e.g., ``according to SBS: \url{https://www.sbs.com.au/news/more-than-100-ford-jobs-cut-in-victoria}''
}

An important next step is to use \OurApproach{} to improve the recall of the benchmark dataset by complementing the answer set with missing answers derived from relaxing the dataset assumptions.
Recognizing relation directionality is an important direction for future work, which requires extending existing benchmark datasets and the addition of more cases where an explicit order is required to retrieve correct answers.
Another direction is to improve predicate matching, which is the weakest component of the proposed approach as identified in our ablation study.
Finally, we believe that unsupervised message passing can be adopted for other tasks that require uncertain reasoning on KGs, such as knowledge base completion, text entailment, summarization, and dialogue response generation.
In particular, we are interested in using this approach to identify relevant concepts also in response to utterances that do not contain a question.
In Chapter~\ref{chap:structure} we showed that there are other interactions besides question answering that are characteristic of information-seeking dialogues.
Next, in Chapter~\ref{chap:browsing} we formulate the task of conversational browsing as an alternative type of interaction that does not require explicit query formulation upfront.

\chapter{Conversational Browsing}\blfootnote{The Cheese Shop sketch \url{https://youtu.be/Hz1JWzyvv8A}}
\label{chap:browsing}

\epigraph{
- Good morning, Sir... What can I do for you, Sir?..\\
-  I want to buy some cheese…\\
- Certainly, sir. What would you like?\\
-  Well, eh, how about a little Red Leicester.\\
-  I'm, afraid we're fresh out of Red Leicester, sir…\\
- Greek Feta?\\
- Uh, not as such.\\
-  Uuh, Gorgonzola?\\
-  No\\
-  Parmesan?\\
-  No...\\
--- The Cheese Shop sketch, Monty Python's Flying Circus, 1972}

In many real-world situations an information seeker is not in a position to adequately assess content of an underlying information source.
Consequently, formulating a query in this case boils down to a best-effort guess solely based on the background knowledge already available to the seeker.
When a collection is sparse and an information seeker is unaware of its content, a more efficient strategy is for an information provider (intermediary) to present the collection content.
This task becomes more challenging as the size of the collection grows.
Foremost, presentation of a large information space requires explicit structures in place that help to organize content of the collection for a more efficient navigation.
Furthermore, exposing the whole content of a vast information collection at once may be too overwhelming for a human cognition.
Therefore, structural properties and information content should be presented interactively, while considering cognitive load on the mechanisms enabling human information processing.

To this end, we explore the interaction patterns manifested in multi-turn information seeking dialogues and use these insights to design a dialogue system that does not require explicit query formulation upfront as in the question answering task.
We collected observations of human participants performing a similar task to obtain inspiration for the system design.
Then, we studied the structure of conversations that occurred in these settings and used the resulting insights to develop a grounded theory, design and evaluate a first system prototype.
Evaluation results show that our approach is effective and can complement query-based information retrieval approaches.
We contribute new insights about information-seeking behavior by analyzing and providing automated support for a type of information-seeking strategy that is effective when the clarity of the information need and familiarity with the collection content are low.

\section{Introduction}
\label{section:intro}


Conversational search has evolved as a new paradigm with the goal of making information retrieval interfaces feel more natural and convenient for their users~\citep{DBLP:conf/chiir/RadlinskiC17}.
Ongoing research and development efforts in this direction are now heavily skewed towards the question answering task~\citep{DBLP:conf/chi/VtyurinaSAC17,DBLP:conf/acl/RajpurkarJL18,DBLP:conf/emnlp/SaeidiBL0RSB018}.
However, there is ample evidence that conversational search interfaces need to support a more diverse set of interactions to be able to assist their users~\citep{DBLP:journals/corr/abs-1812-10720,bates1989design,belkin2016people}.
The limitation of the question answering interaction paradigm is in its inherent bias towards knowledge that the user already has: users need to be able to formulate an appropriate question before they can engage with a question answering interface in a meaningful way~\citep{belkin1982ask}.
A similar issue occurs also in situations when a system poses questions to its user~\citep{DBLP:conf/cikm/ZhangCA0C18}.

In this chapter we focus on the task of information presentation in conversational search interfaces designed to communicate all available knowledge to the user.
While similar to the information presentation task required for traditional spoken dialogue systems, which list available options in response to a user query~\citep{DBLP:conf/eacl/DembergM06}, what we have in mind goes beyond this paradigm.
\citet{DBLP:journals/corr/abs-1709-05298} have proposed to apply the concept of \emph{interactive storytelling}~\citep{DBLP:conf/icids/2018} to conversational search systems for exploratory search~\citep{DBLP:series/synthesis/2009White}.
Interactive storytelling is an extension of computational storytelling that makes the story generation process dynamicaly adopt in response to the user input~\citep{DBLP:journals/aim/RiedlB13}.
We are interested in applying interactive storytelling to explore the content of an information source thereby forming a conversation between a user and a system.
Conversational exploratory search can be useful in a range of knowledge access scenarios, including education and e-commerce.
For example, an intelligent shopping assistant should be able to fluently guide a customer through the whole product catalog, carefully picking up on the user's reactions to form a preference model and adopt the exploration direction that optimizes customer satisfaction.

We view \emph{conversational browsing} as a first step towards the bigger agenda of enabling conversational exploratory search via interactive storytelling~\citep{DBLP:journals/corr/abs-1709-05298}.
It focuses primarily on supporting navigation control, where a user can influence and change the direction of exploration.
Conversational browsing is a task designed to enable conversational exploratory search for structured information sources, such as a database table or a knowledge graph.
The goal of the conversational interface is to unfold the content of the collection to the user in an interactive manner, that is, in response to their chosen exploration direction.
Explicit structure of an information source allows us to model it as a graph abstraction and evaluate different navigation strategies, i.e., the sequences in which the nodes can be visited.

We start with the basic setup in which an ``interactive story'' is to be generated from a single database table.
Our main research question is how to enable efficient information access in a situation where the information goal of the user is implicit or vaguely defined.
Examples include cases in which the user is not familiar with the domain vocabulary, wants to understand the available content and structure of the information source, or is simply looking for inspiration and serendipitous discovery.
In this chapter we describe and evaluate the design of an automated dialogue system that helps users to acquire knowledge about the structure and content of a catalogue through dialog-based interaction without the requirement to specify their information need in advance.
The kind of system we have in mind has to be considerate of the user, in the sense that it should account for:
\begin{itemize}
\item \textit{cognitive load} to determine and regulate a reasonable pace of the information flow; and
\item \textit{user preferences} for the user to regulate the direction of the information flow, i.e., conversation topic.
\end{itemize}
To this end, we follow an end-to-end methodology from collecting and analyzing dialogue transcripts through model design, implementation and evaluation.

To be able to formulate an informed hypothesis of what kind of interaction the envisaged system should provide we seek to get inspiration from human-to-human conversations collected in a controlled laboratory study.
In this study one of the students seeks information (S -- Seeker) and the other one is trying to help using the Austrian Open Data portal (I -- Intermediary):
\begin{enumerate}[nosep]
\item[\textbf{(I)}] \url{data.gv.at} is an Austrian Open Data portal.
\item[\textbf{(S)}] What kind of data can you find there?
\item[\textbf{(I)}] You can search for datasets in economics and politics categories, but also education, sports, culture etc.
\item[\textbf{(S)}] What exactly do you mean?
\item[\textbf{(I)}] Statistics about birth rates, kindergartens locations, public transport, for example.
\item[\textbf{(S)}] What data do you have related to birth rates?
\item[\textbf{(I)}] I can tell you statistics about the places where newborn live or the names they get.
\item[\textbf{(S)}] That sounds all right. I'm curious about the names...
\end{enumerate}

We collected and analyzed this kind of dialogue transcripts (see Section~\ref{section:dataset}) to come up with a general framework, which we formalized into a conversational browsing model.
This model describes a general information-seeking process and is applicable across different use cases, in contrast to  supervised models trained directly on dialogue transcripts~\citep{DBLP:journals/taslp/RieserLK14}.
We design a prototype based on the main concepts of our conversational browsing model and evaluate the prototype in a user study by contrasting it against a traditional conversational search system that follows request-response paradigm.
Thus, our conversational browsing model provides a general framework that not only provides a theoretical understanding of an information-seeking dialogue but also forms the basis for system design.

In summary, our main contributions in this chapter are:
\begin{itemize}
\item a \textbf{dataset} of dialogue transcripts that provide insights on human strategies in information-seeking conversations, in which an information provider takes on a pro-active role;
\item a \textbf{model} that systematizes these insights as a set of requirements, components and functionality they should support to automate such information-seeking conversations; and
\item an \textbf{evaluation} of a proof-of-concept implementation of a conversational browsing system.
\end{itemize}

We find that conversational browsing can be a powerful tool able to mitigate the vocabulary mismatch problem and assist search.
Vocabulary mismatch (aka vocabulary or semantic gap) is a common phenomenon in information retrieval, when two parties, e.g. user in queries and system in collection documents, use different words to describe the same concepts~\cite{van2017remedies}.
Based on the conversational data that we collect, we discover that the essence of conversational browsing interaction lies in the recurrent process of vocabulary exchange that attempts to iteratively reduce the search space and bring an information seeker closer to their information goal removing the burden to formulate the search query.

We proceed by introducing the concepts of exploratory search and interactive storytelling in Section~\ref{section:background}.
Section~\ref{section:methodology} provides an overview of the methodology that we followed to design and evaluate our conversational browsing system.
For our empirical evaluation and data collection steps we instantiated the task of conversational browsing with a concrete use case described in Section~\ref{section:usecase}.
We proceed by describing the setup and the outcomes of a user study organized to collect a dataset of human-to-human conversations (Section~\ref{section:dataset}), which served as a blue-print for our dialogue system design (Section~\ref{section:design}). 
Finally, we describe the evaluation of our conversational browsing prototype in Section~\ref{section:evaluation6}.
We consider two evaluation setups: (1)~a user simulation, which helps us to tune the hyper-parameters of the model and estimate the system performance, and (2)~a user study to test our modeling assumptions with target users.
We conclude by discussing the relation of our findings to previous work in Section~\ref{section:related6}.

\section{Background}
\label{section:background}

One of the major challenges in information seeking is query formulation~\citep{kelly2009comparison}.
The concept of the anomalous state of knowledge (ASK) suggests that it is not reasonable to expect ``that it is possible for the user to specify precisely the information that she/he requires.''~\citep{belkin1982ask}
Recognition of this phenomena is an important step towards considering alternative solutions that may help information seekers in practice.

Most of the on-going work in conversational search is still focused on the question answering task, which is an important interaction type in the context of an information seeking conversation but not the only one a conversational system has to provide support for~\citep{DBLP:journals/corr/abs-1812-10720}.
We argue that conversational agents and search systems should also support exploratory search.

\subsection{Conversational exploratory search}

The goal in exploratory search is to provide guidance for seekers who are exploring unfamiliar information landscapes~\citep{Marchionini:2006:Exploratory,DBLP:series/synthesis/2009White}.
\citet{DBLP:series/synthesis/2009White} distinguish between two main activities within the exploratory search paradigm: \emph{exploratory browsing} and \emph{focused searching}.
Exploratory browsing is an initial step that provides necessary domain understanding required for focused searching activities.
It is related to \citet{DBLP:conf/chiir/RadlinskiC17}'s \emph{system revealment} property: ``The system reveals to the user its capabilities and corpus, building the user's expectations of what it can and cannot do.''
Browsing is one of the information-seeking activities that is defined as ``semi-directed or semi-structured searching"~\citep{ellis1989behavioural}, i.e., the information need is vague and the goals include general collection understanding and serendipitous discovery~\citep{mckay2017manoeuvres}.
The purpose of browsing interface design is to make aspects of the collection apparent to the user and provide the means to traverse (navigate) between different options, which requires making the choice of the interaction modes and definition of a closed set of eligible operations available to the system and its user in advance~\citep{DBLP:conf/semweb/NunesS15}.

\begin{figure*}
\includegraphics[clip,trim=0mm 15mm 0mm 0mm,width=\textwidth]{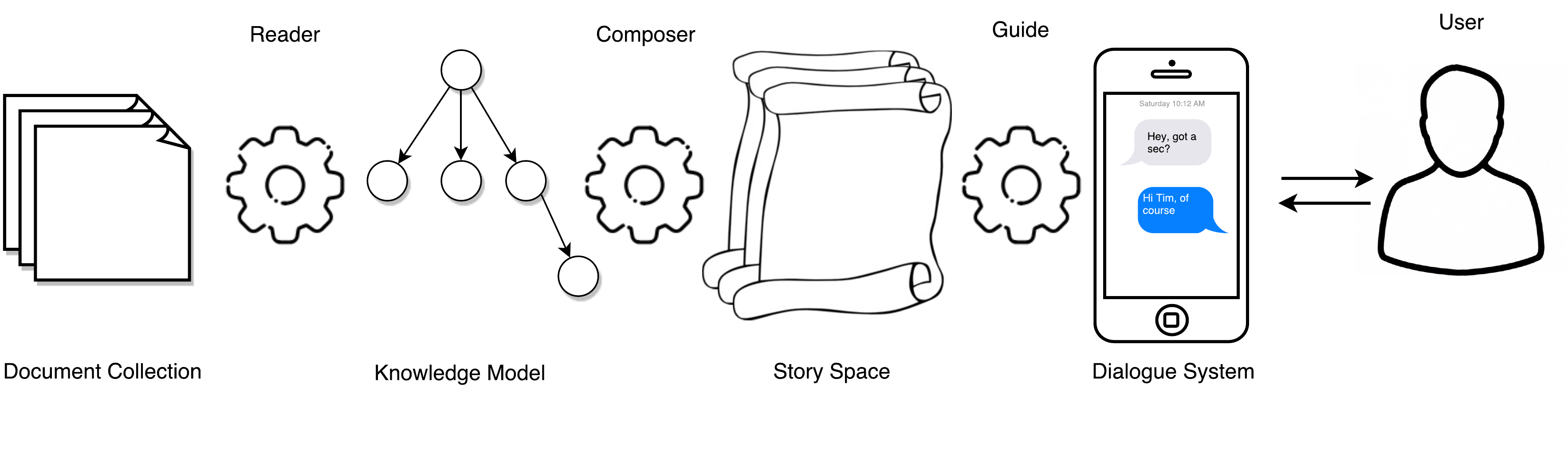}
\caption{Communicating knowledge via an interactive storytelling process.}
\label{fig:architecture}
\end{figure*}

Conversational agents and search systems are becoming increasingly popular~\citep{DBLP:conf/chi/VtyurinaSAC17}.
However, such systems mainly focus on question answering and simple search tasks, i.e., those that are to a large extent solved by web search engines.
Conversational agents and search systems should also support exploratory search~\citep{DBLP:journals/corr/abs-1709-05298}.
A conversational exploratory search system is represented in Figure~\ref{fig:architecture}.
It has a number of key components: Document Collection, Knowledge Model, Story Space, Dialog System and User.
These components are connected through the Reader, Composer, and Guide modules.
The interplay of the system components and modules happens at different stages.

\subsubsection{Knowledge representation} 
Knowledge representation consists of a \textit{Reader} module that extracts concepts and relations from the Document Collection and embeds them into a single Knowledge Model.
The Knowledge Model integrates different elements (words, concepts or entities) and describes relations between them.
The knowledge can be explicitly modeled by means of a taxonomy or ontology (knowledge graph) but it can also be embedded into a latent structure.

\subsubsection{Story generation} 
Story generation consists of the \textit{Composer} module that is able to generate stories by combining elements of the Knowledge Model.
To create a story, the Composer has to select elements (characters, words, facts, concepts, relations), choose their ordering, arrange selected elements in time and/or space.
The set of all possible stories constitutes the Story Space.

\subsubsection{Interactive storytelling} 
Interactive storytelling includes a \textit{Guide} module that helps the User to navigate through the Document Collection via the Story Space.
The Guide can change the current position within a single story or traverse the space across different stories.
Interactive storytelling integrates the Dialogue System to communicate a story to the User and to receive an input from the User. Supporting such a conversation with the User requires natural language (utterance) generation and understanding. Note that the input/output modalities do not have to be restricted to text and speech only and may include images, videos, interactive visualization, virtual reality interactions, etc.






\medskip\noindent%
A conversational exploratory search system should support the following types of the user-system interactions:

\begin{itemize}

\item \textit{Navigation Control} -- a user chooses a direction (branch) for exploration and is also able to influence and change the current direction of the narrative at any point in time;

\item \textit{Feedback} -- a user may provide feedback to the system (positive, neutral, negative) that may help to correct and steer the direction of the story that shall maximize the user satisfaction with the system;

\item \textit{Question} -- a user may pose questions to the system, e.g., a request for a definition, look up query, etc.

\end{itemize}

In this chapter we focus on mechanisms for navigation control in conversational browsing, which is a version of conversational exploratory search specifically designed for structured information sources, such as databases and knowledge graphs.
We ground our model design in observations of information-seeking dialogues, paying special attention to the ability of human information providers to prioritize and structure information into coherent chance to balance the cognitive load of an information seeker.

\subsection{Cognitive load}
\textit{Cognitive information processing} (CIP) theory~\citep{atkinson1968human} is a popular model that describes diverse cognitive processes.
The central idea in CIP is that the human mind can be modeled as an information processor that receives information from the senses (input), processes it, and then produces a response (output).
Learners are viewed as active seekers and processors of information.
CIP focuses mainly on the information processing task, in particular the aspects of memory encoding and retrieval.

\textit{Cognitive load theory} (CLT)~\citep{sweller1988cognitive} builds upon the information processing model.
``Cognitive load'' relates to the amount of information that working memory can hold and operate on at one unit of time.
The capacity of the human working memory is very limited.
When too much information is presented at once, it becomes overwhelmed and much of that information is subsequently lost.
CLT aims at making learning more efficient, calling for communication strategies that take into account cognitive limitations of the human mind.

\textit{Span theory}~\citep{bachelder1977theory} is a behavioral theory describing the relation between performance and span load, a fundamental task characteristic. In particular, several researchers studying the limits of human cognitive abilities point to the average number of objects a human brain can hold in working memory, i.e., the working memory capacity. 
The famous ``magic number'' originally suggested by \citet{miller1956magical} was $7\pm 2$ objects, while more recent research shows that this estimate is too optimistic and suggests a new limit close to 4 objects~\citep{cowan_2001}.

One other process that seems to be limited at about 4 elements is subitizing, the rapid enumeration of small numbers of objects. When a number of objects are flashed briefly, their number can be determined very quickly, at a glance, when the number does not exceed the subitizing limit, which is about 4 objects.
Larger numbers of objects must be counted, which is a slower process. 
 
However widely criticized as a single number not reflecting the task difficulty and individual differences, this number is supported by a remarkable degree of similarity in the capacity limit of working memory observed in a wide range of procedures and is likely to reflect a reasonable distribution mean able to inform chunking decisions for efficient information processing by humans. 
Research also shows that the size, rather than the number, of chunks that are stored in short-term memory is what allows for enhanced memory in individuals.
A chunk is defined as a collection of concepts that have strong associations to one another and much weaker associations to other chunks acting as a coherent, integrated group~\citep{cowan_2001}.
It is believed that individuals create higher order cognitive representations of the items on the list that are more easily remembered as a group than as individual items themselves.

Over the years various readability formulas have been proposed to predict comprehension difficulty of a text passage~\citep{crossley2017predicting}.
For example, one of the early studies~\citep{mc1969smog} reports a negative correlation between reading efficiency and the count of polysyllable words, defined as words of three or more syllables.

The combination of linguistic features capturing two elements of text difficulty (lexical and syntactic complexity) constitutes a good predictor for the time required to process text and comprehension~\citep{crossley2017predicting}, e.g.,
\begin{itemize}
\item texts are more comprehensive if the words are less sophisticated, there are fewer verbs, and lower text cohesion; and
\item a larger number of unique trigrams and proper nouns per sentence slows processing.
\end{itemize}
Another important set of features that describe the intrinsic properties of the reader's mental model, as an essential part and a prerequisite for the success of the text comprehension process, include individual reading ability, background knowledge, etc.

We draw upon results in cognitive science to inform our model and design an effective dialog system for conversational exploratory search.
More specifically, we explore the structure of messages exchanged in an information-seeking conversation -- the number of, and relations between, the concepts contained within a single message, as part of an important mechanism designed to support human information processing abilities with respect to the cognitive properties of the human mind.

\section{End-to-end Methodology for Designing a Dialogue System}
\label{section:methodology}

\begin{figure}[t]
\centering
\includegraphics[width=.6\textwidth]{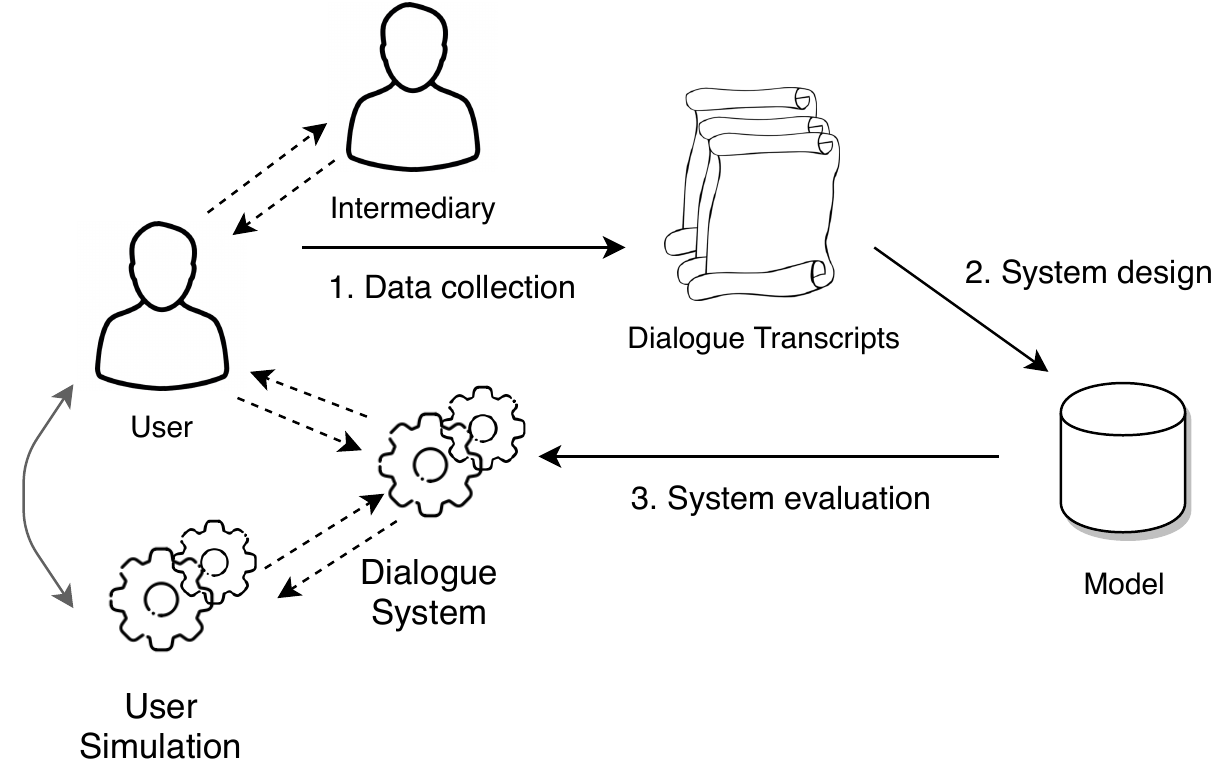}
\caption{End-to-end methodology for designing a dialogue system.}
\label{fig:methodology}
\end{figure}

We follow an end-to-end methodology for designing a dialogue system outlined in Figure~\ref{fig:methodology}.
It is a data-driven approach that helps us to formulate a general theoretical framework for conversational browsing and demonstrate its effectiveness on a sample scenario of open data exploration. 
We follow a mixed method approach to structure the design and evaluation of our conversational browsing system.

We start with a user study to collect samples of conversations and learn from the strategies human participants employ in the context of a conversational browsing task, which results in a dataset of dialogue transcripts (1.~Data collection). 
These empirical observations inform the design of a theoretical framework for conversational browsing that we propose in this chapter (2.~System design). 
The model is formally defined on a higher level of abstraction to separate the details inherent for the particular scenario from the general structure that can be reused across other use cases.
In the final stage (3.~System evaluation), we implement a prototype of a conversational browsing interface following the proposed framework and evaluate it via: (1) a user simulation; and (2) a laboratory study with real users~\citep{Koren:2008:PIF:1367497.1367562}.
We start the final stage by estimating an expected number of turns and tune the hyper-parameters in a user simulation.
Then, we conduct a second user study with real users to evaluate the assumptions behind our model's design.

\section{Use Case: Open Data \& Dataset Search}
\label{section:usecase}

This research was conducted in the context of the CommuniData project\footnote{\url{https://www.communidata.at}} that focused on making open data, i.e., publicly available datasets, more accessible for lay users to support citizen participation and transparency in decision making.
Making data public does not equate enabling citizens to make use of the data~\citep{DBLP:conf/cedem/BenoFUP17}.
Dataset search evolved as a new research direction to address the challenges of data heterogeneity~\citep{DBLP:journals/sigir/KoestenMGSR18,DBLP:conf/chi/KoestenKTS17}.

Google Dataset Search aggregates metadata of 14M datasets from 3k repositories.\footnote{\url{https://toolbox.google.com/datasetsearch}}
One of the first features users asked for was to extend the interface beyond a search box to support a browsing functionality~\citep{DBLP:conf/www/BrickleyBN19}.
A recent study of open data portal logs~\citep{10.1007/978-3-319-60131-1_29} also highlights that open data sources ``are used exploratively, rather than to answer focused questions.''

We started by designing a simple conversational search interface, which allows a user to submit a query and returns a list of matching datasets~\citep{DBLP:conf/esws/NeumaierSV17}.
The chatbot attracted a lot of attention but many users were not able to formulate adequate queries since they were not aware of the structure and terminology of the underlying repository.
In this case such conversational search interfaces turn out to be of little use.
This motivated us to look for alternative types of interaction that could allow users to understand which information a system can provide.




We formulated this task as \emph{conversational browsing}, which is reminiscent of web browsing, which allows users to explore a vast web graph by traversing links between pages.
However, use cases for the conversational browsing functionality are not limited to the dataset search/exploration scenarios.
This type of functionality is relevant to other applications in which a conversational interface relies on a large-scale structured information source, e.g., a database, table or index. 
Such use cases arise in product search or content recommendation scenarios, when a user is willing to learn about the underlying structure of the collection and explore the alternatives to get a comprehensive overview of the available options.

\section{Data Collection}
\label{section:dataset}

We conducted a Wizard of Oz experiment to better understand the types of dialogue interactions that occur in the context of conversational browsing.
Sample tasks illustrating conversational browsing scenario, as detailed in Section~\ref{sec:task}, were completed by 26 student participants grouped in pairs.
The volunteers were recruited among the undergaduate students taking a Data Processing course at the Vienna University of Economics and Business.
All participants had previous experience with web search interfaces, but no previous experience with the web sites they were instructed to access during the user study.
Every participant was assigned an individual working place in a lab equipped with a desktop PC. 
The pairs of students were seated apart from each other and had a separate web-based chat-room with a full-screen chat window, which served as the only channel for communication.\footnote{\url{https://tlk.io/}}
The conversational data\-set that we collected consists of 26 conversation transcripts with a total of 416 messages (minimum: 6, maximum: 36, mean: 16 messages per conversation).\footnote{The data is available at \url{https://github.com/svakulenk0/ODExploration_data}}

\subsection{Task description}
\label{sec:task}

We designed two different tasks using two open data portals with faceted search interfaces so that every participant would have a chance to play both roles.
For one of the two portals they assumed the role of the Intermediary (I), and for the other the role of the Seeker (S). 
Seekers were assigned specific information goals (to find one of the datasets from the portal) but were explicitly instructed not to share the goal with the Intermediary but try to reach it by providing relevant feedback to the Intermediary, that is, choosing from the relevant exploration directions suggested by the Intermediary.
Open data portals are particularly suited for such an experiment, since they provide a ready-made user search interface as well as a machine-readable API to access the same data repository, both publicly available.

The experimental procedure consisted of two phases: 
\begin{enumerate}
\item a learning phase, and 
\item a teaching (or knowledge sharing) phase.
\end{enumerate}
After each phase a quiz was completed to assess the knowledge acquired by the Intermediary and the Seeker.
During the learning phase the Intermediary studied the structure and content of the collection using the web site of the open data portal, which provides a faceted search interface. 
The Intermediary completed the quiz designed to evaluate the extent of the acquired knowledge about the structure and content of the collection by browsing the web site.
After completion of the learning phase, the Intermediary shared the acquired knowledge with the Seeker in a conversation. 
The Seeker completed a quiz to assess the extent to which they acquired knowledge about the structure and content of the collection through dialogue interactions with the Intermediary.

\subsection{Dataset description}
\label{sec:conversation_structure}

The conversational data\-set that we collected in the manner described in Section~\ref{section:dataset} consists of 26 conversation transcripts with a total of 416 messages (minimum: 6, maximum: 36, mean: 16 messages per conversation).
Most of the conversations (24 out of 26) were identified as successful based on the Seeker's explicit feedback and the correct dataset link provided by the Intermediary by the end of a conversation.
However, little additional knowledge beyond the specific information goal provided in the task was shared between the study participants (Seekers indicated that the topics were not discussed in the conversation).

One of the authors annotated each message in every transcript with the speaker identifier (a Seeker or an Intermediary role), clustered similar utterances and annotated them with labels reflecting the function they play in the conversation (dialogue act types), e.g., greeting or question.
In total, we identified 15 distinct utterance types in this dataset shared across different conversations (see the full list with descriptions in Table~\ref{dialog_acts}).
Messages were also annotated with span-level labels to keep the count for the number of concepts per message and their types.

To understand the structure of the conversations collected in the dataset and relations between different dialogue acts, we extracted a model of the conversation flow from the conversation transcripts by feeding the transcript as sequences of annotated utterances into the ProM framework\footnote{\url{http://www.promtools.org}}~\citep{Aalst16}, which is a popular process mining software toolkit. 
The directed graph of the conversation model was constructed and visualized using the Inductive Visual Miner ProM plugin~\citep{DBLP:conf/bpm/LeemansFA14}. 
We provide a snapshot of the core of the extracted conversation process model (see \figurename~\ref{fig:conversation_model}), which describes the information exchange loop used by the conversation partners to traverse the information model in the direction of the information goal.

Many conversations begin with a ``hand-shaking'' message exchange, which may include mutual greetings, introductions and goal statements that provide the context for the rest of the conversation. 
For example, for the Seeker it would include a general question about the content of the information source, while for the Intermediary it would be a short description of the information source, an offer of the information service and a request about the concrete information goal of the Seeker. 

The conversations that we observe are mostly a vocabulary exchange aimed at traversing the information space towards the subset of items containing the information goal. 
The Intermediary (\textbf{I}) lists keywords, which correspond to a set of related concepts in the information space, and the Seeker (\textbf{S}) chooses one or more of these concepts to continue exploration. 
To illustrate, here is a snippet of a sample conversation:

\begin{enumerate}[nosep]
\item[\textbf{(I)}] It is an Open Data source that contains data about various \textbf{topics}: \textit{work}, \textit{culture}, \textit{education}, \textit{population}, \ldots\ What would interest you the most?
\item[\textbf{(S)}] \textit{Population}.
\item[\textbf{(I)}] Okay, is there a specific \textbf{region} for which you would like to find a dataset (\textit{Steiermark}, \textit{Vorarlberg}, \textit{Vienna} etc.)?
\item[\textbf{(S)}] I'm interested in the \textit{population} of \textit{Upper-Austria}.
\end{enumerate}

\begin{landscape}
\begin{table}[tb]
\mbox{}\vspace*{-\baselineskip}
\centering
\ttabbox{%
\begin{adjustbox}{width=\linewidth}
\begin{tabular}{ l rr l l l }
\toprule
\bf Type       & \multicolumn{2}{c}{\bf Occurrences}  & \bf Description                                                        & \bf Example                                                          & \bf Role \\
\midrule

list(keywords)   & 102 & (24.5\%) & Suggest available options for an exploration              & We have data on culture, finance. & A    \\
&&& direction & \\
set(keywords)    & 85 & (20.4\%)  & Choose an exploration direction                                    & I am interested in culture.                     & U    \\
confirm()        & 51 & (12.3\%)  & Confirm an exploration direction                                   & That would be perfect!                                            & UA \\
success()        & 36 & (8.7\%)   & Indicate reaching an information goal                             & Thank you very much!                                             & UA \\
question(data)   & 31 & (7.5\%)   & Indicate a general information need                               & What data do you have?                                           & U    \\
prompt(keywords) & 26 & (6.3\%)   & Request to specify the information goal                            & Any state that interests you?                   & A    \\
reject()         & 19 & (4.6\%)   & Reject an exploration direction                                    & This is not what I am looking for.               & U    \\
greeting()       & 19 & (4.6\%)   & Common start of the conversation                                   & Hello!                                                            & UA \\
bool(data)       & 13 & (3.1\%)   & Report whether requested subset exists & Yes, we have data about this year.~~~\mbox{}                                & A    \\
count(data)      & 13 & (3.1\%)   & Report the size of a subset                                        & There are 314 datasets in CSV.                            & A    \\
link(dataset)    & 9 & (2.2\%)    & Report link to a dataset                                           & There you go: \url{http://data}                      & A    \\
verify()         & 4 & (1.0\%)    & Prompt to confirm an exploration direction                        & Is that what you are looking for?                                & UA \\
more()           & 3 & (0.7\%)    & Request to continue in the same exploration~~~\mbox{}               & Is there only one dataset?                     & U    \\
&&& direction & \\
top(keywords)    & 3 & (0.7\%)    & Report a subset of the most frequent                       & The most popular license is CC.            & A    \\
&&& concepts & \\
prompt(link)     & 2 & (0.5\%)    & Suggest or request the link to a dataset                           & Send me the link, please.                                & UA \\
\bottomrule
\end{tabular}
\end{adjustbox}
}{
\caption{Types of utterances exchanged between the Intermediary and the Seeker.}
\label{dialog_acts}
}
\end{table}
\end{landscape}

\begin{landscape}
\centering
\begin{figure}
\begin{adjustbox}{width=\linewidth}
\includegraphics[trim={0.01cm 0 0 0.01cm},clip,width=\linewidth]{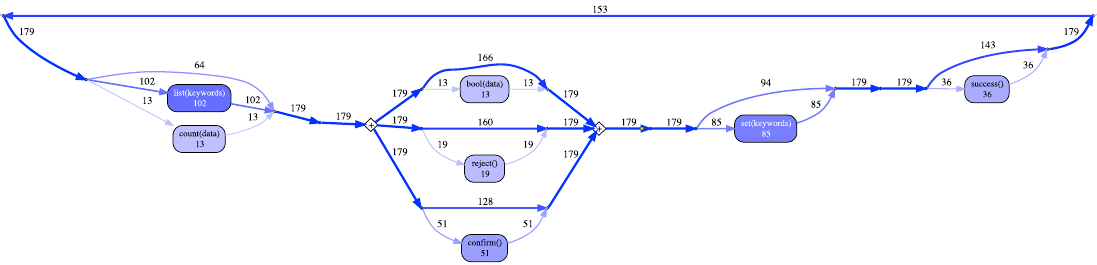}
\caption{Model of the conversational browsing interaction process extracted from the conversation transcripts. Nodes of the graph correspond to the utterance types described in Table~\ref{dialog_acts} and the arrows between them (transitions) show the direction of the conversation flow. The types may follow each other in a sequence, in parallel (joints marked with ``+'' symbol, meaning that the order varies between the conversations), alternate or form loops. The numbers above the arrows indicate the number of times each transition type occurs in the transcripts. Color intensity indicates relative frequency of the transitions and utterance types in the corpus. (Best viewed in color.)}
\label{fig:conversation_model}
\end{adjustbox}
\end{figure}
\end{landscape}

We collect the most frequent patterns of concept types used within the same message and analyze their relations. 
Communicating a subset of entities that belong to the same attribute (or \emph{facet} -- a general category of the attribute) with or without mentioning the name of this attribute as well is the most common pattern we observed. For convenience, we mark such attribute names with boldface: e.g., \textbf{topic}, or \textbf{publisher}. 
Another frequent pattern is listing several attributes or facets within the same message for the Seeker to choose from, e.g., ``I can group the datasets by \textbf{organization}, \textbf{format} or \textbf{topic}.''
A common strategy for the Intermediary is to make an attempt to reduce the subset of items for exploration by prompting the Seeker to select one of the shared attribute values (subset search). We mark values with italics in dialogs: e.g., \textit{education}, or \textit{The City of Vienna}.
When the subset of items is small and more homogeneous, i.e., many datasets have the same values of multiple attributes, the Intermediary starts listing values of unique dataset attributes (linear search), such as title, description, or link.

\medskip

In summary, the action space used by the Seeker has three operations:
\begin{itemize}
\item \textit{select} -- provides positive feedback towards one or more of the exploration directions (facets, or attributes), e.g., ``Yes, \textbf{population} sounds interesting'';

\item \textit{skip} -- provides negative feedback towards one or more of the exploration directions, e.g., ``I do not care about the \textbf{data format}'';

\item \textit{prune} -- provides negative feedback towards a subset of items, e.g., ``Is it about \textit{education}? -- No.''
\end{itemize}



The average number of turns per dialogue in our dataset is~5, with a minimum of~1 and a maximum of~14. 
The one turn dialogues consist of answers to direct questions expressing the information need, i.e., the user query.
It usually takes 2--3 turns when the Intermediary also describes the information source before or after answering the user query.
If the Seeker smoothly follows the options offered by the Intermediary, the minimum number of turns for the conversational browsing scenario is~4; it is at least~6 turns if the Seeker rejects some of the options offered by the Intermediary. 
Inefficient strategies leading to an increase in the number of turns required to satisfy the information need include asking general questions and providing an insufficient number of options.

The majority of messages composed by the human Intermediaries contain up to~8 concepts. 
The maximum number of concepts per message was~16; this was an extreme case in which the Intermediary listed all available categories. 
The number of concepts per message positively correlates with the performance of the interaction;  Seekers were more likely to find one of the options useful when supplied with more options. 

\section{System Design}
\label{section:design}

We define \textit{conversational browsing} as a collaborative exploration search task with asymmetric roles with an uneven distribution of goals and information available to the conversation participants.
One of the conversation participants (the \textit{Intermediary}) has access to the \textit{information model} $I$, while the other participant (the \textit{Seeker}) has access to the \textit{information goal} $G$.
While the goal of the Intermediary is to help the Seeker satisfy the information goal, only the Seeker is in a position to define the goal and/or tell when it is reached.
More formally, the task for the Intermediary is to communicate a subset of $I$ to the Seeker in a sequence of \textit{messages} $M = \langle M_1, \ldots, M_n \rangle$ to form the \textit{knowledge state} $K$ in alignment with $G$ so as to satisfy the success condition $G \subseteq K$.

\begin{figure}[!h]
\centering
\includegraphics[width=.8\textwidth]{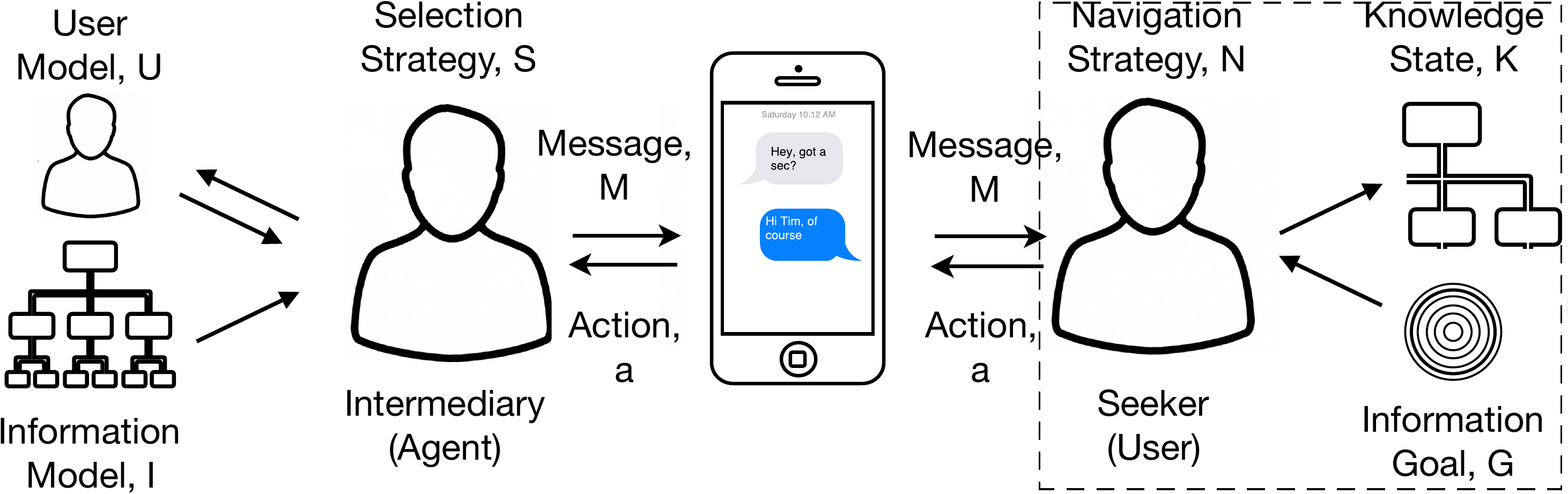}
\caption{Conversational browsing model (CBM). The user model $U$ maintained by the agent is expanded on the right-hand side of the figure consisting of the knowledge state $K$, information goal $G$ and a navigation strategy $N$.}
\label{fig:model}
\end{figure}

We design a \textit{dialogue agent} to play the role of an Intermediary in this task and the \textit{user} to take on the role of the Seeker. 
A model of the conversational browsing task is illustrated in \figurename~\ref{fig:model}.

\noindent%
The actual knowledge state $K$ and the information goal $G$ of the Seeker are not directly observable by the Intermediary. 
Instead, the Intermediary maintains a \textit{user model} $U$ that reflects the Intermediary's belief about the Seeker's state based on the Seeker's \textit{action} $a$. 
The Seeker chooses one of the actions from the \textit{set of available actions} $a \in A$ using the \textit{navigation strategy} defined as a function $N$, which is a generative process also hidden from the Intermediary.

The Intermediary is assumed to be able to adequately model both the user state $U$ and the information model $I$ in order to choose an optimal \textit{selection strategy} $S$ to compose messages $M = \langle M_1, \ldots, M_n \rangle$. 
The goal in this case is to satisfy the success condition $G \subseteq K$ using a minimal number of messages.

\subsection{User model}
\label{section:user-model}

\textit{Cognitive load} corresponds to the amount of information that working memory can hold and operate on at a unit of time $t$. 
We model cognitive load as a function $L(t)$ that defines the bound on the available cognitive resource of the user, which can represent time, memory, attention span, motivation, patience or user fatigue.

If too much information is presented at once, at time $t$ (that is, if $|M_t| \gg l_t$), the user becomes overwhelmed and much of the information is lost.
Therefore, a na\"{\i}ve brute force selection strategy $S$ that simply pushes the entire information model into a single message is likely to fail according to CLT.
We ground our assumptions about the shape of $L(t)$ in results from cognitive science. 
Various experiments suggest the working memory limit to be close to $7\pm2$~\citep{miller1956magical} objects, or even less~\citep{cowan_2001}. 
While these bounds have been debated, we assume them as a reasonable average $l$ to inform our user model. 

The concept of cognitive load motivates the design of a selection strategy $S$ that takes into account the cognitive resource limitation of the human brain, $L(t)$, to make human learning more efficient. 
In particular, it motivates the need for partitioning the message $M$ into a sequence of messages distributed in time $M = \langle M_1, \ldots, M_n \rangle$ with the upper-bound on every message size provided by the cognitive load function such that $|M_{ut}| \leq l_{ut}$.

We assume that the Seeker employs a rational navigation strategy $N$ and is more likely to choose an action $a_{ut}$ that is expected to maximize the knowledge gain with respect to the information goal: $M_{ut+1} \cap G_{ut}$. 
If none of the available actions has any expected value with respect to the information goal $G_{ut}$, the Seeker will choose the action that triggers the default exploration direction selected by the Intermediary based on the structure of the information model.

\subsection{Information model}
\label{section:information-model}

\begin{figure}[t]
\centering
\includegraphics[width=0.7\columnwidth]{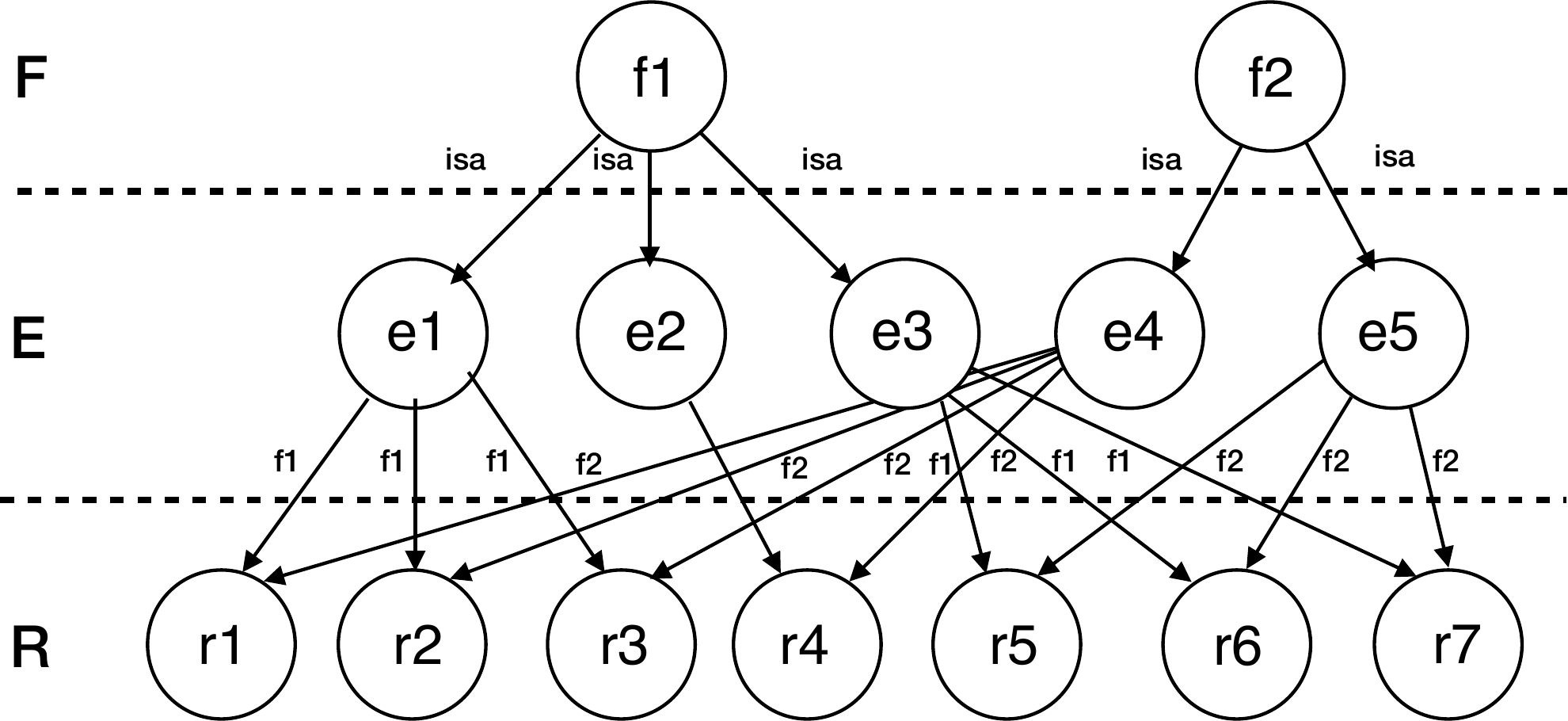}
\caption{Information model with subsets of nodes for attributes $F=\{f1, f2\}$, entities $E=\{e1, \ldots, e5\}$ and items $R=\{r1, \ldots, r7\}$.}
\label{fig:table}
\end{figure}

We assume a relational structure of the information source, in which a set of \textit{items} are characterized by a set of \textit{attributes}. 
This grid-like structure is a common data model used in tables and databases across different domains to characterize a group of homogeneous elements, e.g., movies or other products.
In Figure~\ref{fig:table} we represent this data model as a graph with three distinct sets of nodes: attributes $F$, entities $E$ and items $R$.
Individual items $R$ correspond to the rows of the table or records in a database; and their attributes $F$ correspond to the columns or facets. 
The intersection of a row $r \in R$ and a column $f \in F$ contains at least one of the \textit{entities} $e \in E$, which corresponds to the value of the attribute $f$ for the item $r$: $f(r) = e$. 
%
Entities can provide \emph{unique} identifiers for specific items, e.g., a name or a URL, or can be \emph{shared} between several items, e.g., location or time dimensions. 
Shared entities provide the structure useful for search and browsing of the collection.

We define a ranking function that calculates the score $v_c$, which allows us to compare the importance of every concept $c$ according to the structure of the information model $I$.

\subsection{Selection strategy}
\label{section:selection-strategy}

The selection strategy $S_{ut}$ takes into account the user model $U_{ut}$ and the information model $I$ described in the previous sections. 
It allows the dialogue agent to construct message $M_{ut}$ to be sent to user $u$ at time $t$. 
The task of the selection strategy $S_{ut}$ is to find the optimal order of the messages. 
The objective is to maximize the amount of information per unit of time while respecting the limit of cognitive resource of the user.

The entities are ranked by the number of items they belong to. 
Thus, the unique entities (titles) receive the lowest scores and the most frequent entity in the dataset receives the highest score.
The messages are composed from choosing a subset of top $l$ entities that belong to the same attribute.

\section{System Evaluation}
\label{section:evaluation6}

We advance to Step \emph{3.~System evaluation} in Figure~\ref{fig:methodology} and provide
details of the implementation, and two types of evaluation: a user simulation and a user study.
In a user simulation we evaluate the trade-off between the expected number of dialogue turns and the maximum message size and then evaluate the performance of our system in a user study.

All experiments were performed using the dataset downloaded from one of the open data portals, as an information source, that was also used in Step \emph{1.~Data collection}.
This dataset contains more than 2,000 items described by 74 different attributes.
Using the conversational transcripts collected in Step 1 we identified a set of 5 attributes that were used by human intermediaries to describe the items: title, license, organization, categorization and tags.

\subsection{User simulation}



\added{\paragraph{Baseline.} The baseline for comparison with the proposed conversational browsing interface was implemented to provide basic search functionality with a standard BM25 ranking function. The search system was exposed via a text-chat interface, where for each user query the generated response is shown below the original query in the same text-chat window. This setup is similar to the demonstration system proposed and evaluated earlier~\cite{DBLP:conf/esws/NeumaierSV17}. We chose to use a text-based interface to avoid errors from a speech recognition component, which still remains one of the major sources of errors in state-of-the-art spoken dialogue systems~\cite{swarup2019improving}.}

\paragraph{Setup.} We evaluate the robustness of our approach to conversational browsing by simulating several user models with different information goals. 
We simulate the Seeker in the following manner.
In every run a new information goal of the Seeker is initialized by picking one of the items (all of its entities) from the database uniformly at random. 
We assume the knowledge state of the Seeker is updated every time the Intermediary sends a message without any loss in the perception channel: $K_{t+1} = K_{t} \cup M_t$. 

The performance metric we used for evaluation was the number of turns the Intermediary needs to satisfy the latent information goal of the Seeker.
A user simulation was applied to tune the cognitive resource capacity $l$, which is the upper bound on the number of concepts the Intermediary can send to the Seeker within a single message $M$: $|M| < l$.
Integer values in the range 3..8 were considered, guided by the related work in cognitive load theory~\citep{miller1956magical,cowan_2001} as well as our own observations drawn from the analysis of the user study in Section~\ref{section:dataset}. 

\begin{table}[ht]
\centering
\ttabbox{%
\begin{tabular}{cccc}
\toprule
\multirow{2}{*}{Cognitive resource $l$} & \multicolumn{3}{c}{Number of turns per dialog} \\
\cmidrule{2-4}
 & Minimum & Average & Maximum \\
\midrule
3 & 5 & 18 & 67 \\
4 & 2 & 15 & 50 \\
5 & 5 & 12 & 80 \\
\rowcolor{Gray}
6 & 2 & 11 & 38 \\
7 & 2 & 9 & 29 \\
8 & 2 & 9 & 27 \\
\bottomrule
\end{tabular}
}{
\caption{Simulation results for different values of the cognitive resource $l$ (500 independent runs).}
\label{simulation}
}
\end{table}

\paragraph{Results.} The results were aggregated across 500 independent simulation runs used for the metric to converge and are listed in Table~\ref{simulation}.
The minimum number of turns to satisfy the simulated information need is~2, when the first message contains an entity that is able to uniquely identify the item $G$ and the second message contains all the entities that belong to the item $G$. 
The average number of turns required to reach $G$ monotonically decreases with the increase of the parameter $l$.
All the simulations were run in parallel.

The simulation results show that a greedy heuristic maximizing the out-degree in our information model performs reasonably well in the selection strategy for conversational browsing but is sensitive to the value of parameter $l$. 
\noindent%
Based on these estimates we chose the value of hyperparameter~$l = 6$ as an estimate of the cognitive resource limit in our user model that determines the maximum number of concepts per message. 
In this case, a simulated user will require~11 actions, on average, to reach any item in the dataset using our conversational browsing system.

\subsection{User study}

\begin{figure}[t]
\includegraphics[clip,trim=0mm 0mm 0mm 0mm, width=\columnwidth]{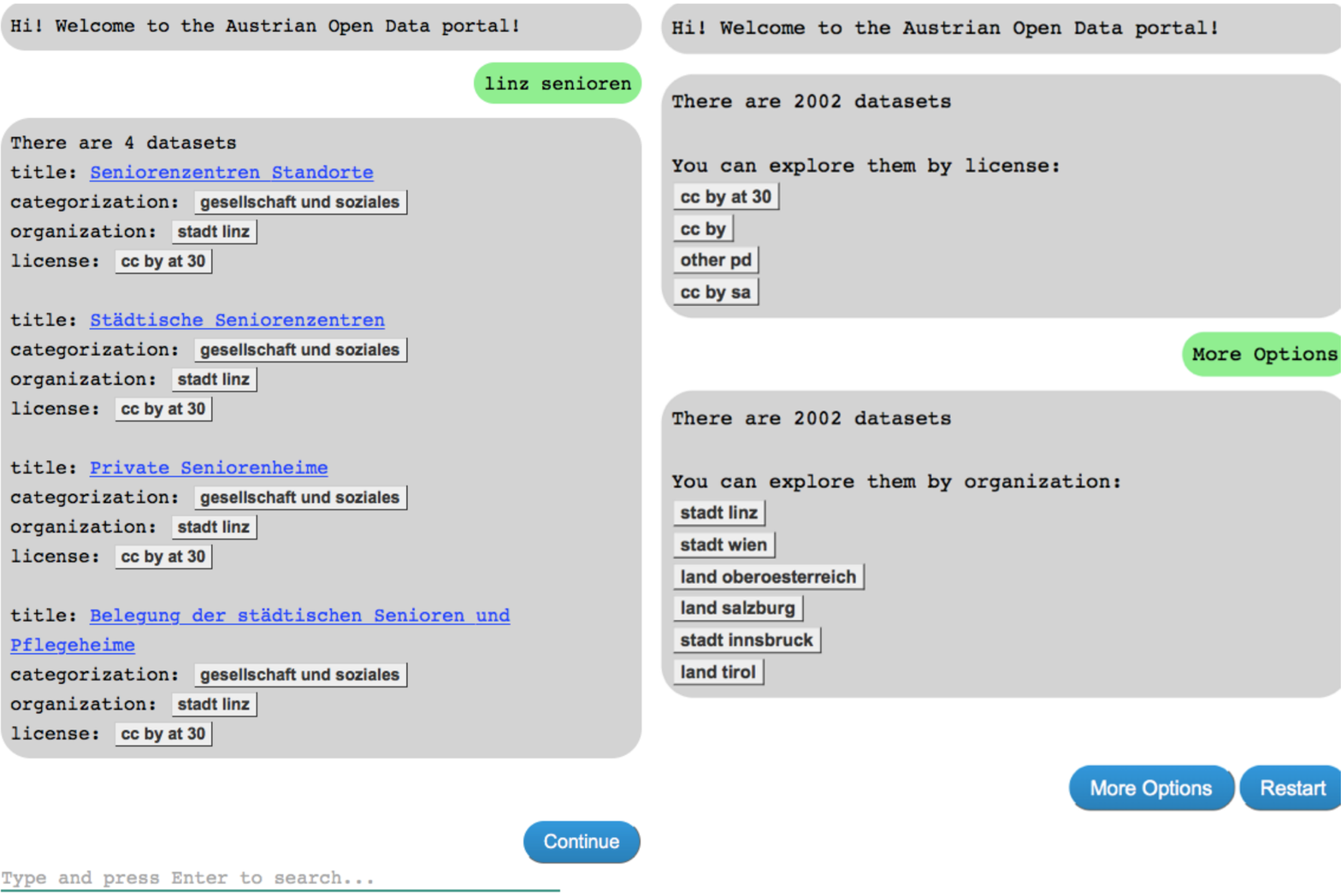}
\caption{Two types of interaction with the system: conversational search (left) versus conversational browsing (right).}
\label{fig:screenshots}
\end{figure}

For the user study we designed two conversational interfaces (see \figurename~\ref{fig:screenshots}).
The first one (left) provides a typical conversational search functionality: the user query is used to produce a ranked list of the matching items retrieved from the index.
The alternative interface (right) implements conversational browsing functionality by interactively revealing the subsets of the most discriminative attributes based on the user feedback.

The goals of browsing, as an information-seeking strategy, can vary from general collection understanding and learning to exploratory search~\citep{bates1989design}.
It is very challenging to evaluate the success of learning and the level of understanding.
We focus on the latter goal of exploratory search instead, defined as the ability to discover relevant information via browsing.
Moreover, in this setup we can directly compare the results achieved using the proposed conversational browsing interface on the same search tasks that can be completed using the query-based conversational search interface, which serves us as a baseline.
\paragraph{Setup.} A total of 24 participants took part in the experimental evaluation of our conversational browsing approach.
The volunteers were recruited among the university students and none of them participated in our data collection study.
All participants had previous experience with basic web search interfaces, such as keyword- and faceted search, but no previous experience with the repositories, web sites or our conversational interfaces used in this user study.


Each participant filled out a questionnaire that included a competency question for assessing their prior domain knowledge, accommodated search results for two search scenarios, and asked for participant's feedback at the end of the experiment. 
In this way we collected two types of feedback: 
\begin{inparaenum}[(1)]
\item subjective feedback by the participants reflecting on their experience using the systems; and \item objective feedback reflecting the average performance on the search task using different systems\end{inparaenum}.
%

%
\begin{table}[ht]
\centering
\ttabbox{%
\begin{tabular}{ l c c c c }
\toprule
 & \multicolumn{2}{c}{Task}                             &       \\ 
\cmidrule{2-3}
System       & (1) Immigration Vienna~~~ & (2) Retirement Linz &~~~& Total \\ 
\midrule
Search & 0.33                  & 0.08               && 0.21  \\ 
Browse~~~\mbox{} & 1.00                  & 0.33               && 0.67  \\
Total  & 0.67                  & 0.21               &       \\ 
\bottomrule
\end{tabular}
}{
\caption{User study success rates: proportion of the users who successfully completed the search and browsing tasks.}
\label{user_study}
}
\end{table}

To design the sample information seeking scenarios we picked two items from the dataset at random and formulated the tasks for the user based on these items. 
We carefully phrased the search task so as to reflect the vocabulary mismatch problem, which often occurs in real-world settings, by rephrasing some of the keywords in the title and other attributes of the target items:

\begin{enumerate}
\item \emph{Population by country of birth since 2011 municipal districts Vienna:\footnote{\url{https://www.data.gv.at/katalog/dataset/0a0f2617-3609-42ca-97bc-2f8a8be98cbf}} locate datasets that can provide information about immigration in Vienna}; and
\item \emph{Private retirement homes:\footnote{\url{https://www.data.gv.at/katalog/dataset/8421a66f-dc80-4bd3-8253-de532bc5b67c}} locate datasets that can provide important information especially for the older generation of adults living in and around Linz}.
\end{enumerate}


We evaluate performance on the tasks using a success rate that corresponds to the number of participants who manage to successfully complete the task by finding at least one of the correct datasets and analyzing the number of turns it took users to complete the task to compare it with the expected performance from our simulation. 
It took our user simulation between~5 and 9~turns to reach the item for every item in the pool of correct results, with~6 and 8~turns on average, respectively, for the different tasks.

\begin{figure}[t]
\includegraphics[width=\columnwidth]{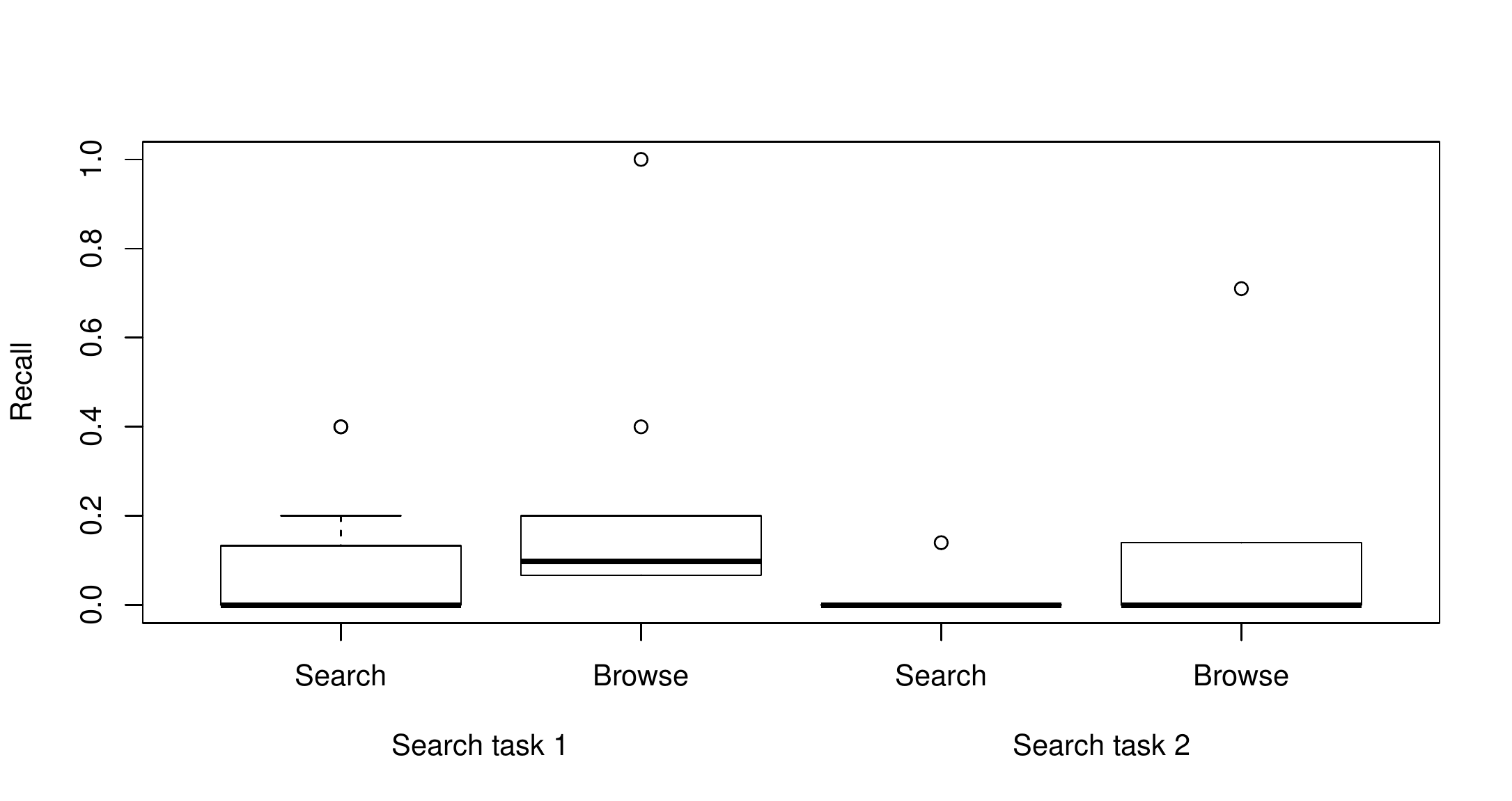}
\caption{Distribution of recall results from the user study.}
\label{fig:ncorrect}
\end{figure}

\paragraph{Results.} On average, participants performed better using our conversational browsing system in comparison with the basic search functionality. 
Only 5~out of 24~participants managed to complete the search task using the baseline system, in comparison with a success rate of more than 50\% for the conversational browsing system (16~out of 24). 
Table~\ref{user_study} displays the task success rate, i.e. the ratio of users who successfully completed the task.
The recall for both tasks was also higher for the conversational browsing system than for the search interface (see \figurename~\ref{fig:ncorrect}).
\added{We compared the means of the recall values using the Tukey test~\cite{thompson2006foundations}. The difference in recall between Search and Browsing interfaces is 1.08, with Browsing interface resulting in higher average recall. The 95\% confidence interval of this difference is between -0.24 and 2.41 points.}

\begin{figure*}[!t]
\centering
\includegraphics[width=\textwidth]{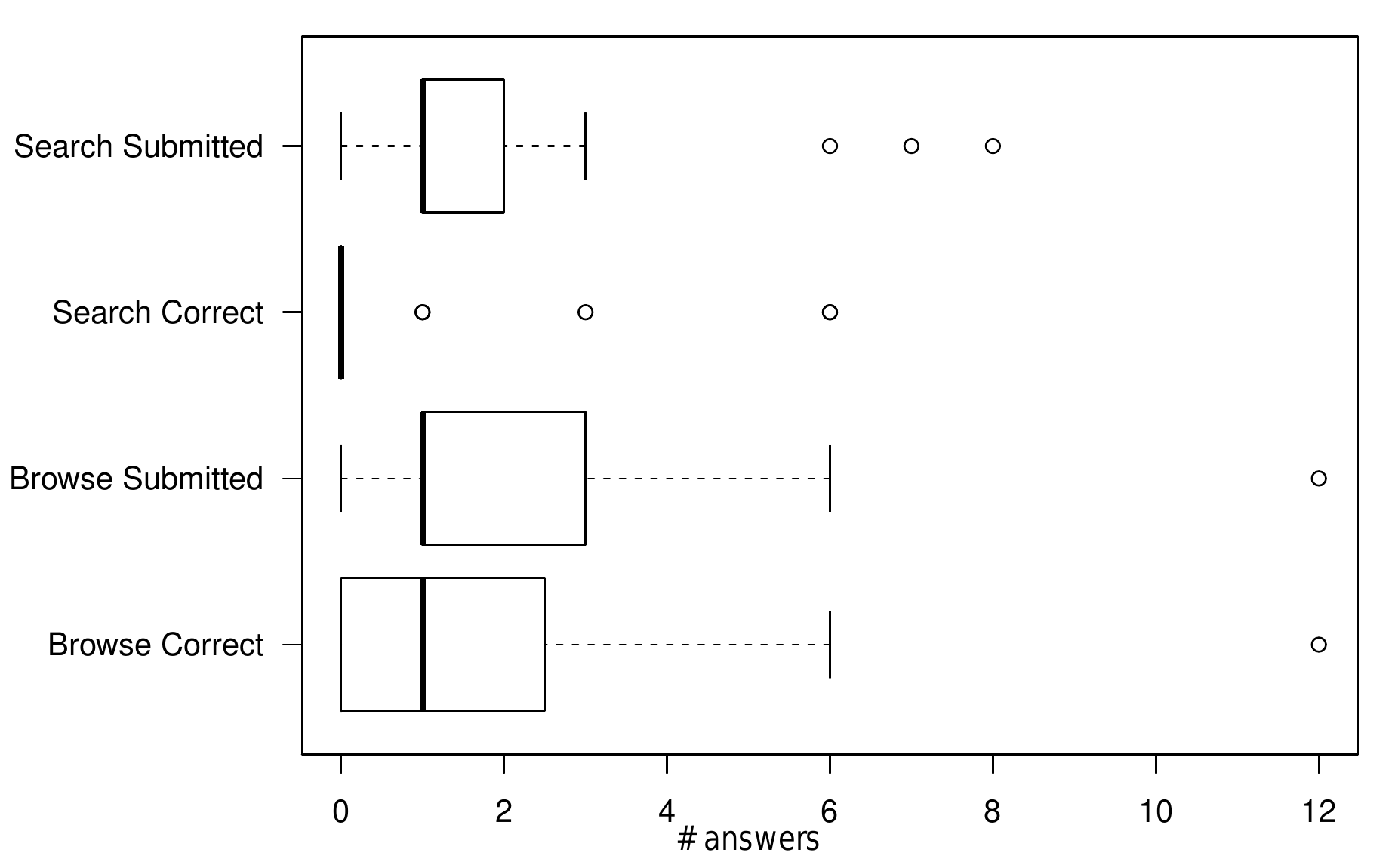}
 \caption{
Distribution of the submitted and correct answers for search and browsing interfaces.
 }
\label{fig:incorrect}
\end{figure*}

\added{Moreover, by analysing incorrect answers submitted by the participants we found that the users of the conversational search prototype tend to submit incorrect answers much more often than the users of the conversational browsing prototype (Figure~\ref{fig:incorrect}). This is a statistically significant result with p-value = $9 \times 10^{-6}$. This observation suggests that the conversational browsing interface was able to provide users with a better understanding of the collection content, sufficient to make a better distinction between relevant and non-relevant results. When using the proposed browsing system, participants submitted more answers and more of these answers were correct, in comparison with the answers submitted when using the baseline search system.}

\added{In addition, we made a qualitative analysis of the logs to extract the most successful strategies and contrasted them with the approaches other participants employed when using the provided interfaces to complete each of the tasks. Participants who managed to find correct answers using the search interface had to reformulate their search queries at least 5 times before hitting the correct result space.
That is why many participants submitted less relevant results while they could find them much faster giving up when fine-grained search queries did not return any results. In the browsing scenario, however, many participants could immediately find the logical path leading to the relevant search space where the correct answers are located. They would still frequently explore other facets to get a feeling of their semantics and the full search space. The most successful browsing experiment explored the variety of facets at first and then picked the answers from several different paths with a restart.
Query formulation is a difficult task and automated approaches, such as query expansion, are prone to failures since they rely on the availability of external knowledge, such as thesauri or query logs~\cite{van2017remedies,DBLP:conf/www/CuiWNM02,DBLP:journals/ipm/AzadD19}.
We see browsing as a viable alternative to explicit query formulation and demonstrate how it can be implemented in a conversational setting.}

The second task (``Retirement Linz'') turned out to be much harder to complete than the first one (``Immigration Vienna'') as evident from the difference in success rates. 
All participants succeeded when using conversational browsing for the first task, and a third of them -- for the second task, in comparison with the third of the participants for the first task and a single person only for the second task, when using the baseline search system. 

We created a pool of results submitted by participants for each of the tasks to enrich our subset of results marked as correct.
Two independent annotators marked correct answers in the pool with an inter-annotator agreement of 0.95 and resolved disagreements by discussing the content of the datasets.
Statistics for each of the tasks completed via the browsing interface are summarized in Table~\ref{task_stats} with the number of all unique results submitted by the users (row 1), the number of correct results among those marked as relevant with respect to the task by the annotators (row 2), the average number of turns produced by the user simulation (row 3), the average number of turns in conversations with human evaluators (row 3) and the number of restarts initiated by human evaluators (row 4).

For the more difficult task the fraction of incorrect results submitted is higher.
Also the users took more dialogue turns and restarts to complete the more difficult task (16 versus only 2 for the simpler task).
The number of turns predicted in the simulation is also higher for the more difficult task, but the gap is much bigger for the real users, which is likely due to the restarts the users take when they are not sure that they are navigating in the right direction.

\newcommand{\SingleD}[1]{\phantom{0}#1}

\begin{table}[t]
\centering
\ttabbox{%
\begin{tabular}{ l@{}c  c }
\toprule
\multirow{2}{*}{Statistics}                    & \multicolumn{2}{c}{Task}                              \\ 
\cmidrule{2-3}
                              & (1) Immigration Vienna~~~ & (2) Retirement Linz \\ 
\midrule                              
All results                   & 19                    & 28                 \\ 
Correct results               & 15                    & \SingleD{7}                  \\ 
\#Turns simulated & \SingleD{6}                     & \SingleD{8}                  \\ 
\#Turns user study~~~\mbox{} & \SingleD{8}                     & 21                 \\ 
\#Restarts                    & \SingleD{2}                     & 16                 \\ 
\bottomrule
\end{tabular}
}{
\caption{Conversational browsing results for two evaluation tasks.}
\label{task_stats}
}
\end{table}

\noindent%


\begin{figure}[!t]
\includegraphics[width=\columnwidth]{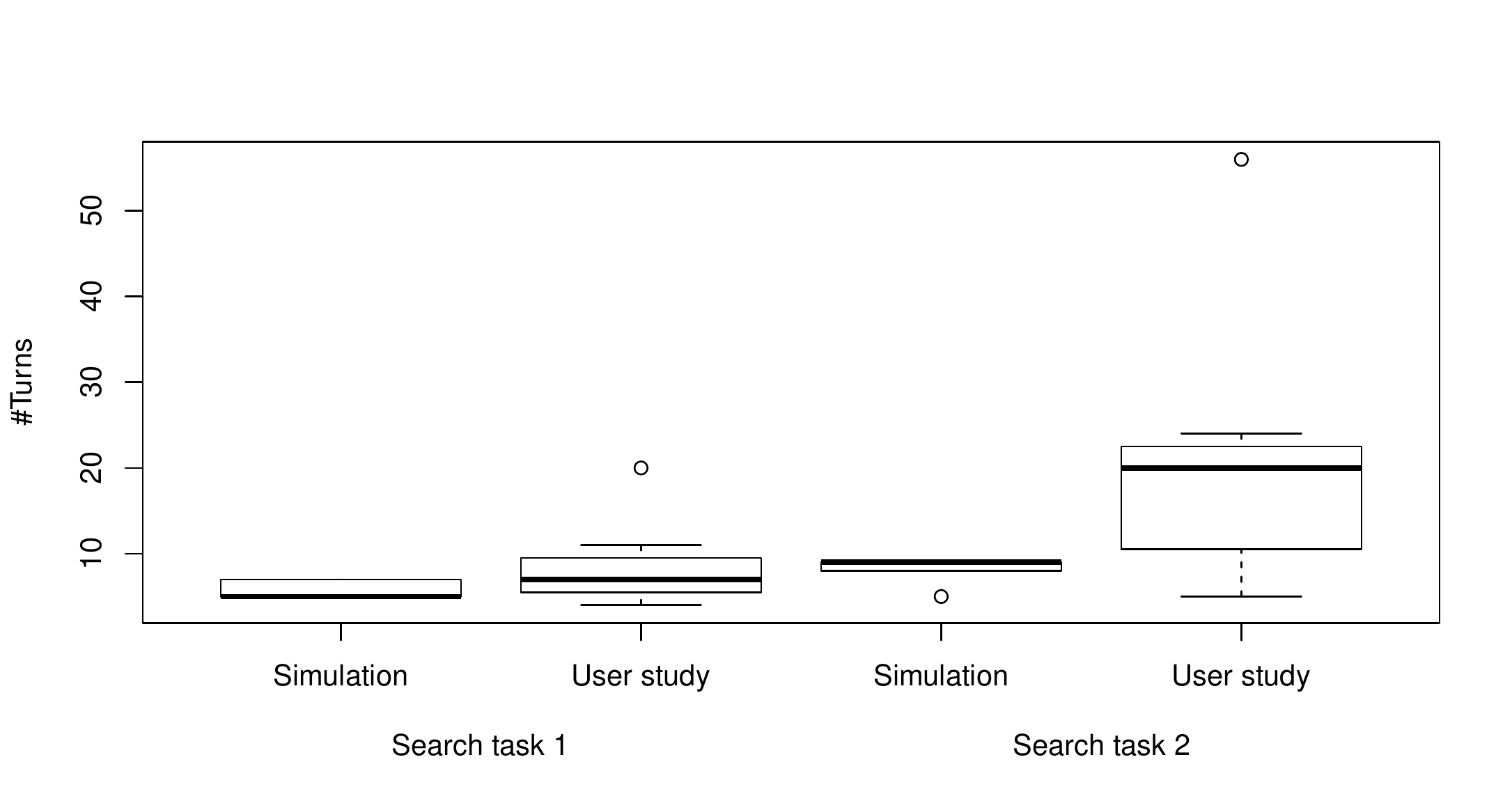}
\caption{Number of turns taken to complete the conversational browsing task (simulation and human participants).}
\label{fig:nturns}
\end{figure}

\figurename~\ref{fig:nturns} shows that the items ranked high with our ranking function, i.e., the items with the most frequent attributes, are much easier to find than the items with less frequent attributes.
The average number of turns for the first search task was~8 (vs.\ 6~turns in the user simulation) and 21~for the second task (vs.\ only 8~in simulation).

\subsection{Discussion}

The user study results showed that the conversational browsing functionality helps users to mitigate the vocabulary mismatch problem and find relevant information, even in the case of limited domain knowledge.
We observed a striking difference in search performance: the success rates of the browsing interface are three times higher than the query-based search interface on the same search tasks (see Table~\ref{user_study} and \figurename~\ref{fig:ncorrect}).
Among the positive feedback for the conversational browsing system were the ease of use and clarity, and the possibility to explore the data when the search criteria are not clear.

The majority vote, however, showed the opposite result in favor of the baseline: 70\% of the study participants preferred the conversational search interface rather than conversational browsing, when they were explicitly asked for their subjective feedback with the question ``Which system did you like more?''.
We attribute this result to several factors:
(1) all participants had previous \textbf{experience} with query-based search engines, which are similar to the functionality and the interaction type that our conversational search interface provides;
(2) the participants received a description of the information \textbf{goal}, which they could use to formulate the search query faster than browse the entire collection; 
(3) finally, the participants were not able to adequately assess their search performance since the correct results were not provided, which in turn led to a \textbf{misconception} about the system performance that likely influenced the preference choice.

More experiments are needed to evaluate usability and integration of browsing components into the conversational search interfaces.
Also, the selection strategy should be able to integrate alternative ranking functions beyond the information-theoretic objective only.
The results of the user study showed that some of the attributes identified as highly discriminative for the given dataset, such as data license, could not help users to decide on relevance.
User preferences, such as perceived attribute relevance, can be either collected in a separate survey or a user feedback form, or harvested from the logs of the conversational search system directly.


Our experimental results support previous findings and claims that dialogue systems can be an effective instrument for information retrieval also without the need to explicitly formulate the query, which can be especially relevant in situations promoting serendipitous discovery and general collection understanding~\citep{oddy1977information}.
We complement previous work in this direction by providing an extensive description of the approach we used and various aspects of its evaluation extended with an analysis of the challenges that arise in the design and evaluation of this kind of systems.

\section{Related Work}
\label{section:related6}

% ...


\added{The problem of query formulation was previously analysed and explained in terms of the Anomalous State of Knowledge (ASK)~\cite{belkin1982ask}.
It is difficult for an information seeker to precisely describe something that was not encountered before. Often an information seeker is in a situation when he is ``not able to formulate a precise query, and yet will recognize what he has been looking for when he sees it''~\cite{oddy1977information}. We built upon the idea of surfacing the underlying search space and bring it into the conversational settings by proposing a set of techniques that can allow users to explore the content of the document collection without an access to a reach graphical user interface or navigation control mechanisms.
Finally, we developed and successfully evaluated a dialogue-based interface, where all interactions between a user and a system are confined to a single text-chat window.}

\added{\citet{DBLP:conf/dgo/ZhangM05} proposed a new interface, which they called Relation Browser++ (RB++), that combined search and browsing functionalities. They also reported lower error rates in comparison with the traditional search interface used as a baseline. RB++ has a rich graphical interface with several panels displaying the facets available for filtering and the result set. In our experiments we showed that a browsing interface can be also implemented in a much more constrained space of a single text-chat window.
Meaning that we attempt to reproduce their results of improving search interfaces by introducing browsing functionality but we restrict all interactions to a single chat window rather than two separate windows for filters and results.
This modification is important because we aim to gradually move away from desktop applications to mobile and speech interfaces which have only a single communication channel.
These constraints however also require more interaction turns between the system and the user because the channel width is narrower, i.e., a mobile phone screen can not fit as much information as the desktop screen, i.e., ability of the system and the user to successfully interact while navigating the vast information space becomes even more crucial.
This limitation we imposed on the width of the communication channel is due to the cognitive limit theory, which motivated our approach in the first place.}

\added{Our initial experiments suggest that the proposed conversational browsing interface is effective at surfacing shallow queries, i.e., search results that require a few traversal steps, but also fails to provide a reliable guidance to the user when looking for less obvious directions. By decoupling browsing and search functionalities we can complement the results of the previous work~\cite{DBLP:conf/dgo/ZhangM05} and confirm that the lower error rate is indeed due to providing browsing functionality rather than the combination of search with browsing. In future work we would like to further explore how search and browsing functionalities can be better combined within a conversational setting.}

Conversational browsing is also similar to the information presentation subtask of a dialogue system designed to optimise the display of available options to a user~\citep{DBLP:conf/eacl/DembergM06,DBLP:journals/taslp/RieserLK14}.
However, conversational browsing does not assume an initial user query, i.e., available options always equate to the whole information space.
With the amount of information that can be potentially communicated to a user getting larger, a major design challenge arises with respect to taking in account cognitive limitations of the human brain for partitioning the information space into messages and using structural properties of the information space to allow a more efficient traversal.

Conversational browsing is conceptually different from a task-oriented dialogue, where an agent tries to pin-point an item or an information subspace relevant to the user's query~\citep{DBLP:conf/chiir/RadlinskiC17}.
In this respect, conversational browsing is hard to optimize, since there is no single correct answer.



\section{Conclusion}
\label{section:conclusion6}

We introduced a novel conversational dataset illustrating an asymmetric collaborative information-seeking scenario, in which an Intermediary, having access to an information source, plays a pro-active role by interactively revealing and dynamically adjusting the possible exploration directions based on the feedback from an information Seeker.
This scenario, which we cast as the \emph{conversational browsing task}, is appropriate when the Seeker is not sufficiently familiar with the domain of interest to  formulate their information need as a concise search query, or prefers to explore available options.

We proposed a formalization of conversational browsing as an interactive process in which the Intermediary guides the Seeker in discovering the relevant attributes (facets) and filtering conditions (entities) to single out a subset within the information source model that contains the information goal of the Seeker. 
Our experiments indicate that conversational browsing is a viable paradigm able to mitigate challenges in query formulation and assist users in conversational search.


The dataset and the model that we proposed indicate much broader implications for conversational system design than we could utilize in the first set of experiments.
Despite these limitations, we believe that our initial results showcase conversational browsing as a useful component for conversational search, which is able to complement the already established question answering task and encourage development for the set of more advanced interaction patterns with a dialogue system.

While similar ideas were already discussed decades ago~\citep{oddy1977information}, they were abandoned at the time, not matched by adequate technology for natural language understanding~\citep{belkin2016people}.
We believe that it is time to revisit these ideas.
The combination of novel techniques for semantic parsing and information retrieval with more advanced information-seeking models constitutes a promising direction for future work.

Thus, we showed how conversational browsing can support request-based interactions intiated by the information provider, which fills the research gap identified in the QRFA model described in Chapter~\ref{chap:structure}.
Our ultimate goal is to design a single conversational interface that integrates both browsing and question answering functionalities that naturally complement each other fusing into a single interaction model supporting various information-seeking strategies.
An important direction for future work is to extend the information model to handle arbitrary graph structures beyond a single table, such as a web graph or a knowledge graph as in Chapters~\ref{chap:coherence} and \ref{chap:qa}.
Further research is needed to integrate semantic coherence criteria based on the graph structure as presented in Chapter~\ref{chap:coherence}.
There is also room for a learning component able to learn from interactions with users to improve the overall performance and maintain personalized user models. 

\chapter{Conclusions}
\label{chap:conclusion}

\epigraph{ I believe the dreamers come first, and the builders come second. A lot of the dreamers are science fiction authors, they’re artists... They invent these ideas, and they get catalogued as impossible. And we find out later, well, maybe it’s not impossible. Things that seem impossible if we work them the right way for long enough, sometimes for multiple generations, they become possible. \\ --- Jeff Bezos, Heinlein Prize award ceremony, 2016}

\added{The first two chapters of the thesis aim at understanding the structure of a conversation from a general perspective, while the last two chapters utilise the insights about both structural and contextual dependencies (relations) between utterances in a conversation, and propose concrete approaches to designing conversational search interfaces. As for the first two chapters, the structure of an information-seeking dialogue was examined from two orthogonal perspectives. The QRFA model introduced in Chapter 3 extracts patterns of structural similarities between conversations about different topics and from various domains (e.g., bus schedules and restaurant reservations), while the approach to measuring semantic coherence from Chapter 4 focuses on the structure of relations between specific semantic concepts mentioned in a conversation that can allow to distinguish a topic shift, i.e., different conversations, within the same domain (Ubuntu).}

\added{Furthermore, the QRFA model revealed that (1) question answering is the key component essential for being able to define the information need and retrieve the answer; (2) often question answering alone is not sufficient to retrieve the answer, e.g., in the situations, when the information need is vague or undefined, or/and the information seeker is unfamiliar with the collection content. Therefore, in Chapter 5 we examine the current state of the art for question answering over knowledge graphs and contribute an original approach that alleviates the major bottle-neck of the existing systems for complex QA, namely query generation used for answer retrieval. Chapter 5 also heavily builds upon the empirical results discussed in Chapter 4: maintaining alternative interpretations associated with uncertainties from matching natural language to labels in a knowledge graph and the scalability limitations of the breadth-first search.
Finally, Chapter 6 provides the first insights on the promising direction of extending conversational search systems with browsing functionality, which does not require explicit query formulation and is designed to mimic alternative types of interactions captured in the QRFA model.}

In this thesis, we have reported on the results obtained from several complementary studies, to analyse different requirements and components to design advanced conversational search systems able to support advanced dialogue-based interactions, namely: 
\begin{itemize}
\item understanding the structure and processes behind conversations;
\item maintaining semantic coherence of conversations;
\item complex question answering;
\item conversational browsing.
\end{itemize}
We believe that all these requirements should be covered within a comprehensive conversational system.

\added{Some of the key contributions presented in this thesis also promise to spark important directions outside of the knowledge-based conversational search frame. The process mining approach for extracting a compact theoretical model from empirical data that we proposed can be applied more broadly for developing grounded theories by capturing patterns of similarity and variance in process data~\cite{langley1999strategies}. Existing process mining algorithms normally prioritize precision and recall over simplicity, which often results in producing ``spaghetti models'' that are able to fit the data present in the logs but are not interpretable by humans.}

\added{The message-passing approach introduced in Chapter 5 is an important step towards natural language interfaces for knowledge graphs. We show a viable alternative to translating a natural language question into a structured query. Our approach operates directly on the compressed version of the graph and is able to compute multiple alternative question interpretations in parallel via matrix multiplications. These results can hopefully provide inspiration for developing other algorithms that leverage knowledge graph structure in a scalable fashion without explicit query formulation.}

Design of conversational search systems requires integration of advances from several research disciplines, including natural language processing (NLP), information retrieval (IR), knowledge management (KM), and human-computer interaction (HCI).
We analysed the structure of information-seeking dialogues through the lens of cognitive load theory and showed how neural language models and indexed entity catalogs can be used to retrieve answers from a knowledge graph.
Our findings uncover the limitations of the state-of-the-art approaches and propose alternatives.
Below, we provide answers to the research questions posed in Chapter~\ref{chap:intro} and outline possible directions for future work.

\section{Main Findings}

In Chapters~\ref{chap:structure} and \ref{chap:coherence}, we analysed relations that hold between utterances in a conversation from two orthogonal perspectives -- in terms of a domain-independent functional role and domain-specific semantic coherence -- to derive conversational knowledge models that describe structural properties of a generic information-seeking dialogue.

Chapter~\ref{chap:structure} introduced conversation mining as a framework that allows us to collect and aggregate frequent patterns across conversational transcripts, treating each as an individual instance of a single underlying generative process shared between the conversation participants.
We used conversation mining to analyse the general structure of an information-seeking dialogue, and answer the following question:

\textbf{RQ 1} What is the general structure of an information-seeking dialogue?

Our results unveil the types of interactions typical for information-seeking dialogues, which we consider as functional requirements for a conversational search system.
We discovered a common structure that persists across information-seeking dialogues from different domains.
The QRFA model we proposed is simple and interpretable, describing the interaction patterns from a high level of abstraction.

In Chapter~\ref{chap:coherence} we described a novel evaluation setup for the semantic coherence task that tests the ability of a knowledge model, both vector embeddings and a knowledge graph, to predict concept relevance with respect to the conversation context.
We raised the following question:

\textbf{RQ 2} What are the relations between concepts mentioned in the course of a conversation and how can we detect them?

Our results show that proximity relations in a knowledge graph and a vector space are indicative of entities co-occurring in a conversation.
Adjacent entities that are 1 or 2 hops away are much more likely to occur within the same conversation than entities that are further away in the graph.
It is an important property that should inform the design of a dialogue system based on relations stored in a knowledge graph.
Our analysis revealed that the major source of errors when establishing semantic coherence using a knowledge graph are incorrect entity linking and relation sparsity.
We also showed that language models carry semantic relations that are missing from a knowledge graph.

Chapter~\ref{chap:qa} introduced a novel approach for question answering over knowledge graphs.
We addressed the following question:

\textbf{RQ 3} How to design a system able to answer complex questions using information stored in a knowledge graph?

Our approach to question answering incorporates three stages, question parsing, linking, and answer retrieval, that are based on approximate reasoning under the uncertainty inherent in natural language question interpretation.
We show that our approach is able to improve upon the state-of-the-art performance results, especially in terms of recall and computation speed, by leveraging efficient knowledge compression technologies and pre-trained language models for linking questions to a knowledge graph.
Our results also demonstrate that combining evidence from both structured (knowledge graphs) and unstructured (text) data sources has a great potential to mitigate knowledge sparsity, increase support and interpretability of semantic relations.

In Chapter~\ref{chap:browsing} we introduced a new conversational browsing task that releases users from the burden of having to formulate a precise search query when accessing an information source.
To come up with the conversational system design we organized a laboratory study to collect dialogue transcripts that exemplify human strategies in the context of an information-seeking task, in which an information provider assists an information seeker by enumerating available exploration directions.
Next, we systematically analysed the dialogue transcripts to formalise a conversational browsing model as a hypothesis for the basic structure and properties that the conversational system should have to support users with limited domain knowledge and collection understanding.
We verified the model assumptions and the proposed setup by running a user simulation and another laboratory study, in which human intermediaries were replaced by our system prototypes.

\textbf{RQ 4} How to design a conversational system able to support information retrieval without the need to explicitly formulate a search query?

In contrast with the question answering task, users assume a passive role by providing feedback to the requests from a conversational system.
According to the aforementioned settings and concepts from cognitive load theory the task is to partition the information model into coherent messages that should be communicated to the user interactively, i.e., in a sequence conditioned on the user input.
In the absence of a user model we proposed to base the optimality criteria on the structure of the information model, i.e. more informative concepts should be communicated first to reduce the average traversal time.
We showed that this type of interaction allows for a sufficient control of the exploration direction and discovery of the content of an information source.

Our long term goal is to enable conversational exploratory search via interactive storytelling.
The research work presented in this thesis achieves some important milestones in this direction.
Our findings provide a better understanding of the structure of human information-seeking dialogues.
We also discovered promising approaches for development and evaluation of knowledge-based conversational search systems.

\section{Future Work}

We believe that it is worthwhile to continue development of each of the proposed tasks (conversation mining, coherence measurement, question answering and conversational browsing) individually but also integrate them as components into a single dialogue system enabling a wide set of interactions to support conversational search.
It is also interesting to verify and further extend our results by applying the proposed frameworks to other conversational datasets and annotation types.

An end-to-end evaluation of a dialogue system remains a big challenge.
None of the approaches proposed in this thesis incorporates natural language generation (NLG) components, which is an important direction considered for the future work.
NLG evaluation requires a large number of human participants since automatic evaluation metrics, such as BLEU~\cite{DBLP:conf/acl/PapineniRWZ02} that is also used in machine translation, fall short in capturing synonyms and paraphrases~\cite{DBLP:conf/emnlp/LiuLSNCP16}.
Developing a good dialogue evaluation model is likely to be as hard as developing the dialogue agent itself~\cite{DBLP:conf/emnlp/LiuLSNCP16}.
In this situation user studies remain irreplaceable for understanding conversation success criteria.
To ensure the quality of empirically collected data it is crucial to establish realistic settings able to provide sufficient context for an interaction.
Experiments should be centered around a use case scenario that is particularly relevant for the study participants.
This setup can help to reproduce the kind of information-seeking behaviour, which occurs in natural environment and driven by a genuine information need.

To ensure maximum user engagement the future work towards interactive storytelling should ideally involve an everyday activity in which potential users may engage voluntarily, such as self-education and learning.
Wikipedia as a reference corpus and DBpedia or Wikidata as a knowledge graph are the largest collections available at the moment.
More data can also be harvested from on-line news sources or the cultural heritage domain.

The next steps include development of the three consecutive stages: story (narrative) composition, story generation and interactive storytelling.
Story composition involves producing a sequence of concepts as the building blocks of a plot (storyline).
Then, a plot as an abstract representation can be used to generate text of the story.
Finally, interactive storytelling will produce dialogues in which the story composition and generation are conditioned on the user input.

We believe that approaches presented in this thesis, such as adversarial dialogue generation and message passing in a knowledge graph, can be further adopted for the interactive storytelling subtasks.
This line of research will also require more work on knowledge integration from both structured and unstructured data sources.
In particular, machine reading and summarisation approaches are important to harvest knowledge from textual documents.

Conversational search interfaces should support interactive information retrieval functionality spanning across multiple dialogue turns, similar to the question negotiation process, which helps both the system and the user to better understand the collection content and the information need.
Other relevant subtasks that were not considered in the scope of this thesis include sequential (conversational) question answering and question generation.

Another important element that we did not discuss yet in the context of this thesis is continuous learning and updating the conversational model.
It is crucial to develop mechanisms that allow a conversational system to correct and further extend its behaviour model based on interactions with its users.
These updates should not be limited to extending only the knowledge model component of the system by adding new or missing concepts and relations between them but also allow the system to discover and adopt new ways and modes of interaction altogether.









\backmatter

\bibliographystyle{plainnat}
\bibliography{refs}

\end{document}